\documentclass{aastex701}

\usepackage{graphicx}
\usepackage{booktabs}
\usepackage{array}
\usepackage{placeins}
\usepackage{makecell}
\providecommand{\tightlist}{\setlength{\itemsep}{0pt}\setlength{\parskip}{0pt}}
\newcommand{\papertablesize}{\footnotesize}
\newcommand{\tblleft}[2]{\parbox[t]{#1}{\raggedright #2}}
\newcommand{\papertablenote}[1]{\vspace{2pt}\par\begingroup\footnotesize\noindent\textit{Note.} #1\par\endgroup}

\graphicspath{{./figures/}}
\shorttitle{Extinction-calibrated QSO Candidates}
\shortauthors{Cao et al.}

\begin{document}

\title{A Gaia-linked High-purity QSO Candidate Catalog in Selected Fields with Extinction-binned Calibration and Spectrum-informed Training}

\author[orcid=0000-0002-9494-0946,gname=Zi-Huang,sname=Cao]{Zi-Huang Cao}
\affiliation{National Astronomical Observatories, Chinese Academy of Sciences, Beijing 100012, China}
\affiliation{University of Chinese Academy of Sciences, Beijing 100049, China}
\email{zhcao@nao.cas.cn}

\author[orcid=0000-0002-8402-3220,gname=Zhao-Xiang,sname=Qi]{Zhao-Xiang Qi}
\affiliation{Shanghai Astronomical Observatory, Chinese Academy of Sciences, Shanghai 200030, China}
\affiliation{University of Chinese Academy of Sciences, Beijing 100049, China}
\email{zxqi@shao.ac.cn}

\author[orcid=0000-0003-3243-464X,gname=Juan-Juan,sname=Ren]{Juan-Juan Ren}
\affiliation{National Astronomical Observatories, Chinese Academy of Sciences, Beijing 100012, China}
\affiliation{University of Chinese Academy of Sciences, Beijing 100049, China}
\email{jjren@nao.cas.cn}

\author[orcid=0000-0003-1353-9040,gname=Bo,sname=Zhang]{Bo Zhang}
\affiliation{Shanghai Astronomical Observatory, Chinese Academy of Sciences, Shanghai 200030, China}
\email{zb@shao.ac.cn}

\author[orcid=0000-0002-8669-5370,gname=Dongwei,sname=Fan]{Dongwei Fan}
\affiliation{National Astronomical Observatories, Chinese Academy of Sciences, Beijing 100012, China}
\affiliation{University of Chinese Academy of Sciences, Beijing 100049, China}
\affiliation{National Astronomical Data Center, Beijing 100101, China}
\email{fandongwei@nao.cas.cn}

\author[orcid=0000-0002-9346-0211,gname=Shi-Long,sname=Liao]{Shi-Long Liao}
\affiliation{Shanghai Astronomical Observatory, Chinese Academy of Sciences, Shanghai 200030, China}
\affiliation{University of Chinese Academy of Sciences, Beijing 100049, China}
\email{shilongliao@shao.ac.cn}

\author[orcid=0009-0006-3524-105X,gname=Yuzhou,sname=Wang]{Yuzhou Wang}
\affiliation{National Astronomical Observatories, Chinese Academy of Sciences, Beijing 100012, China}
\email{ngc224@mail.ustc.edu.cn}

\author[orcid=0000-0001-5298-2833,gname=Yong-Heng,sname=Zhao]{Yong-Heng Zhao}
\affiliation{National Astronomical Observatories, Chinese Academy of Sciences, Beijing 100012, China}
\affiliation{University of Chinese Academy of Sciences, Beijing 100049, China}
\email{yzhao@bao.ac.cn}

\author[orcid=0000-0003-2179-3698,gname=Yong,sname=Zhang]{Yong Zhang}
\affiliation{National Astronomical Observatories, Chinese Academy of Sciences, Beijing 100012, China}
\affiliation{University of Chinese Academy of Sciences, Beijing 100049, China}
\email{zhangyong@bao.ac.cn}

\author[orcid=0009-0009-6931-2276,gname=Meng-Xin,sname=Wang]{Meng-Xin Wang}
\affiliation{National Astronomical Observatories, Chinese Academy of Sciences, Beijing 100012, China}
\email{mxwang@nao.cas.cn}

\author[orcid=0000-0002-3143-9337,gname=Yihan,sname=Tao]{Yihan Tao}
\affiliation{National Astronomical Observatories, Chinese Academy of Sciences, Beijing 100012, China}
\email{y.tao@nao.cas.cn}

\author[orcid=0000-0002-0747-0078,gname=Gao-Yuan,sname=Zhang]{Gao-Yuan Zhang}
\affiliation{Zhejiang Lab, Hangzhou 311121, China}
\email{gaoyuan.zhang@zhejianglab.org}

\author[gname=Yong,sname=Yu]{Yong Yu}
\affiliation{Shanghai Astronomical Observatory, Chinese Academy of Sciences, Shanghai 200030, China}
\affiliation{University of Chinese Academy of Sciences, Beijing 100049, China}
\email{yuy@shao.ac.cn}

\author[orcid=0000-0001-7865-2648,gname=A-Li,sname=Luo]{A-Li Luo}
\affiliation{National Astronomical Observatories, Chinese Academy of Sciences, Beijing 100012, China}
\affiliation{University of Chinese Academy of Sciences, Beijing 100049, China}
\email[show]{lal@nao.cas.cn}
\correspondingauthor{A-Li Luo}

%

\begin{abstract}
We present an extinction-calibrated, Gaia-source-level QSO candidate catalog for selected fields, designed as a high-purity input catalog for fiber-spectroscopic follow-up rather than as an all-sky QSO census. The deployed selector uses Gaia astrometry and photometry together with optical/infrared catalog features and \(E(B-V)\)-binned threshold calibration; spectra enter only during training through a source-grouped spectrum-teacher model. The sample definition is layered: a four-field core domain ladder provides the main validation baseline, four application/stress-test fields probe portability, and COSMOS is treated separately as an Extreme Deep boundary case. At the recommended conservative operating point, calibrated to validation-set purity 0.98, the P3 spectrum-informed catalog selector reaches measured test-set purity 0.9809 and spectroscopic-label completeness 0.8869 within the frozen Gaia-linked benchmark, whereas the Gaia official QSO probability yields spectroscopic-label completeness of 0.4493 under the same threshold protocol. The evaluation protocol excludes all downstream validation/test Gaia \texttt{source\_id} values from teacher fitting and checkpoint selection, and teacher probabilities are used only for downstream training rows. Relative to the earlier P2 teacher, P3 yields a modest mean completeness gain across five seeds, with a small decrease in purity and a small increase in false positives; the gain is most evident in several higher-extinction and faint-source diagnostics. The released product is structured as a catalog and empirical selection-function data product with source identifiers, field-layer assignments, input-coverage flags, calibrated scores, threshold flags, validation metadata, and provenance/QC fields. In COSMOS, the Gaia-linked parent set is much shallower than COSMOS2020; the robust 39-object subset is therefore interpreted as a purity-oriented priority list, not a completeness measurement.

\end{abstract}

\keywords{\uat{Quasars}{1319} --- \uat{Catalogs}{205} --- \uat{Sky surveys}{1464} --- \uat{Astronomy data analysis}{1858} --- \uat{Classification}{1907} --- \uat{Photometry}{1234}}

\section{Introduction}\label{introduction}

Quasar candidate selection is a central requirement in modern survey astronomy. High-purity QSO samples support spectroscopic follow-up, large-scale-structure analyses, time-domain searches, obscured and high-redshift AGN studies, and the construction of well-characterized candidate catalogs for fields where spectroscopy is incomplete. Large QSO catalogs such as DESI quasar target samples, Gaia-unWISE/Quaia, and ZTF-based catalogs demonstrate the scientific reach of wide-area QSO selection, from spectroscopic targeting to clustering and time-domain studies \citep{chaussidon_2023,storey_fisher_2024,nakoneczny_2025}. The practical requirement is therefore not simply to assign a class label. A selector for follow-up or catalog construction must define an operating point, expose its purity/completeness trade-off, and diagnose how the rule changes with extinction, Galactic latitude, magnitude, source density, and contaminant population.

Historically, this requirement has driven the field from explicit color-box target selection to probabilistic and multi-survey classification. The SDSS quasar target-selection and photometric-quasar catalogs established the use of optical color space, morphology, radio matches, and training-set density estimates for large-area QSO candidate selection \citep{richards_2002,richards_2004}. The XDQSO/XDQSOz line of work then recast selection as a probabilistic generative problem in flux space and made the color degeneracies between quasars and stars explicit \citep{bovy_2011,bovy_2012}. BOSS and eBOSS target selection further demonstrated that a survey-ready quasar catalog is shaped by science requirements, spectroscopic efficiency, redshift distributions, and selection-function control rather than by classification accuracy alone \citep{ross_2012,myers_2015}. These classical baselines remain relevant because the main statistical tension is unchanged: selecting more QSOs inevitably exposes the catalog to stellar, galaxy, reddening, and photometric-systematic contaminants.

The same statistical tension is reinforced by the astrophysics of quasars, which argues against relying on a single observable family. QSO colors depend on continuum slope, broad emission lines moving through filters, IGM absorption, host-galaxy light, and dust reddening, producing redshift-dependent overlaps with stellar loci and compact galaxies \citep{richards_2001,vanden_2001,krawczyk_2013,hickox_2018}. Mid-infrared data from WISE, unWISE, and CatWISE provide strong AGN contrast and have enabled large infrared-selected AGN/QSO catalogs \citep{wright_2010,lang_2014,secrest_2015,schlafly_2019,marocco_2021}. Time-domain work from Stripe 82, damped-random-walk variability modeling, and ZTF shows that variability is another powerful quasar signature \citep{kelly_2009,macleod_2010,schmidt_2015,nakoneczny_2025}. However, infrared blending, host-dominated or obscured AGN, limited multi-epoch coverage, and survey-dependent depth mean that color, infrared, morphology, and variability must be treated as complementary evidence rather than interchangeable solutions.

Against this multi-observable background, the relevant comparison baseline is now more demanding than the historical color-selection problem. DESI quasar target selection uses Legacy imaging, WISE, and spectroscopic validation at survey scale, building on the DESI survey design and Legacy Surveys imaging system \citep{desi_2016,dey_2019,chaussidon_2023}. Gaia DR3 provides official extragalactic and non-stellar classification products, including QSO and galaxy probabilities, through the Gaia DR3 extragalactic and Apsis/DSC classification products \citep{gaia_2022,delchambre_2022}. Gaia-CatWISE and Gaia-unWISE catalogs show that Gaia astrometry/photometry combined with infrared data can produce powerful all-sky QSO samples and cosmology-ready catalogs \citep{hughes_2022,storey_fisher_2024}. Recent machine-learning and representation-learning searches, including high-redshift Legacy+WISE selection, few-shot type-II QSO classification, contrastive high-redshift candidate discovery, and large ZTF variability catalogs, further show that QSO selection is now a multimodal candidate-ranking and selection-function problem \citep{ye_2024,byrne_2024,nakoneczny_2025}. In this context, comparison with a simple color-selection baseline is insufficient; Gaia official QSO probability is treated below as a strong external reference.

The remaining gap addressed here is candidate-realistic, domain-aware selection for selected fields where the all-sky operating point is not necessarily optimal. Low and intermediate Galactic latitudes, anti-center structure, variable extinction, crowded stellar foregrounds, faint-end photometric noise, and compact galaxy-like contaminants can alter both purity and completeness. Dust maps and reddening recalibrations have long shown that optical selection is spatially structured rather than uniform \citep{schlegel_1998,schlafly_2011,planck_2014}. Anti-center and outer-disk structures, together with molecular-cloud fields such as Taurus/Perseus and Orion/Monoceros, provide physically motivated stress tests for foreground complexity \citep{dame_2001,penarrubia_2007,newberg_2010,xu_2015}. These effects are especially relevant when the goal is to deliver a candidate list for follow-up rather than to publish a fully all-sky, selection-function-limited catalog. A global threshold on an otherwise strong probability score can be too conservative in some domains and too permissive in others.

To keep this scope explicit, several operational terms used below are defined in an astronomical catalog context. A ``candidate-realistic'' or ``deployable'' selector is one that can be applied before a new spectrum is taken: it may use Gaia astrometry and photometry, optical/infrared measurements, variability summaries, field metadata, and quality flags, but it may not use the target's future spectrum or Gaia's official QSO classifier probability as an inference-time feature. ``Source-grouped'' means that the Gaia \texttt{source\_id}, rather than an individual spectrum, catalog row, or cross-match entry in an auxiliary table, is the statistical unit for matching, splitting, teacher construction, and evaluation. This choice reflects a common survey-data issue: the same astrophysical source may have repeated spectra or multiple catalog matches, and treating those rows as independent would overstate both the training evidence and the test performance. A ``spectrum teacher'' is therefore a training-time use of existing SDSS/DESI spectra to provide a softer supervisory signal; it is not a data requirement for future candidates. Finally, an ``operating point'' is the adopted score threshold, described by purity, completeness, false positives, false negatives, and target density rather than by a ranking metric alone.

With these definitions fixed, the exploration strategy becomes a sequence of selection-function tests. The field ladder asks whether one source-level rule remains stable as Galactic latitude, extinction, source density, and foreground population change. The comparison with Gaia asks whether the proposed catalog score recovers additional spectroscopically confirmed QSOs beyond a Gaia-only reference selection, under the same threshold protocol. The \(E(B-V)\)-binned thresholds ask whether the same score needs different decision boundaries in different reddening regimes. The contaminant and morphology analyses then ask where the remaining false positives enter, and whether they can be controlled without removing too many true QSOs. In this framework, the machine-learning components are used to build and audit an astronomical selection function, not to replace the sample-definition and validation logic.

We therefore test a deliberately constrained hypothesis: a deployable multimodal QSO selector can use spectroscopic information during training to improve over a calibrated Gaia-only QSO selection at matched high purity, while retaining only catalog-level observables at inference. We build a fixed, source-grouped benchmark over a domain ladder spanning low-extinction high latitude, higher-extinction anti-center, and intermediate-extinction bridge fields. Spectroscopic labels from SDSS, DESI, and LAMOST provide the training and validation backbone, while Gaia, Legacy, WISE/unWISE, and CatWISE-like catalog observables form the candidate-realistic feature space \citep{sdss_2022,desi_2025,lamost_2015,yao_2022,dey_2019,wright_2010,schlafly_2019,marocco_2021}. Cross-survey matches are handled at the Gaia \texttt{source\_id} level, and repeated spectra are resolved with a union-best policy so that multiply observed sources do not dominate the training or evaluation. This source-level design follows the broader astronomical cross-identification and model-evaluation principle that the statistical unit of splitting must match the scientific object being classified \citep{budavari_2008,raschka_2018}.

To implement this hypothesis, the model design follows the knowledge-distillation framework but keeps the deployed inference path simple \citep{hinton_2015}. Spectra are used as training-time teachers; the final student uses candidate-realistic catalog features and does not require spectra at inference. This choice is motivated by recent progress in astronomical multimodal learning, including AstroM3 and large-scale multimodal astronomy datasets, while avoiding the operational cost of requiring spectra or image-level products for every future candidate \citep{astrom_2024,multimodal_2024}. Gaia official classifier probabilities are excluded from the student feature matrix and are used only as an external reference calibrated under the same validation/test protocol, ensuring that the Gaia comparison remains independent of the proposed model.

For the gain tests below, the labels P2 and P3 are used only to distinguish two frozen spectrum-teacher coverage stages. P2 is the earlier priority-complete, source-grouped union-best teacher used as the v0.3 stable baseline. P3 is the later expanded teacher built after the DESI domain-ladder spectra were completed and merged with SDSS under the same union-best source policy. In the v0.3.2 fixed-split evaluation, both versions use the same downstream catalog-student architecture, the same candidate-realistic inference inputs, the same frozen benchmark split, the same validation-derived threshold protocol, and the same exclusion of downstream validation/test source IDs from teacher fitting. Therefore, a P3-P2 comparison measures the effect of expanding and refreezing the training-time spectrum teacher, not the effect of adding spectra, Gaia classifier probabilities, test-set teacher targets, or a new feature family at candidate-scoring time.

The resulting benchmark is designed for transparency. Thresholds are learned on the validation set and evaluated on the frozen test set. The primary calibration uses three \(E(B-V)\) bins, motivated by the role of reddening in optical selection and by the need for simple subgroup-aware operating points \citep{schlegel_1998,schlafly_2011}. Studies of probability calibration and threshold fairness warn that a single global threshold can produce group-dependent operating points, while overly flexible post-hoc calibration can be unstable in limited data regimes \citep{guo_2017,kleinberg_2017,pleiss_2017,woodworth_2020,ding_2024}. We therefore use \(E(B-V)\)-binned threshold calibration as the main method and retain Galactic latitude and sky region as diagnostic axes. Robustness is evaluated through fixed-benchmark seed repeats, bootstrap confidence intervals, and subgroup checks, following standard model-evaluation practice \citep{raschka_2018,wager_2022}. More speculative extensions, such as morphology-aware hard-negative reranking, are treated as Discussion material and are interpreted with Gaia IPD/PSF-LSF and Legacy Tractor morphology references rather than promoted directly into the main selection rule \citep{gaia_2020,rowell_2020,dey_2019}.

This design yields three contributions. First, we provide a candidate-realistic comparison between Gaia official QSO probabilities and multimodal students under matched high-purity operating points. Second, source-grouped spectrum-teacher expansion gives a modest improvement in the stricter high-purity regime on the frozen benchmark, with the largest measured gains in several higher-extinction and faint-end diagnostics rather than uniformly across all fields. Third, we characterize the remaining high-purity limitations: stellar contaminants account for much of the sky-transfer failure mode, while compact galaxy-like sources remain a persistent high-purity tail, motivating Gaia IPD and Legacy Tractor morphology as future or optional filters rather than as the default selection rule.

Finally, the scope is intentionally conservative. We do not infer an all-sky unbiased QSO density field, nor do we treat Gaia probabilities as ground truth. The benchmark tests whether a deployable selector can recover more spectroscopically confirmed QSOs than Gaia at comparable purity in controlled science and stress-test fields. COSMOS is added with a different evidentiary role: it is an Extreme Deep layer used to test whether the Gaia-linked catalog can provide a high-priority follow-up list beyond the calibrated benchmark domain. The resulting product is therefore both a high-purity candidate catalog and a documented selection rule: field membership, score provenance, threshold policy, validation subset, and quality-control flags are kept explicit so that users can reproduce the recommended conservative operating point or choose the auxiliary denser flag for follow-up. Section \ref{data-and-field-selection} defines the input surveys, reference catalogs, and field layers used for the present catalog sample. Section \ref{catalog-construction} describes the source-grouped catalog-construction pipeline, including cross-matching, label construction, spectrum-teacher supervision, distillation, and calibrated selection flags. Section \ref{performance} evaluates the resulting selection against spectroscopic labels and Gaia's official QSO classifier, with emphasis on high-purity operating points and field-dependent behavior. Section \ref{catalog-description} describes the released catalog products and recommended usage. Section \ref{applications} illustrates applications in selected fields, fiber-spectroscopic target preparation, and the COSMOS Extreme Deep case. Section \ref{systematics} summarizes the main systematics, contaminants, capability boundaries, and limitations that define the present selection function; Section \ref{discussion} discusses the broader implications and future extensions; and Section \ref{summary} summarizes the main results.

\section{Data and Field Selection}\label{data-and-field-selection}

The astronomical inputs and field-selection design are defined before the algorithmic preprocessing because the scientific interpretation depends on the provenance of each measurement and on the role assigned to each sky region. For this reason, standard survey products, value-added literature catalogs, and controlled evaluation fields are described separately before the model construction is introduced.

\subsection{Catalog Selection Rationale}\label{catalog-selection-rationale}

The catalog set was chosen to support the candidate-realistic definition above: every feature used by the deployed student must be available for objects without spectra, while spectroscopy is reserved for labels and training-time supervision. Gaia DR3 provides the object backbone, astrometry, broad-band photometry, and the official QSO-probability reference used for external comparison \citep{gaia_2022,delchambre_2022}. Optical and infrared survey products provide the color, depth, morphology, and masking information required to separate quasars from stars and compact galaxies, following the long history of SDSS color selection, the XDQSO flux-space framework, and WISE-based AGN selection \citep{richards_2002,bovy_2011,bovy_2012,wright_2010,dey_2019}. Spectroscopic surveys provide the supervised reference labels and the spectra used by the teacher model, but spectra are not required at inference.

This rationale defines the data model used throughout the paper. Standard survey releases provide the reproducible measurement system and the spectroscopic label backbone. Published value-added catalogs provide external context, candidate-priority information, literature comparison sets, and contaminant diagnostics; they are not treated as ground truth for the main QSO-vs-non-QSO split unless an accepted spectroscopic label is available. This separation fixes the provenance of each input quantity and prevents literature classifications from being folded into the training labels without explicit spectroscopic confirmation.

\subsection{Standard Survey Releases}\label{standard-survey-releases}

With the catalog roles established, Gaia DR3 provides the source-level coordinate and identity frame. The current v1.0 field extraction contains 22,988,081 Gaia DR3 rows across the eight active fields, with the largest source densities in the lower-latitude anti-center, Taurus/Perseus, and Orion/Monoceros fields. Gaia's official QSO probability is retained only as an external reference, not as a student feature.

Spectroscopic supervision comes from SDSS DR18, DESI DR1, and LAMOST DR10. The current-field SDSS DR18 spectroscopic extraction is complete at 800/800 planned tiles, with 687,175 rows and no failed tiles in the data-readiness status. The auxiliary DESI DR1 zcatalog table contributes 2,281,195 rows in the active fields, and the auxiliary LAMOST DR10 low-resolution table contributes 1,409,958 rows. These surveys are complementary: SDSS and DESI provide deep extragalactic coverage in several high- and mid-latitude regions, while LAMOST contributes dense stellar and contaminant information in low-latitude and anti-center fields.

Because the supervised backbone combines multiple surveys, the SDSS-DESI merge is used deliberately rather than as a simple concatenation. DESI supplies the larger modern spectroscopic sample, broad target-selection coverage tied to Legacy imaging, and many high-quality spectra in fields relevant to the present benchmark. SDSS contributes an independent historical selection function, well-tested quasar and stellar classifications, and overlap with earlier QSO-targeting work. Combining the two therefore increases source-level label coverage, widens the range of magnitudes and colors represented in the teacher cache, and provides overlap checks for repeated or conflicting classifications. The trade-off is that the merged sample inherits two different target-selection functions, reduction pipelines, wavelength responses, quality flags, and class-assignment conventions. The merged spectroscopic set is therefore treated as a supervised training and validation backbone, not as an unbiased census of the QSO population.

Candidate-observable photometry and morphology come from Legacy Surveys DR9 and WISE-family products. Legacy DR9 Tractor/sweep products provide optical fluxes, inverse variances, masks, depths, and morphology-like quantities \citep{dey_2019}. WISE, unWISE, CatWISE, and WISE-SCoS-like products provide mid-infrared and optical-infrared information that is known to be highly informative for AGN/QSO selection \citep{wright_2010,lang_2014,schlafly_2019,marocco_2021}. In the current-field summary, the auxiliary WISE-SCoS table contains 1,595,457 rows. Legacy DR9 sweep products are validated separately with complete-marker and quarantine checks before they are promoted into the final feature table.

\subsection{Published Reference and Value-added Catalogs}\label{published-reference-and-value-added-catalogs}

Beyond the standard releases, the current auxiliary-catalog build includes published reference and value-added catalogs. These products do not constitute a single homogeneous data source. They fall into four roles: an all-sky optical--infrared photometric-redshift context catalog, a variability-based QSO/RR Lyrae catalog, a high-redshift QSO-candidate catalog, and a LAMOST spectroscopic QSO reference catalog. We keep them separate because each catalog encodes a different selection function, depth, contaminant population, and intended use. In the main benchmark, these value-added classifications are treated as external context unless the same Gaia source also has an accepted spectroscopic label in the frozen supervised sample.

Within this set, WISE-SCoS provides the broad optical--infrared context. The WISE x SuperCOSMOS photometric-redshift catalog introduced by \citet{bilicki_2016} cross-matches WISE mid-infrared photometry with SuperCOSMOS optical photographic measurements to build a large-area galaxy sample over roughly three quarters of the sky, with photometric redshifts and infrared--optical colors useful for large-scale-structure and extragalactic source studies. In the present field extraction, the auxiliary WISE-SCoS table contains 1,595,457 rows. We use it as a source of candidate-observable context and as a galaxy-like comparison population: it helps indicate whether a selected object lies in a region of optical--infrared color and photometric-redshift space populated by galaxies or AGN-like extragalactic sources. It is not used as a QSO truth catalog, and its photometric redshift is not used to replace spectroscopic redshift in the supervised benchmark.

The time-domain comparison axis is supplied by the northern-hemisphere QSO/RR Lyrae catalog. \citet{hernitschek_2016} used Pan-STARRS1 3pi sparse multi-epoch (grizy) photometry, WISE color information, and structure-function variability features to classify variable point sources as likely QSOs or RR Lyrae. The catalog records quantities such as the variability significance \texttt{chi2\_hat}, the best-fit variability amplitude and timescale parameters \texttt{omega\_best} and \texttt{tau\_best}, PS1 mean magnitudes, the WISE \texttt{W1-W2} color, and the probabilities \texttt{p\_QSO} and \texttt{p\_RRLyrae}. Outside the Galactic plane, that work reports QSO/RR Lyrae samples with purity of about 75\% and completeness of about 92\%. In our current fields, the clipped NH QSO/RR auxiliary table contains 2,138,209 rows. Its main value here is diagnostic: agreement with high \texttt{p\_QSO} supports consistency with a variability-selected QSO population, while high \texttt{p\_RRLyrae} flags a physically distinct and observationally challenging contaminant class for low-latitude or time-variable candidates.

The high-redshift comparison role is filled by the Polsterer QSO-candidate catalog. \citet{polsterer_2013} targeted high-redshift quasars, especially (z\textgreater4.8), using SDSS DR6 photometry, a nearest-neighbor photometric-redshift estimator, and a multi-stage nearest-neighbor classifier designed to reject lower-redshift objects and cool stars. The released catalog includes the SDSS object identifier, position, estimated redshift, redshift uncertainty, stage-wise classifier ratios, and SDSS \texttt{psfmag\_u} through \texttt{psfmag\_z}. In their spectroscopic test, the method recovered 75 of 147 known (z\textgreater4.8) quasars and misclassified 34 of 32,210 known cool stars as quasars; the final released list contains 121,909 high-redshift QSO candidates, with an estimated detection performance of about 50\%. Our clipped Polsterer auxiliary table contains 17,413 rows. We use it to test whether the present Gaia--multiwavelength selector overlaps an older SDSS-color high-redshift selection function and to identify candidates that may be scientifically interesting despite not being optimized for the full-redshift, high-purity operating point adopted here.

The local spectroscopic-QSO check comes from the LAMOST phase-I QSO reference catalog. The parent LAMOST quasar catalog contains quasars selected from the LAMOST DR1--DR5 phase-I regular surveys and is documented through the LAMOST quasar-survey series: DR1 in Ai et al.~(2016, AJ, 151, 24), DR2--DR3 in Dong et al.~(2018, AJ, 155, 189), and DR4--DR5 in Yao et al.~(2019, ApJS, 240, 6). It includes spectroscopic quantities such as visual-inspection redshift, warning flags, signal-to-noise summaries, broad-line measurements, and BAL-related flags. The current-field auxiliary table contains 7,245 rows. In this paper it is used mainly as a known-QSO recovery and consistency check in LAMOST-accessible sky regions, complementing the larger LAMOST DR10 spectroscopic backbone described above. It does not alter the frozen SDSS/DESI/LAMOST benchmark definition; any promotion of additional LAMOST-reference labels is handled through the same source-grouped duplicate and conflict checks as the other spectroscopic inputs.

Together, these auxiliary catalogs define comparison axes rather than primary labels. WISE-SCoS tests galaxy-like optical--infrared context, NH QSO/RR tests variability and RR Lyrae contamination, Polsterer tests a high-redshift SDSS-color selection tradition, and the LAMOST phase-I catalog tests recovery of known LAMOST quasars. They provide external checks on the candidate catalog, but they do not define the primary QSO-vs-non-QSO training labels unless a source also satisfies the accepted spectroscopic-label rules used for the frozen benchmark.

\subsection{Field Taxonomy, Sample Definition, and Current-field Coverage}\label{field-taxonomy-sample-definition-and-current-field-coverage}

With the input roles fixed, the field selection is organized as a purpose-driven design rather than as a simple list of sky regions. The core domain ladder contains four controlled fields: a low-extinction high-latitude field, two intermediate-extinction bridge/northern fields, and a higher-extinction anti-center field. Its primary role is validation: it tests whether the P3 gain is stable across a deliberately ordered latitude and extinction sequence. The application/stress-test layer contains NEP/WISE, Taurus/Perseus, DESI-SDSS mid-latitude, and Orion/Monoceros fields. These fields are weighted toward stress testing, because they probe scientifically interesting regions with less uniform spectroscopic labeling and more heterogeneous foreground or survey conditions.

COSMOS requires a separate third role, which we call the Extreme Deep layer. Its depth, wavelength coverage, and external AGN/QSO resources extend beyond the calibrated Gaia-linked benchmark domain. It is therefore not folded into the core P3-vs-P2 evidence chain; instead, it tests whether the catalog score identifies a conservative Gaia-visible subset with independent X-ray, radio, spectroscopic, or deep-photometric support in a classical extragalactic deep field.

The layered design gives each field group an explicit role in the evidence chain: the core ladder supports the robustness claim, the application fields test portability under more difficult conditions, and COSMOS tests the boundary of a Gaia-linked selector in an extreme-depth multiwavelength regime. Table \ref{tab:field-layer-design} summarizes these roles. The \texttt{Layer} and \texttt{Representative\ fields} columns define the field taxonomy used throughout the paper. The validation, stress-test, and extrapolation columns indicate the intended weight of each layer in the present analysis; these entries are design priorities rather than measured performance rankings. The final column states how each layer enters the evidence chain, and is the main column to consult when interpreting which fields support the benchmark claim, which fields test application behavior, and which field defines the current extrapolation boundary.

\begin{table}[htbp]
\centering
\caption{Field-layer design.}
\label{tab:field-layer-design}
\papertablesize
\setlength{\tabcolsep}{2.5pt}
\renewcommand{\arraystretch}{1.20}
\begin{tabular*}{0.98\textwidth}{@{\extracolsep{\fill}}llllll@{}}
\toprule
Layer & \tblleft{1.55in}{Representative fields} & \makecell[c]{Vali-\\dation} & \makecell[c]{Stress\\test} & \makecell[c]{Extra-\\polation} & \tblleft{1.75in}{Main role in this paper} \\
\midrule
Core & \tblleft{1.55in}{High-lat.; mid-ext. north; mid-ext. bridge; anti-center} & High & Medium & Low & \tblleft{1.75in}{P3-vs-P2 robustness across the domain ladder.} \\
Application & \tblleft{1.55in}{NEP/WISE; Taurus/Perseus; DESI-SDSS mid; Orion/Monoceros} & Medium & High & Medium & \tblleft{1.75in}{Portability and stress test in complex fields.} \\
Extreme Deep & COSMOS & Low & High & High & \tblleft{1.75in}{Gaia-linked boundary test; high-purity priority list.} \\
\bottomrule
\end{tabular*}
\papertablenote{Entries in the validation, stress-test, and extrapolation columns are design priorities for the present analysis, not measured performance ranks.}
\end{table}

To make the layer design quantitative, the current-field extraction contains 138,362 spectroscopically labeled Gaia-linked objects in the active fields, including 22,390 QSOs and 115,972 non-QSOs. The labeled sample is intentionally uneven across fields: NEP/WISE and the DESI-SDSS mid-latitude field already have usable supervised coverage, whereas Taurus/Perseus and Orion/Monoceros currently have no supervised label set in the first-analysis table and are therefore treated as candidate-catalog stress tests rather than performance-evaluation fields. Table \ref{tab:field-layer-evidence-summary} gives the same design in numerical form. The core domain ladder has fewer Gaia sources than the application layer but a higher source-grouped label coverage rate, 1.11\% versus 0.36\%, and a higher QSO fraction among labels, 18.0\% versus 13.5\%. Accordingly, the core layer carries the main validation weight. COSMOS has a much smaller Gaia-linked denominator, 7,968 sources over about 2 deg\(^2\), and its benchmark-label columns are not comparable to the large-field layers; its evidence comes from the deep-field resources discussed below.

\begin{table}[htbp]
\centering
\caption{Field-layer sample statistics.}
\label{tab:field-layer-evidence-summary}
\papertablesize
\setlength{\tabcolsep}{2.4pt}
\renewcommand{\arraystretch}{1.18}
\begin{tabular*}{0.98\textwidth}{@{\extracolsep{\fill}}lccccccc@{}}
\toprule
Layer & Fields & Area & \makecell[c]{Gaia\\sources} & \makecell[c]{Labeled\\sources} & \makecell[c]{QSO\\labels} & \makecell[c]{Label\\coverage} & \makecell[c]{QSO\\fraction} \\
\midrule
Core domain ladder & 4 & 1629.7 deg$^2$ & 7,311,366 & 81,347 & 14,677 & 1.11\% & 18.0\% \\
\tblleft{1.15in}{Application/ stress-test} & 4 & 1632.4 deg$^2$ & 15,676,715 & 57,015 & 7,713 & 0.36\% & 13.5\% \\
\tblleft{1.15in}{Extreme Deep (COSMOS)} & 1 & 2.0 deg$^2$ & 7,968 & -- & -- & -- & -- \\
\bottomrule
\end{tabular*}
\papertablenote{Label coverage is the fraction of Gaia-linked sources with accepted source-grouped spectroscopic labels in the frozen benchmark. Dashes in the COSMOS row mean not applicable to the frozen benchmark; COSMOS external support is summarized in Sections \ref{cosmos-as-an-extreme-deep-multimodal-testbed} and \ref{capability-boundaries-and-the-cosmos-deep-field-layer}.}
\end{table}

\begin{figure}[htbp]
\centering
\includegraphics[width=\textwidth]{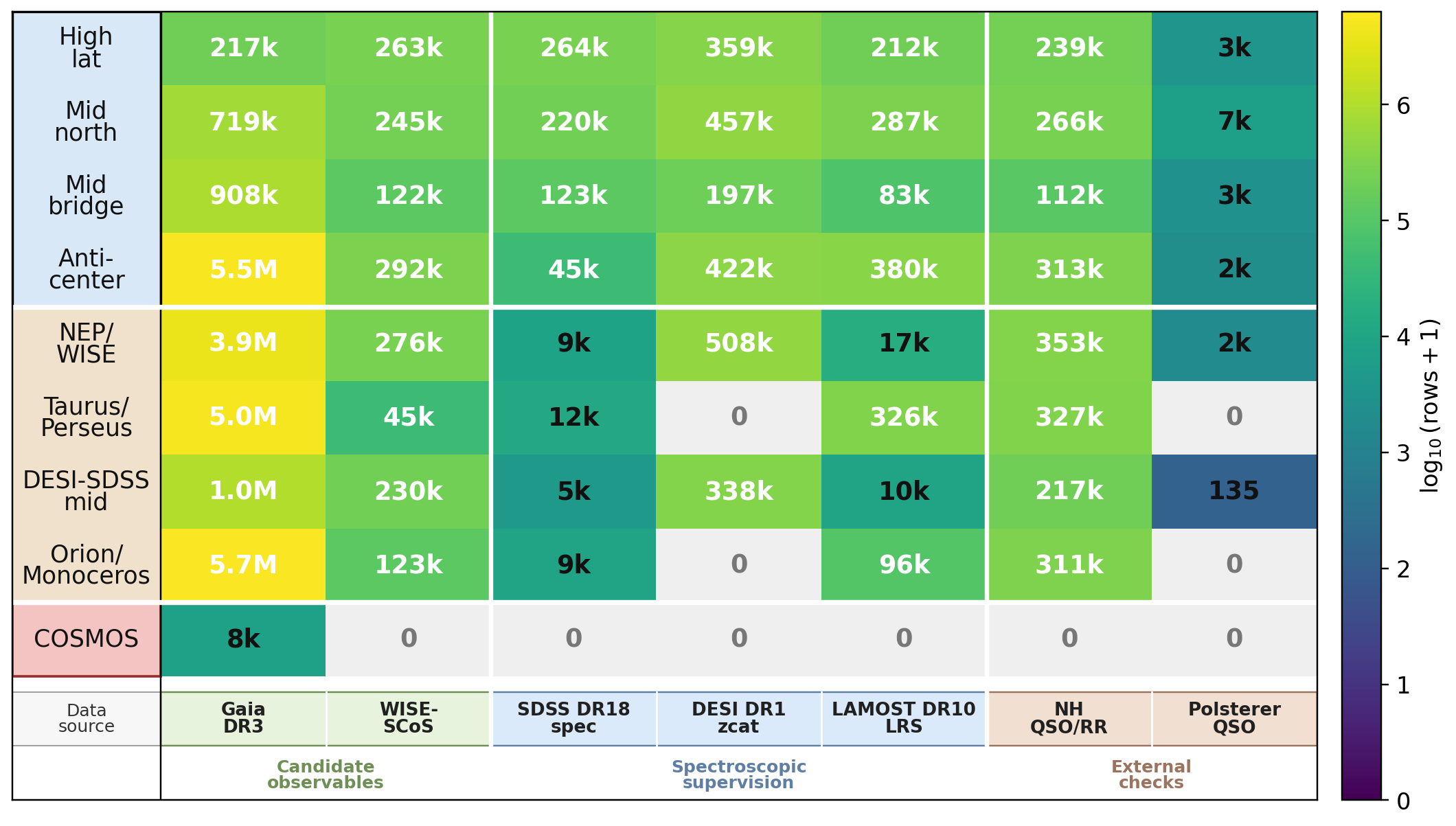}
\caption{Catalog coverage by field and source product for the current field-layer design. Rows are grouped by the layer definitions in Table \ref{tab:field-layer-design}: the black-outlined blue block is the core domain ladder, the beige block is the application/stress-test layer, and the red-toned COSMOS label marks the Extreme Deep layer. Cell colors give log-scaled row counts and the overplotted labels give the approximate counts, so the figure is read as a data-availability map rather than as a performance plot. The bottom band groups the columns by their role in the analysis: Gaia DR3 and WISE-SCoS are candidate-observable inputs, SDSS DR18, DESI DR1, and LAMOST DR10 LRS provide spectroscopic labels, teacher-cache support, or complementary validation, and NH QSO/RR and Polsterer QSO provide external comparison and contaminant diagnostics. The COSMOS row is intentionally different from the large-field rows: it shows the Gaia-linked source universe of about 8k sources, while the COSMOS-specific support data are summarized outside this matrix and include 964,506 COSMOS2020 FARMER rows, about 4k Chandra COSMOS-Legacy X-ray sources, about 11k VLA 3 GHz radio sources, and about 44k COSMOS-region spectroscopic rows. Zero-valued cells therefore mean that an auxiliary catalog is not part of the clipped field package or has no matched rows in that layer, not that a download is active or incomplete. Together with Table \ref{tab:cosmos-extrapolation-summary} and Figure \ref{fig:capability-envelope}, the COSMOS row defines a Gaia-linked Extreme Deep test case with external support, not a completeness denominator for all COSMOS AGN.}
\label{fig:catalog-coverage-by-field}
\end{figure}

Figure \ref{fig:catalog-coverage-by-field} then visualizes the same layer design in data-availability space and groups the plotted sources by their role in the analysis. The bottom data-source band places the survey or catalog names inside the corresponding role colors, while the role labels below the band separate candidate observables, spectroscopic supervision, and external checks. Gaia DR3 and WISE-SCoS represent source-level, spectrum-free candidate information; SDSS DR18 spectroscopy and the DESI DR1 zcatalog form the dominant spectroscopic supervision and validation backbone for the frozen benchmark; and LAMOST DR10 LRS supplies complementary spectroscopy, especially in low-latitude and anti-center regions. The smaller LAMOST phase-I QSO-reference catalog is not plotted as a separate column to avoid overstating its role relative to SDSS and DESI. Direct inputs that are not naturally represented as source-level row-count auxiliary tables, such as Legacy DR9 sweep products, are described in the data-source text rather than mixed into this row-count matrix. The last row shows COSMOS within the same main auxiliary-catalog matrix, but its interpretation is different: the Gaia column gives the Gaia-linked candidate universe, while the red-toned field label marks the Extreme Deep layer, where the external data are much deeper than the calibrated benchmark domain. The COSMOS-specific validation resources are therefore not plotted as additional columns in the matrix; they are summarized in the caption and used in the follow-up diagnostics below.

This separation is especially relevant for the COSMOS row. Its zeros indicate that the main eight-field auxiliary-catalog matrix and the COSMOS extreme-depth package are different data products, not that COSMOS lacks external information. The scientifically relevant COSMOS evidence comes from the X-ray, radio, deep-photometric, and spectroscopic resources discussed in Sections \ref{cosmos-as-an-extreme-deep-multimodal-testbed} and \ref{capability-boundaries-and-the-cosmos-deep-field-layer}.

For practical use, the current field table summarizes the field geometry, layer assignment, and labeled-sample state. The coordinate and area columns define the sky windows used for clipping auxiliary catalogs. For the first eight large fields, the entries are rectangular Galactic-coordinate windows. COSMOS is listed separately as a 0.8 deg radius deep-field cone; the tabulated Galactic-coordinate ranges are the empirical range spanned by the Gaia cone, and the area is the circular footprint rather than a rectangular sky window. The \texttt{Gaia\ DR3} column gives the current source universe in each field, while \texttt{labeled}, \texttt{QSO}, and \texttt{f\_QSO} describe the frozen spectroscopic benchmark coverage available after source grouping. This distinction separates high-density candidate fields from fields with supervised evaluation labels: Taurus/Perseus and Orion/Monoceros have large Gaia source counts but no current supervised label set, so their blank \texttt{f\_QSO} and median-extinction entries mark fields reserved for candidate-catalog stress testing rather than benchmark performance measurement. COSMOS has extensive external spectroscopy in the deep-field package, but it is not included in the frozen supervised benchmark; dashes in its benchmark-label columns mean not applicable to this benchmark definition, not the absence of QSO-supporting external data. Its external validation statistics are reported in Sections \ref{cosmos-as-an-extreme-deep-multimodal-testbed} and \ref{capability-boundaries-and-the-cosmos-deep-field-layer}. The COSMOS median \(E(B-V)\) value is computed for Gaia sources with a local extinction/Legacy match and is listed only to indicate the low foreground-reddening regime of the cone. The full catalog-coverage table is retained as a machine-readable companion table.

\begin{table*}[htbp]
\centering
\caption{Current field geometry and benchmark-label coverage.}
\label{tab:current-field-geometry}
\papertablesize
\setlength{\tabcolsep}{2.0pt}
\renewcommand{\arraystretch}{1.16}
\begin{tabular*}{\textwidth}{@{\extracolsep{\fill}}llccccrrcc@{}}
\toprule
Field & Layer & $l$ range & $b$ range & \makecell[c]{Area\\(deg$^2$)} & Gaia DR3 & Labeled & QSO & $f_{\rm QSO}$ & \makecell[c]{Median\\$E(B-V)$} \\
\midrule
High latitude & core & 180-210 & +45 to +75 & 444.9 & 216,665 & 8,815 & 1,300 & 0.147 & 0.017 \\
\tblleft{0.95in}{Mid-extinction north} & core & 180-210 & +25 to +40 & 378.4 & 719,341 & 9,610 & 2,046 & 0.213 & 0.037 \\
\tblleft{0.95in}{Mid-extinction bridge} & core & 165-195 & +25 to +40 & 378.4 & 908,041 & 24,749 & 5,922 & 0.239 & 0.049 \\
Anti-center & core & 165-195 & +10 to +25 & 427.9 & 5,467,319 & 38,173 & 5,409 & 0.142 & 0.060 \\
NEP/WISE & application & 80-110 & +20 to +35 & 398.0 & 3,947,248 & 47,839 & 6,675 & 0.140 & 0.043 \\
Taurus/Perseus & application & 160-190 & -25 to -10 & 427.9 & 5,028,865 & 0 & 0 & -- & -- \\
DESI-SDSS mid & application & 120-150 & +25 to +40 & 378.4 & 1,021,714 & 9,176 & 1,038 & 0.113 & 0.037 \\
\tblleft{0.95in}{Orion/ Monoceros} & application & 190-220 & -25 to -10 & 427.9 & 5,678,888 & 0 & 0 & -- & -- \\
COSMOS & \tblleft{0.75in}{Extreme Deep} & 235.8-237.9 & +41.3 to +42.9 & 2.0 & 7,968 & -- & -- & -- & 0.019 \\
\bottomrule
\end{tabular*}
\papertablenote{The first eight fields are rectangular Galactic-coordinate windows. COSMOS is a 0.8 deg radius cone; the listed Galactic ranges are the empirical range of Gaia sources in that cone. Dashes denote quantities not applicable to the frozen benchmark definition.}
\end{table*}

With these sample definitions fixed, the main results below use the frozen source-grouped benchmark, not the broader application-field coverage table. This prevents newly downloaded or partially validated auxiliary catalogs from changing the evidence chain after the benchmark has been fixed. In the current field table, the COSMOS row is included to make the field-layer design complete, but it is read differently from the first eight rows: it defines the Gaia-linked Extreme Deep source universe and low-foreground-reddening footprint, whereas the COSMOS2020, X-ray, radio, and spectroscopic resources are used as external support for candidate prioritization rather than as a homogeneous benchmark label set.

\section{Catalog Construction}\label{catalog-construction}

With the data and field layers defined, this section describes how the field-defined survey inputs are converted into a deployable QSO-candidate catalog. The construction is organized as an audit trail from astronomical measurements to a target-selection flag. The source-input layer defines what is known before follow-up; the catalog-representation layer turns heterogeneous survey measurements into a single source-level feature table; the training-supervision layer uses existing spectra only to improve the learned score; and the candidate-scoring layer produces the quantities that appear in the released catalog. The technical model details are included because they determine the selection function, but the deployed product remains a catalog-level selector rather than a spectrum- or image-required classifier.

\begin{figure}[htbp]
\centering
\includegraphics[width=\textwidth]{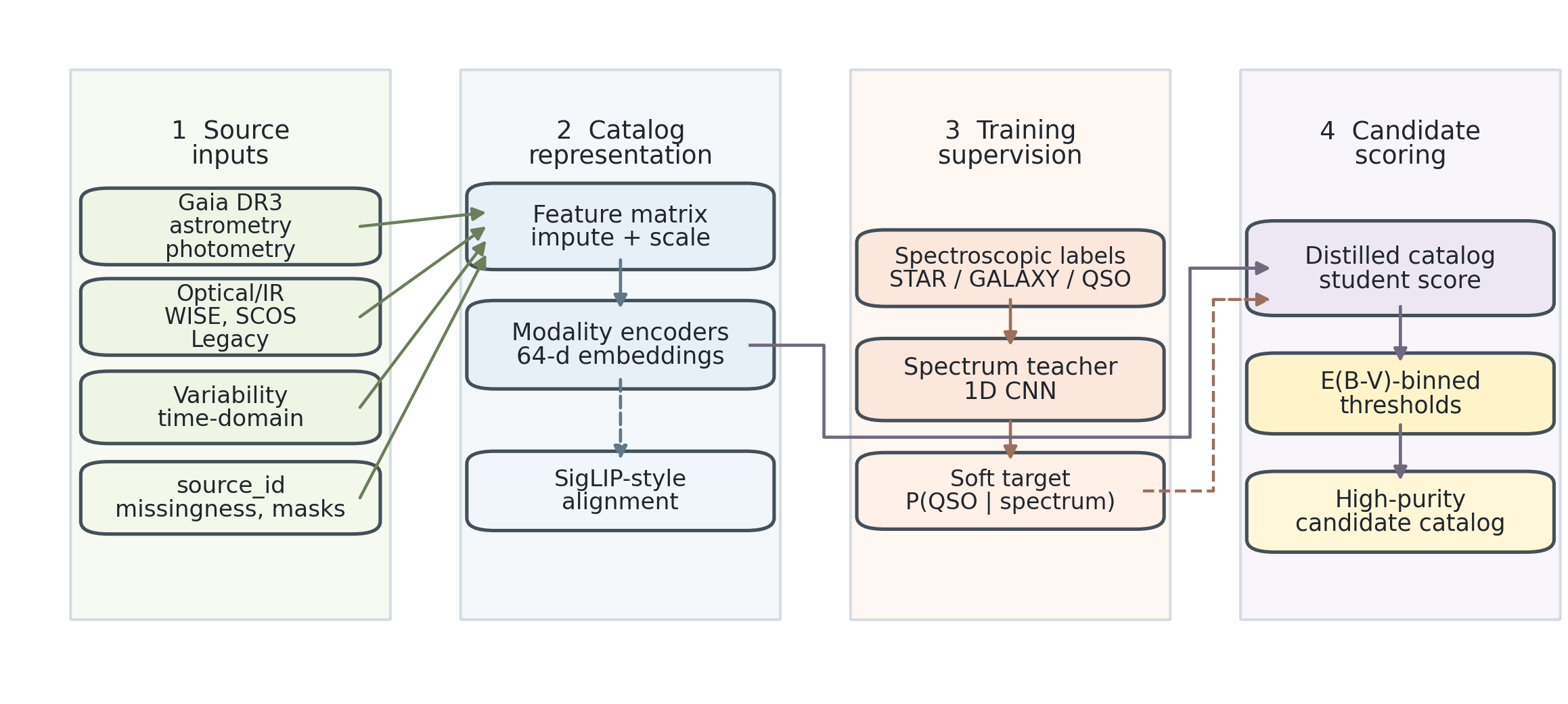}
\caption{Catalog-construction workflow used to turn survey measurements into a deployable QSO-candidate score. The four blocks separate information by its role in the released catalog. Source inputs contain Gaia astrometry and photometry, optical/infrared matches, optional variability information, source identifiers, missingness indicators, and masks. Catalog representation converts those heterogeneous measurements into an imputed, scaled, source-level feature matrix and learns modality-aware embeddings with same-source SigLIP-like alignment. Training supervision is restricted to sources with spectra: source-grouped STAR/GALAXY/QSO labels provide the hard target and the one-dimensional spectrum teacher provides fixed soft QSO targets. Candidate scoring is the only path used for previously unobserved sources; the distilled catalog student assigns QSO scores from catalog features alone, and \(E(B-V)\)-binned thresholds convert the scores into high-purity candidate flags. Solid arrows mark information available at inference, whereas dashed arrows mark training-only transfer, emphasizing that spectra improve the catalog selector but are not required for scoring new candidates.}
\label{fig:methods-overview}
\end{figure}

Figure \ref{fig:methods-overview} therefore gives the logic of the construction before the implementation details. The key design choice is to separate information that is available for every future candidate from information that is available only for labeled or previously observed sources. Gaia, optical/infrared, and optional variability measurements define the deployable source-input and catalog-representation path. Spectroscopic labels and spectrum-teacher probabilities are used only in the training-supervision path. The final candidate-scoring path therefore assigns QSO scores and selection flags from catalog measurements alone, which is essential if the product is to serve as a pre-spectroscopic target catalog for LAMOST-like or DESI-like follow-up. In this interpretation, solid arrows in the figure correspond to quantities available when selecting new fiber targets, while dashed arrows correspond to information transferred from existing spectroscopic samples during training.

Viewed as a roadmap, the teacher first learns the QSO/STAR/GALAXY distinction from existing spectra, where the class information is strongest. The student then learns a deployable QSO score from Gaia and multiwavelength catalog measurements, with the teacher contributing only train-split soft targets. The released catalog is produced by the student and calibrated thresholds, so a future source does not need a spectrum to be scored.

The same four-level structure also defines the statistical unit of the catalog. The Gaia \texttt{source\_id} is the object-level identity used for cross-matching, splitting, duplicate-spectrum resolution, and final scoring. This keeps repeated spectra and repeated auxiliary-catalog matches assigned to a single source-level split. It also keeps the selection function auditable: a user can trace each released candidate from source inputs, through feature construction and training-only supervision, to the final \(E(B-V)\)-calibrated threshold flag.

The remaining subsections follow Figure \ref{fig:methods-overview} from left to right. Sections \ref{source-level-feature-matrix-construction} and \ref{cross-matching-splitting-and-label-construction} define the source-level feature table and labels. Sections \ref{feature-block-representation-learning}--\ref{candidate-scoring-and-distillation} describe representation learning, the spectrum teacher, and the distilled catalog student. Section \ref{thresholds-cuts-and-catalog-selection-flags} then describes how validation-derived thresholds turn the continuous score into released catalog flags.

\subsection{Source-level Feature Matrix Construction}\label{source-level-feature-matrix-construction}

The first construction step is to reduce all catalog products to a common source-level representation before model training. The sky selections are defined in Galactic coordinates, but object-level matching and feature tables are carried in the Gaia DR3 identity frame whenever a Gaia \texttt{source\_id} is available. Survey-specific columns are standardized into reproducible feature families: Gaia astrometric and photometric quantities; optical and infrared fluxes, inverse variances, and colors; survey depth and mask indicators; and optional morphology/IPD diagnostics used only in the error-analysis and ablation sections.

After this source-level alignment, the downstream tabular representation is built from three physical feature blocks. The Gaia block contains parallax, proper motion, RUWE, excess-noise, visibility-period, and broad-band color information. The optical/infrared block contains WISE-SuperCOSMOS/Legacy Survey matched quantities, including match separations, extinction, calibrated magnitudes, colors, and photometric-redshift side quantities. The variability block contains match separations, damped-random-walk time-scale summaries, band-averaged magnitudes, and optical colors when time-domain information is available. For each numerical feature, missing values are imputed with the training-set median and scaled by the training-set interquartile range, with a paired missingness indicator appended to the feature matrix. Modality masks record whether Gaia, optical/infrared, and variability information is available for each source.

To keep the Gaia comparison independent, the student feature audit explicitly excludes Gaia official \texttt{classprob\_*} classifier products. Gaia official QSO probability is used only as a calibrated external reference under the same validation/test protocol. The proposed student may use Gaia measurements, but it does not import Gaia's own classifier decision as a model feature. Published QSO-candidate probabilities and catalog-level literature classifications are likewise excluded from the default student features unless they enter only as comparison references or post-hoc diagnostics.

\subsection{Cross-matching, Splitting, and Label Construction}\label{cross-matching-splitting-and-label-construction}

With the source-level feature table defined, cross-matching is deterministic and source-grouped. When multiple catalog detections or spectra map to the same Gaia source, the merged table keeps a single source-level row for splitting and evaluation. This is the same statistical unit used by the downstream model, so repeated observations are assigned consistently before train, validation, and test splits are frozen. Local QC checks track row counts, duplicated \texttt{source\_id} values, missing feature blocks, stale staging files, FITS complete markers, and quarantined downloads before any product is promoted into the benchmark.

After matching, the supervised task is QSO versus non-QSO. Spectroscopic QSO labels are positives. Spectroscopic STAR and GALAXY labels are retained as the non-QSO class and are also tracked separately for contaminant diagnostics. Published QSO-candidate catalogs, photometric-redshift catalogs, and variable-star auxiliary catalogs are not promoted to training labels by themselves; they are treated as external context unless a source also has an accepted spectroscopic label.

Repeated and conflicting spectra are then handled at the Gaia-source level. The train, validation, and test split unit is \texttt{source\_id}. For teacher construction, repeated spectra are resolved with a source-grouped union-best policy: each source contributes at most one selected spectrum to the teacher cache. The selection is based on survey and quality metadata with deterministic tie-breaking, not on the downstream student score. This policy prevents heavily observed sources from receiving extra training weight and makes SDSS-DESI-LAMOST overlap checks interpretable.

The same source-grouped rule controls the SDSS-DESI merge. Survey-specific classifications are first mapped to the common STAR, GALAXY, and QSO label vocabulary used by the teacher. Repeated spectra of the same Gaia source are not allowed to appear in different train, validation, or test splits. When more than one accepted spectrum is available, the union-best rule selects a single representative spectrum using survey provenance and quality metadata; the survey of origin and overlap state are retained as provenance fields. This design uses SDSS and DESI as complementary label sources while avoiding an artificial increase in sample size from repeat observations. Its limitation is that the final teacher cache is not survey-balanced: it reflects the available high-quality spectra after source grouping and quality filtering, so survey-specific selection functions remain part of the empirical training distribution.

\subsection{Feature-block Representation Learning}\label{feature-block-representation-learning}

The representation-learning step then organizes the multimodal catalog student around modality-specific encoders, following the same broad principle as CLIP-like representation learning: measurements from different views of the same astrophysical source should be mapped into compatible latent spaces \citep{radford_2021}. Because the present inputs are tabular survey measurements rather than images and natural-language text, the implementation uses a lightweight SigLIP-like tabular pretraining stage rather than a literal image-text CLIP model \citep{zhai_2023}.

Operationally, this stage tests whether the Gaia, optical/infrared, and variability descriptions of the same source provide mutually consistent information that can stabilize the later QSO score. It does not introduce an external class label and it does not recast the problem as an image-text task. Instead, it regularizes the catalog representation when some feature blocks are missing, noisy, or available for only part of the sky.

During this pretraining stage, separate encoders map the Gaia, optical/infrared, and variability feature blocks to 64-dimensional embeddings. For mini-batches in the training split, available modality pairs from the same source are treated as positive pairs, while other sources in the batch provide implicit negatives. The loss is a symmetric sigmoid binary cross-entropy over the pairwise similarity matrix for the Gaia-optical/infrared, Gaia-variability, and optical/infrared-variability pairs. Pretraining uses only source co-occurrence across survey modalities and does not use QSO labels, literature catalog classes, Gaia official classifier probabilities, or spectrum-teacher probabilities.

This SigLIP-like stage serves two roles. First, it provides a regularized initialization for downstream catalog classifiers when label fractions are small. Second, it provides a controlled comparison against no-pretraining and modality-dropout baselines. The main P2/P3 claim in this paper is still made with fixed supervised labels, fixed teacher logits, and the calibrated student score; the contrastive pretraining is therefore treated as a representation-learning component and baseline family rather than as an additional source of labels.

\subsection{Spectrum-teacher Supervision}\label{spectrum-teacher-supervision}

The spectrum-teacher component supplies the training-only supervision in two stages. First, a spectrum teacher is trained on the source-grouped spectra. Second, a deployable student is trained with hard labels and teacher-softened targets using only catalog-level candidate features. This follows the knowledge-distillation idea that a richer training-time model can transfer information to a simpler inference-time model \citep{hinton_2015}, while preserving the operational requirement that future candidates do not need spectra.

In astronomical terms, the teacher provides a controlled way to use spectra that already exist. A spectroscopic class label records the adopted STAR, GALAXY, or QSO class, whereas the teacher probability also carries information about ambiguous spectra, low signal-to-noise cases, and spectral similarity near class boundaries. The student is then asked to approximate that information from pre-spectroscopic catalog measurements. Accordingly, the teacher guides training but is not part of the released candidate-scoring requirement.

The teacher model is a three-class STAR/GALAXY/QSO classifier trained on source-grouped spectra. Spectra are resampled to a common 1024-pixel logarithmic wavelength grid from 3600 to 9000 \AA. Each training example has two input channels: robustly normalized flux and a valid-pixel mask. The flux normalization subtracts the median, divides by an interquartile or standard-deviation scale when needed, clips the result to the range {[}-8, 8{]}, and sets invalid pixels to zero. Spectra with too few valid pixels are rejected before cache construction.

Architecturally, the teacher network is a one-dimensional convolutional neural network with four convolutional blocks. The channel sequence is 2, 32, 64, 128, and 192, with GELU activations, batch normalization, max pooling in the first three blocks, and adaptive average pooling before the embedding layer. The pooled representation is passed through a dropout layer, a 128-dimensional linear embedding, layer normalization, and a final three-class classifier. Training uses class-weighted cross-entropy, AdamW optimization, gradient clipping at norm 5, a learning rate of 0.001, weight decay of 0.0001, batch size 128, and validation by QSO average precision. The maximum training length is 24 epochs, with a minimum of 8 epochs and patience of 5 epochs.

For the version comparison, the earlier P2 teacher contained 12,345 source-grouped union-best spectra after the priority download and QC freeze. It is retained as the previous stable baseline because it already supported the main Gaia-comparison result. The P3 update keeps the same source-grouping and union-best rules but expands the available teacher coverage substantially, especially through DESI spectra in the domain-ladder fields. This makes P3 a broader training prior while preserving the same inference-time catalog product.

For the v0.3.2 fixed-split evaluation, the teacher split is tied to the frozen downstream Gaia \texttt{source\_id} ledger. Any source assigned to the downstream validation or test split is forced into the teacher test split and is excluded from teacher fitting and checkpoint selection. The P2 and P3 teachers overlap the downstream benchmark in 5,234 and 43,606 sources, respectively, but only the downstream-training overlaps are eligible as student soft-target rows: 3,700 for P2 and 30,454 for P3. Teacher probabilities on downstream validation/test sources are retained only for diagnostic reporting, not as training targets.

The P3 union-best teacher used in the fixed benchmark contains 100,143 usable source spectra: 95,663 DESI spectra and 4,480 SDSS spectra. The class distribution is 66,898 STAR, 25,164 GALAXY, and 8,081 QSO. The source-grouped teacher split contains 70,101 training sources, 15,021 validation sources, and 15,021 test sources. This composition has both advantages and limitations. DESI provides the statistical power needed for a stable spectrum teacher, while SDSS anchors the cache to a long-established quasar-selection reference. At the same time, the teacher is DESI-dominated after quality filtering and is therefore a DESI+SDSS source-grouped teacher, not an equally weighted fusion of the two surveys. The best checkpoint is selected at epoch 16. On the teacher test split, the model reaches 0.977 accuracy, QSO purity 0.958, QSO completeness 0.989, QSO F1 0.973, and QSO PR-AUC 0.9955. These teacher metrics document the training-time label source; downstream candidate selection is evaluated separately with the catalog student.

\subsection{Candidate Scoring and Distillation}\label{candidate-scoring-and-distillation}

The candidate scorer is the deployable counterpart to the spectrum teacher. The distilled student uses the same catalog feature blocks and modality masks described above. Each modality is encoded by a small multilayer perceptron with 128 hidden units, 64-dimensional modality embeddings, layer normalization, GELU activations, and dropout of 0.15. The concatenated modality embeddings and modality masks are passed to a binary QSO-versus-non-QSO classifier. During training, modality dropout is applied more strongly to the optical/infrared and variability blocks than to the Gaia block, so the model is exposed to realistic missing-modality patterns.

Its training objective combines two terms. The first is a hard-label binary cross-entropy loss on the supervised spectroscopic labels, with positive-class weighting determined by the labeled training subset. The second is a teacher-soft-target binary cross-entropy loss on downstream training rows with available spectrum-teacher QSO probabilities. The default teacher term has weight 0.5. Downstream optimization uses AdamW with learning rate 0.0007, weight decay 0.0001, batch size 4096, and early stopping by validation QSO average precision. No spectrum, hard label, teacher logit, or duplicated source associated with downstream validation/test Gaia \texttt{source\_id} values is used in student fitting. The teacher probabilities are fixed before the seed study; the five-seed robustness experiment varies only the downstream student initialization and mini-batch order.

This separation is central to the catalog use case. Spectra are used to train and calibrate the teacher, but the deployable score for new candidates is computed from catalog-level measurements alone. The proposed catalog can therefore be applied in sky regions where Gaia, optical/infrared, and optional variability measurements are available even when no spectrum is available for the target source.

For reference, Table \ref{tab:model-components} summarizes the technical role of each model component. The table is intended as a compact guide to what information each component uses and whether that information is available at candidate-scoring time.

\begin{table}[htbp]
\centering
\caption{Model components and inference-time roles.}
\label{tab:model-components}
\papertablesize
\setlength{\tabcolsep}{3.0pt}
\renewcommand{\arraystretch}{1.20}
\begin{tabular*}{0.98\textwidth}{@{\extracolsep{\fill}}llll@{}}
\toprule
Component & \tblleft{1.50in}{Input information} & \tblleft{2.90in}{Training role and settings} & \makecell[c]{Scoring\\use} \\
\midrule
\tblleft{1.15in}{SigLIP-like tabular pretraining} & \tblleft{1.50in}{Gaia, optical/IR, variability blocks; modality masks} & \tblleft{2.90in}{Aligns same-source modality embeddings; initializes catalog encoders. Uses 64-dimensional embeddings and a symmetric sigmoid pair loss without labels.} & Indirect \\
Spectrum teacher & \tblleft{1.50in}{Source-grouped spectra; spectroscopic STAR/GALAXY/QSO labels} & \tblleft{2.90in}{Produces fixed soft QSO targets from spectra. Uses a 1D CNN, 1024 wavelength bins, class-weighted cross-entropy, validation QSO AP for selection, and downstream validation/test source-id exclusion in v0.3.2.} & No \\
\tblleft{1.15in}{Distilled catalog student} & \tblleft{1.50in}{Catalog feature matrix; hard labels; train-only teacher soft targets} & \tblleft{2.90in}{Learns the deployable QSO score from catalog measurements. Uses MLP fusion, hard-label BCE plus teacher BCE, teacher weight 0.5, and five downstream seeds.} & Yes \\
Threshold calibration & \tblleft{1.50in}{Validation-set catalog-student scores; \(E(B-V)\) bins} & \tblleft{2.90in}{Converts scores into high-purity flags. Recommended purity \(\ge 0.98\); auxiliary \(\ge 0.97\); \(E(B-V)\) bins \(<0.02\), \(0.02\)--\(0.05\), and \(\ge 0.05\).} & Yes \\
\bottomrule
\end{tabular*}
\papertablenote{The scoring-use column indicates whether the component itself is required when scoring previously unobserved candidates. ``Indirect'' means that only learned encoder weights are carried forward.}
\end{table}

\subsection{Thresholds, Cuts, and Catalog Selection Flags}\label{thresholds-cuts-and-catalog-selection-flags}

With scores defined, thresholds are learned on the validation set and evaluated once on the frozen test set. The recommended operating point is the conservative purity \(\ge 0.98\) threshold, chosen because a candidate catalog intended for follow-up is more sensitive to false positives than a pure ranking benchmark. A denser auxiliary purity \(\ge 0.97\) flag is retained for users with larger follow-up capacity, but robustness claims in this paper are made at the conservative 0.98 target. The main calibration uses three \(E(B-V)\) bins, \(<0.02\), \(0.02\)--\(0.05\), and \(\ge 0.05\), motivated by the long-established role of dust reddening in optical selection \citep{schlegel_1998,schlafly_2011}. Galactic latitude and sky region are retained as diagnostic axes, not as additional threshold-training variables.

The same calibration rules are applied to Gaia and to the proposed student. This makes the comparison a matched operating-point comparison rather than a comparison against an arbitrary default Gaia threshold. Robustness is assessed with five downstream seeds, bootstrap confidence intervals for P3-P2 gain, matched-threshold Gaia comparisons, and \(E(B-V)\) calibration diagnostics. Morphology/IPD and hard-negative analyses are reported as explanatory checks for the false-positive tail, not as part of the default selection rule, because their gains are not yet strong enough to justify the associated completeness loss.

\subsection{Selected-field Candidate Footprints}\label{selected-field-candidate-footprints}

\begin{figure}[htbp]
\centering
\includegraphics[width=\textwidth]{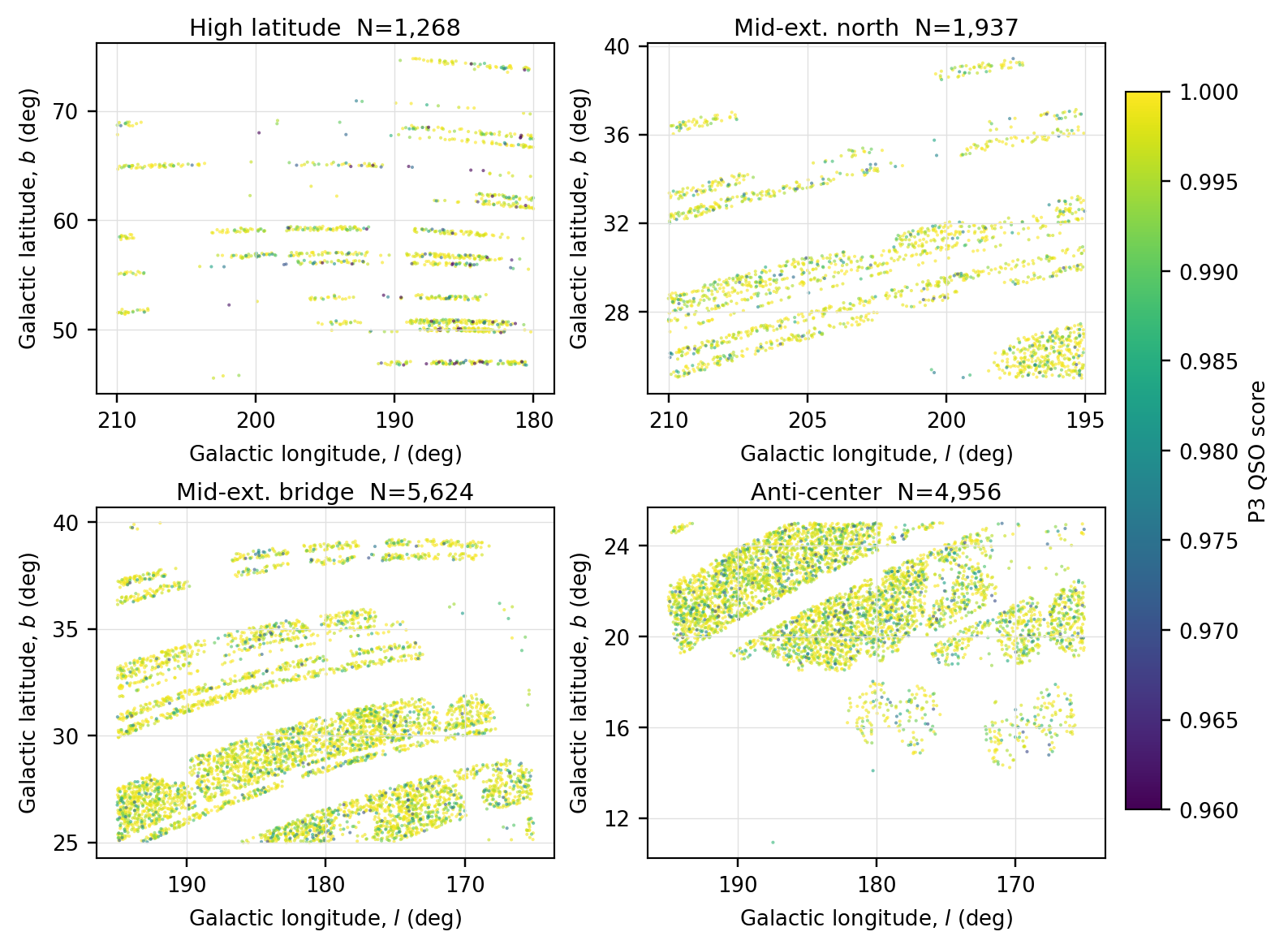}
\caption{Selected-field footprints of the high-purity P3 QSO candidate catalog in the four core domain-ladder fields. Points are plotted in Galactic coordinates and colored by the deployable P3 QSO score. The panels show the released target-list distribution after the conservative validation-calibrated threshold is applied; they are therefore descriptive catalog products rather than measurements of the intrinsic QSO surface density. The structured appearance follows the overlap of the selected sky windows, Gaia-linked source availability, auxiliary photometric coverage, and the calibrated score threshold. These bands should therefore be interpreted as footprint/coverage structure in the released target list, not as intrinsic QSO surface-density features.}
\label{fig:qso-selected-field-footprints}
\end{figure}

The released high-purity QSO catalog also defines a concrete target-list footprint in the selected core fields. Figure \ref{fig:qso-selected-field-footprints} shows the Galactic-coordinate distribution of the conservative P3 candidates in the four fields that carry the main validation baseline, and Table \ref{tab:qso-selected-field-footprint-summary} summarizes the corresponding candidate counts. These are selected-field target-list densities, not an all-sky surface-density measurement. They reflect the adopted field windows, Gaia-linked parent sample, auxiliary-catalog coverage, spectroscopic training set, and \(E(B-V)\)-binned threshold policy.

\begin{table}[htbp]
\centering
\caption{Selected-field high-purity QSO candidate footprint summary.}
\label{tab:qso-selected-field-footprint-summary}
\papertablesize
\setlength{\tabcolsep}{3.0pt}
\renewcommand{\arraystretch}{1.18}
\begin{tabular*}{0.98\textwidth}{@{\extracolsep{\fill}}lrrrrr@{}}
\toprule
Field & \makecell[c]{Area\\(deg$^2$)} & \makecell[c]{Candidates} & \makecell[c]{Density\\(deg$^{-2}$)} & \makecell[c]{Median\\$G$} & \makecell[c]{Median\\$E(B-V)$} \\
\midrule
High latitude & 444.9 & 1,268 & 2.85 & 20.21 & 0.017 \\
Mid-extinction north & 378.4 & 1,937 & 5.12 & 20.36 & 0.036 \\
Mid-extinction bridge & 378.4 & 5,624 & 14.86 & 20.27 & 0.048 \\
Anti-center & 427.9 & 4,956 & 11.58 & 20.22 & 0.058 \\
\bottomrule
\end{tabular*}
\papertablenote{The area values follow the rectangular Galactic-coordinate windows in Table \ref{tab:current-field-geometry}. Densities are catalog target-list densities under the conservative P3 threshold and should not be interpreted as population-level QSO surface densities.}
\end{table}

\section{Performance}\label{performance}

Having fixed the sample, score, and thresholds, performance is reported as a catalog operating-point problem. Purity is the fraction of selected test-set candidates that are spectroscopic QSOs, and completeness is the fraction of spectroscopic QSOs recovered by the calibrated selection rule. The main comparison is against Gaia official QSO probability after applying the same validation-set threshold calibration, so the performance tables compare deployable selection rules rather than uncalibrated scores. Existing spectroscopic and literature catalogs enter in two different ways: SDSS, DESI, and LAMOST provide the supervised validation backbone, while value-added candidate catalogs and Gaia probabilities provide external comparison and diagnostic context.

The Gaia comparison is central to this performance analysis. Gaia DR3 already provides an all-sky, mission-level extragalactic classification product, and Gaia-based QSO catalogs have become strong baselines for all-sky quasar work and cosmological applications \citep{gaia_2022,delchambre_2022,hughes_2022,storey_fisher_2024}. The relevant question is therefore whether a catalog-only student can recover additional spectroscopically confirmed QSOs at the same high-purity operating point, rather than whether it outperforms a simple color cut. This matched-operating-point comparison also reflects the intended use of the catalog as an input list for fiber spectroscopy, where purity, completeness, and field-specific target density directly affect how limited fiber positions are allocated.

\subsection{Parameter-space Coverage and Candidate Exploration}\label{parameter-space-coverage-and-candidate-exploration}

\begin{figure}[htbp]
\centering
\includegraphics[width=\textwidth]{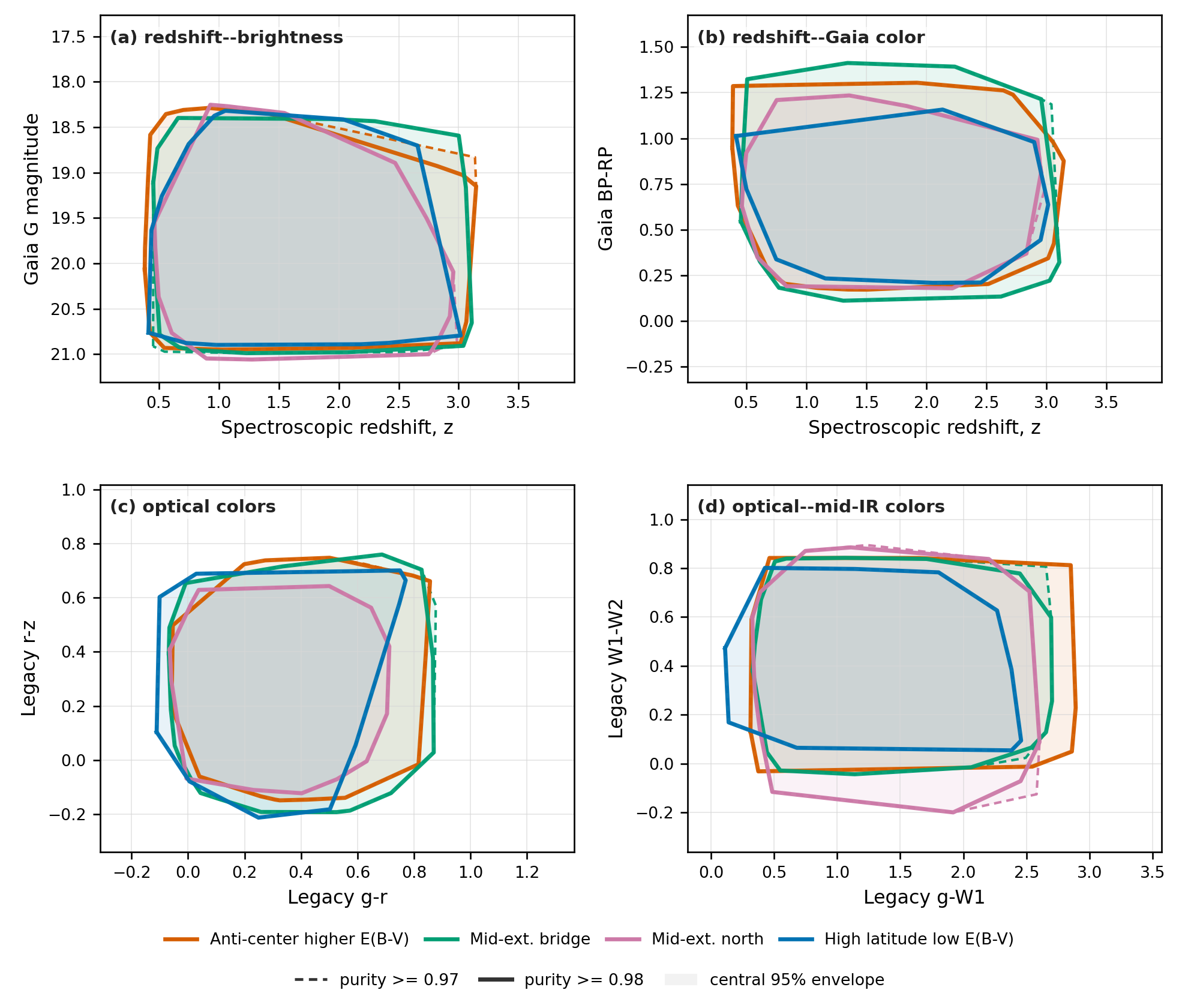}
\caption{Parameter-space coverage of selected, spectroscopically confirmed QSOs in the four core domain-ladder fields for the P3 catalog student. This validation view uses only sources that are already confirmed as QSOs by the frozen spectroscopic labels, so redshift is a measured spectroscopic coordinate and is not assigned to new candidates. The four panels project the selected QSO locus into redshift--Gaia magnitude, redshift--Gaia color, Legacy optical color-color space, and Legacy optical--mid-infrared color space; the Gaia \(G\) axis is inverted so that brighter sources appear higher. Colors identify the benchmark sub-fields. Dashed contours show the denser P3 operating point calibrated to purity \(\ge 0.97\), and solid contours show the conservative operating point calibrated to purity \(\ge 0.98\). Each contour encloses the central 95\% of selected confirmed QSOs in that projection, emphasizing the main locus rather than outliers. Tightening the operating point mainly trims the edge of the locus rather than moving the selected QSOs into a different region of color--magnitude--redshift space. The plotted confirmed-QSO counts are 740, 801, 294, and 176 at purity \(\ge 0.97\), and 693, 754, 284, and 174 at purity \(\ge 0.98\), for the anti-center, mid-extinction bridge, mid-extinction north, and high-latitude fields, respectively.}
\label{fig:selected-qso-parameter-envelopes}
\end{figure}

\begin{figure}[htbp]
\centering
\includegraphics[width=\textwidth]{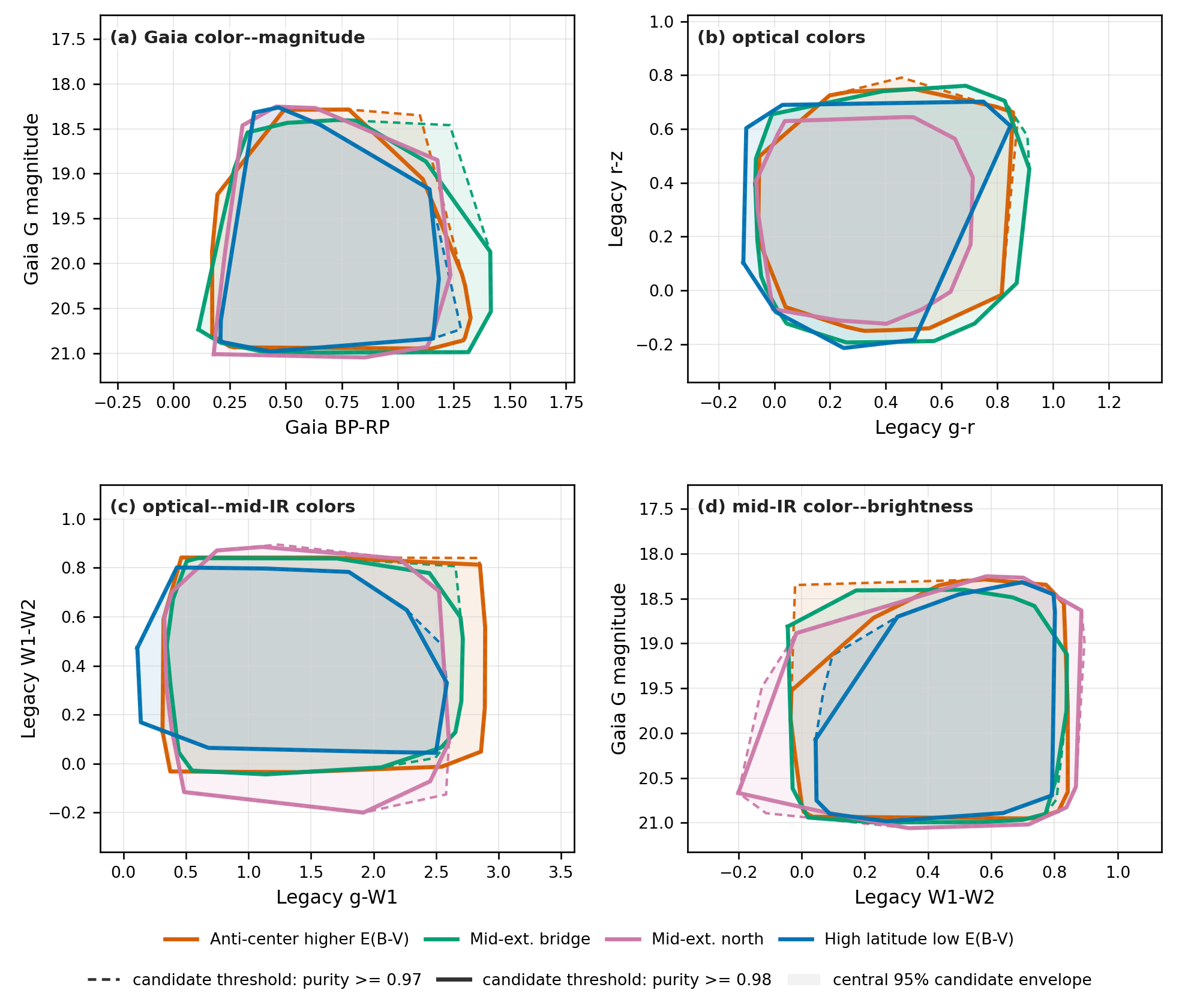}
\caption{Catalog-quantity footprint of threshold-selected candidates in the four core domain-ladder fields before new spectroscopy is obtained for the P3 catalog student. This candidate view uses only quantities available at target-selection time, so no redshift coordinate is plotted and neither spectroscopic nor photometric redshift is assigned to the candidates. The panels show Gaia color--magnitude space, Legacy optical color-color space, Legacy optical--mid-infrared color space, and mid-infrared color versus Gaia brightness. Colors identify the benchmark sub-fields. Dashed contours show the denser P3 selection calibrated to purity \(\ge 0.97\), and solid contours show the conservative selection calibrated to purity \(\ge 0.98\). The contours enclose the central 95\% of selected candidates in each projection and are drawn before separating true QSOs from contaminants; they therefore describe the target-list footprint rather than a confirmed-QSO-only locus. Compared with Figure \ref{fig:selected-qso-parameter-envelopes}, this view shows where follow-up targets enter the catalog in observable color and magnitude space. The selected-candidate counts are 756, 822, 298, and 185 at purity \(\ge 0.97\), and 703, 771, 286, and 182 at purity \(\ge 0.98\), for the anti-center, mid-extinction bridge, mid-extinction north, and high-latitude fields, respectively.}
\label{fig:selected-candidate-parameter-envelopes-noz}
\end{figure}

Before reporting scalar metrics, Figures \ref{fig:selected-qso-parameter-envelopes} and \ref{fig:selected-candidate-parameter-envelopes-noz} give complementary projections of the same calibrated selection. The former anchors the selection to spectroscopically confirmed QSOs, where redshift is known and the benchmark labels support direct validation. The latter shows the corresponding candidate footprint in catalog quantities available before new spectroscopy, where redshifts are still to be measured. Together they separate the part of parameter space already supported by spectroscopic truth from the space in which follow-up targets will be selected.

For the spectroscopic-validation view in Figure \ref{fig:selected-qso-parameter-envelopes}, the sample is restricted to selected sources already confirmed as QSOs by the frozen spectroscopic labels in the four core domain-ladder fields. The redshift coordinate is therefore spectroscopic; it is not a model-predicted redshift and it is not assigned to newly selected candidates. This figure tests whether the calibrated rule recovers known QSOs across the scientifically relevant parts of the benchmark rather than only in a narrow color, magnitude, or redshift locus. Redshift and apparent magnitude describe the validation reach and follow-up depth of the selected confirmed-QSO sample, as in SDSS/BOSS/eBOSS and DESI QSO target and catalog analyses \citep{schneider_2010,ross_2012,myers_2015,chaussidon_2023}. Gaia G and BP-RP are included because Gaia defines the source frame and supplies the primary external all-sky QSO reference classifier used in this paper \citep{gaia_2022,delchambre_2022,storey_fisher_2024}. Legacy optical colors trace the classical QSO color locus and its redshift-dependent degeneracies with stars and compact galaxies, while optical--mid-infrared colors provide one of the most widely used AGN/QSO contrast channels beyond optical selection \citep{richards_2001,richards_2002,vanden_2001,bovy_2011,bovy_2012,wright_2010,secrest_2015,schlafly_2019,marocco_2021}. The dashed contours show the denser P3 selection calibrated to purity \(\ge 0.97\), while the solid contours show the conservative P3 selection calibrated to purity \(\ge 0.98\). Tightening the operating point mainly trims the boundary of the selected confirmed-QSO locus rather than moving the catalog to a qualitatively different part of parameter space. This supports the interpretation that the conservative threshold is a stricter target-priority rule, while also showing that the empirical support remains field-dependent.

The companion view in Figure \ref{fig:selected-candidate-parameter-envelopes-noz} represents the parameter space of targets before new spectroscopy is obtained. At that stage, the relevant information is the deployable catalog measurement set--Gaia, optical, and infrared photometry--and the candidate redshifts are not yet measured. For that reason, the figure contains no redshift axis, and no spectroscopic or photometric redshift is assigned to the candidate sources for the purpose of this visualization. This is the view most relevant to an input catalog for fiber spectroscopy: future candidates will enter with Gaia, optical, and infrared measurements, while their redshifts and final classifications remain follow-up measurements. The color--magnitude and color--color panels show where the high-priority candidate list lies relative to the same \(E(B-V)\)-calibrated operating points, and they provide the practical search space for follow-up targeting. At the conservative purity \(\ge 0.98\) threshold, the selected-candidate counts are 703, 771, 286, and 182 for the anti-center, mid-extinction bridge, mid-extinction north, and high-latitude fields, respectively. These numbers are target-list sizes under the benchmark selection rule, not QSO surface densities. COSMOS is not folded into these envelopes: its Gaia-linked candidate universe is much shallower than the full COSMOS2020 population, and the present COSMOS result is supported by external X-ray, radio, spectroscopic, and deep-photometric checks rather than by a local purity/completeness benchmark.

\subsection{Denser High-purity Reference Point (Purity Target 0.97)}\label{denser-high-purity-reference-point-purity-0.97}

At the denser operating point, Table \ref{tab:performance-purity097} reports the frozen test-set result. This threshold is included to characterize target-list density and the purity--completeness trade-off, but it is not the main robustness point used below. The \texttt{Slice} column identifies the evaluation subset: \texttt{all} denotes the full frozen test set, while later diagnostic tables use physically motivated subsets defined by sky region, extinction, magnitude, or foreground complexity. The \texttt{N} and \texttt{QSO} columns give the fixed evaluation sample size and the number of spectroscopically confirmed QSOs in that sample. \texttt{Purity} is the fraction of selected candidates that are true spectroscopic QSOs, and \texttt{Comp.} abbreviates spectroscopic-label completeness, the fraction of source-grouped spectroscopic QSOs in the frozen benchmark test split recovered by the calibrated selection rule. \texttt{Selected} is the number of sources passing the validation-calibrated threshold, while \texttt{FP} and \texttt{FN} give the corresponding false positives and missed spectroscopic QSOs. \texttt{AP} denotes average precision, a threshold-free ranking diagnostic that summarizes whether spectroscopic QSOs are preferentially assigned high scores before any particular threshold is chosen. The main catalog interpretation comes from the thresholded purity, spectroscopic-label completeness, and selected-candidate counts, because these are the quantities that determine the efficiency of a follow-up target list.

\begin{table*}[htbp]
\centering
\caption{Performance at the denser high-purity operating point.}
\label{tab:performance-purity097}
\papertablesize
\setlength{\tabcolsep}{4.0pt}
\renewcommand{\arraystretch}{1.15}
\begin{tabular*}{0.98\textwidth}{@{\extracolsep{\fill}}llrrrrrrrrr@{}}
\toprule
Model & Slice & $N$ & QSO & Purity & Comp. & F1 & AP & Selected & FP & FN \\
\midrule
Gaia official & all & 12129 & 2148 & 0.9686 & 0.5177 & 0.6748 & 0.9450 & 1148 & 36 & 1036 \\
No teacher & all & 12129 & 2148 & 0.9761 & 0.9120 & 0.9430 & 0.9923 & 2007 & 48 & 189 \\
Hard-label student & all & 12129 & 2148 & 0.9711 & 0.9372 & 0.9538 & 0.9920 & 2073 & 60 & 135 \\
P2 teacher & all & 12129 & 2148 & 0.9716 & 0.9404 & 0.9558 & 0.9922 & 2079 & 59 & 128 \\
P3 teacher & all & 12129 & 2148 & 0.9757 & 0.9362 & 0.9556 & 0.9923 & 2061 & 50 & 137 \\
\bottomrule
\end{tabular*}
\papertablenote{Thresholds are calibrated on the validation set at target purity \(\ge 0.97\) and applied once to the frozen test set. P2 and P3 denote source-grouped union-best spectrum-teacher coverage stages. Comp. is spectroscopic-label completeness within the frozen Gaia-linked benchmark, AP is average precision, Selected is the number of selected sources, and FP/FN are false positives/false negatives relative to the source-grouped spectroscopic labels.}
\end{table*}

Table \ref{tab:performance-purity097} first establishes the role of Gaia as a strong but conservative baseline. After the same validation-calibrated thresholding protocol is applied, Gaia official QSO probability reaches high purity, 0.9686, but selects only 1,148 sources and recovers 1,112 of the 2,148 spectroscopic QSOs in the test set. The corresponding spectroscopic-label completeness is 0.5177, leaving 1,036 false negatives. This behavior defines a conservative reference layer, but it is insufficiently complete for a follow-up program whose goal is to recover a large fraction of available spectroscopic QSOs at high purity.

The catalog-only models substantially change this trade-off. All four catalog-student rows select about 2,000 sources at similar or higher purity than Gaia, reducing the number of missed QSOs by roughly a factor of seven to eight. The no-teacher and hard-label students are already effective, with spectroscopic-label completeness above 0.91, showing that Gaia plus optical/infrared and auxiliary catalog measurements contain much of the separability needed for the task. For the catalog product, the main performance gain is not obtained by requiring spectra at inference, but by learning a deployable catalog-level score from source-matched multiwavelength measurements.

Within the catalog-student family, the denser point mainly shows the purity--completeness trade-off rather than a decisive teacher-version gain. Relative to the hard-label student, P3 has slightly lower completeness but higher purity and fewer false positives. Relative to P2, P3 reduces false positives from 59 to 50 and keeps nearly the same F1, but it also recovers fewer spectroscopic QSOs at this target density. This table therefore provides context for the candidate-density trade-off; the main P3-P2 completeness claim is evaluated at the more conservative purity \(\ge 0.98\) point.

\subsection{Recommended Conservative Operating Point (Purity Target 0.98)}\label{recommended-conservative-operating-point-purity-0.98}

\begin{table*}[htbp]
\centering
\caption{Performance at the conservative operating point.}
\label{tab:performance-purity098}
\papertablesize
\setlength{\tabcolsep}{4.0pt}
\renewcommand{\arraystretch}{1.15}
\begin{tabular*}{0.98\textwidth}{@{\extracolsep{\fill}}llrrrrrrrrr@{}}
\toprule
Model & Slice & $N$ & QSO & Purity & Comp. & F1 & AP & Selected & FP & FN \\
\midrule
Gaia official & all & 12129 & 2148 & 0.9738 & 0.4493 & 0.6148 & 0.9450 & 991 & 26 & 1183 \\
No teacher & all & 12129 & 2148 & 0.9829 & 0.8575 & 0.9160 & 0.9923 & 1874 & 32 & 306 \\
Hard-label student & all & 12129 & 2148 & 0.9815 & 0.8655 & 0.9198 & 0.9920 & 1894 & 35 & 289 \\
P2 teacher & all & 12129 & 2148 & 0.9830 & 0.8617 & 0.9184 & 0.9922 & 1883 & 32 & 297 \\
P3 teacher & all & 12129 & 2148 & 0.9809 & 0.8869 & 0.9315 & 0.9923 & 1942 & 37 & 243 \\
\bottomrule
\end{tabular*}
\papertablenote{The validation-calibrated target purity is \(\ge 0.98\). Column definitions are the same as in Table \ref{tab:performance-purity097}; in particular, Comp. denotes spectroscopic-label completeness within the frozen Gaia-linked benchmark. P2 and P3 are the fixed teacher coverage stages used for the recommended operating point and the main robustness claims.}
\end{table*}

At the recommended conservative operating point, Table \ref{tab:performance-purity098} gives the stricter and more constraining comparison. Raising the target purity from 0.97 to 0.98 makes the comparison closer to a conservative fiber-follow-up use case, where each contaminant can displace a real QSO candidate. Under this threshold, the Gaia official reference selection remains highly pure but becomes strongly incomplete: it selects 991 sources, of which 965 are spectroscopic QSOs and 26 are false positives, leaving 1,183 spectroscopic QSOs unrecovered. This is the expected behavior of a conservative all-sky astrometric reference, but it is not dense enough for a follow-up program that needs a larger high-purity target list in the selected fields.

The catalog students change the operating point rather than merely improving a ranking score. The no-teacher and hard-label students already recover more than 85\% of spectroscopic QSOs at measured purity above 0.98, showing that Gaia measurements plus optical/infrared catalog features contain most of the deployable information needed for this benchmark. This context matters for interpreting the spectrum teacher: the teacher is not replacing a simple catalog model, but refining the high-score decision boundary of an already strong catalog-only selector.

Within this stricter regime, the P3 teacher gives a modest completeness and F1 gain over P2 at the recommended high-purity operating point, at the cost of a small purity decrease and a few additional false positives. P3 selects 54 more true QSOs than P2, reducing false negatives from 297 to 243 and increasing spectroscopic-label completeness from 0.8617 to 0.8869, while F1 increases from 0.9184 to 0.9315; the cost is five additional false positives and a measured purity change from 0.9830 to 0.9809. Because both models remain above the pre-specified 0.98 purity target, this is a science-driven target-selection trade-off rather than a relaxation of the catalog's high-purity requirement. Relative to the hard-label student, P3 recovers 46 additional QSOs with two additional false positives. The gain is therefore not only a higher AP value or a larger selected list; it is a thresholded selection-function improvement at the point where the catalog is intended to be used.

This comparison has a specific interpretation. P2 and P3 share the same downstream catalog-student architecture, inference-time feature set, frozen split, and threshold protocol. The P3-P2 difference therefore tests the effect of expanded and refrozen spectrum-informed supervision under a fixed downstream design, rather than isolating a new-architecture effect.

Relative to Gaia, the practical effect is larger. At this operating point P3 recovers 1,905 spectroscopic QSOs, whereas Gaia recovers 965. The additional recovery comes with 11 more false positives than Gaia in the frozen test set. This supports the main interpretation of the catalog as a Gaia-complement recovery layer: Gaia supplies a conservative reference selection, while the P3 catalog student uses multiwavelength features, source-grouped spectrum-teacher supervision, and \(E(B-V)\)-binned thresholds to recover many Gaia-missed QSOs while remaining within the high-purity regime.

\begin{figure}[htbp]
\centering
\includegraphics[width=\textwidth]{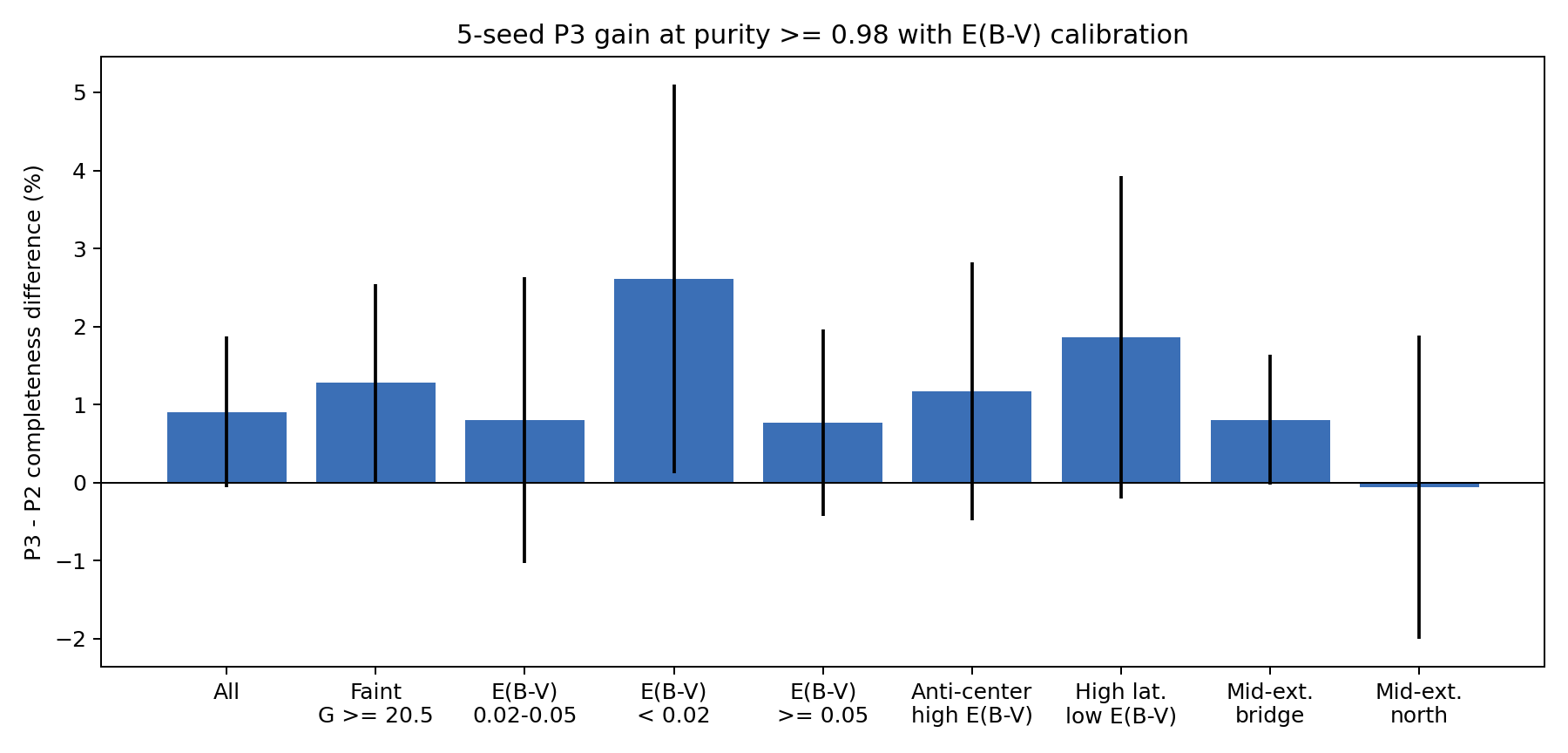}
\caption{Seed-to-seed robustness of the P3--P2 completeness difference at the conservative purity \(\ge 0.98\) operating point with \(E(B-V)\)-binned threshold calibration. Each blue bar is the mean completeness difference, P3 minus P2, over five downstream student seeds; black error bars show the seed-to-seed scatter. Values are plotted in percentage points, so 1\% on the y-axis is an absolute completeness difference of 0.01 rather than a relative 1\% change. The zero line marks equal completeness between P3 and P2 at the same calibrated high-purity threshold. The \texttt{all} bar summarizes the full frozen test set, and the remaining bars separate magnitude, extinction, and sky-region slices. Across these diagnostics, P3 gives a positive mean gain for the full sample and for several difficult regimes, especially the faint, higher-extinction, anti-center, and mid-extinction bridge slices, while some smaller slices remain seed-scattered or unresolved.}
\label{fig:p3-p2-five-seed}
\end{figure}

\subsection{Seed, Bootstrap, and Subgroup Robustness}\label{seed-bootstrap-and-subgroup-robustness}

To test whether the P3 gain is stable, Table \ref{tab:five-seed-robustness} summarizes a robustness test rather than an independent performance benchmark. The test asks whether the P3-P2 difference seen in Table \ref{tab:performance-purity098} persists under repeated downstream student training and whether the sign and scale of the gain remain meaningful in physically motivated subsets. The experiment uses the five seeds \texttt{42,\ 0,\ 66,\ 12,\ 123}, following the comparison protocol used for the downstream student runs. The benchmark split, feature set, threshold-calibration procedure, and P2/P3 teacher logits are held fixed; only the downstream student initialization and mini-batch order vary. The resulting table therefore tests the effect of the expanded P3 spectrum teacher under the same downstream design without introducing new data splits, new candidate features, or a different inference-time model.

The table reports signed differences at the recommended conservative operating point. Positive \(\Delta\) completeness means that P3 recovers a larger fraction of spectroscopic QSOs than P2. Negative \(\Delta\)FN means that P3 misses fewer true QSOs, which is favorable. Positive \(\Delta\)FP means that P3 selects more contaminants, which is unfavorable unless the completeness gain is worth the small purity cost. The mean \(\pm\) sd entries summarize the mean and seed-to-seed scatter over the five downstream runs, and the Slice column shows whether the same behavior appears in the full test set, faint sources, high-extinction bins, and selected sky regions.

The robustness result indicates that P3 is not uniformly better in every part of the sky. Instead, it provides a modest but repeatable completeness gain in the stricter high-purity regime, with the largest practical value in harder subsets and a small purity cost. At purity \(\ge 0.98\), the five-seed all-sample completeness gain is \(0.0090\pm0.0097\), corresponding to about 0.9 percentage points on average, with a mean \(\Delta\)FN of -19.4 and a mean \(\Delta\)FP of +4.0. The anti-center high-extinction, high-extinction, faint-source, and mid-extinction bridge slices show positive mean behavior, while smaller slices have larger seed scatter and are interpreted more cautiously. This supports the interpretation that the expanded spectrum teacher helps most where the catalog-only decision boundary is harder, without implying a uniform improvement in every subset.

\begin{table}[htbp]
\centering
\caption{Five-seed P3--P2 robustness at the conservative operating point.}
\label{tab:five-seed-robustness}
\papertablesize
\setlength{\tabcolsep}{2.4pt}
\renewcommand{\arraystretch}{1.18}
\begin{tabular*}{0.98\textwidth}{@{\extracolsep{\fill}}lcccrr@{}}
\toprule
Slice & \makecell[c]{$\Delta$ comp.\\mean$\pm$sd} & \makecell[c]{$\Delta$ purity\\mean$\pm$sd} & \makecell[c]{$\Delta$ F1\\mean$\pm$sd} & $\Delta$FP & $\Delta$FN \\
\midrule
all & $0.0090\pm0.0097$ & $-0.0019\pm0.0007$ & $0.0043\pm0.0053$ & 4.0 & -19.4 \\
dark G $\ge$20.5 & $0.0128\pm0.0127$ & $-0.0037\pm0.0008$ & $0.0064\pm0.0077$ & 2.6 & -10.0 \\
\(E(B-V)\ge0.05\) & $0.0077\pm0.0119$ & $0.0004\pm0.0010$ & $0.0046\pm0.0071$ & -0.2 & -7.6 \\
\tblleft{1.35in}{anti-center high-extinction} & $0.0117\pm0.0165$ & $-0.0004\pm0.0014$ & $0.0067\pm0.0099$ & 0.4 & -9.4 \\
\tblleft{1.35in}{high-latitude low-extinction} & $0.0186\pm0.0207$ & $-0.0090\pm0.0067$ & $0.0051\pm0.0099$ & 1.8 & -3.4 \\
\tblleft{1.35in}{mid-extinction bridge} & $0.0080\pm0.0083$ & $-0.0014\pm0.0021$ & $0.0039\pm0.0037$ & 1.2 & -6.8 \\
\bottomrule
\end{tabular*}
\papertablenote{Values are signed P3--P2 differences averaged over five downstream student seeds at target purity \(\ge 0.98\), not absolute performance values. Positive \(\Delta\) completeness is favorable, negative \(\Delta\)FN is favorable, and positive \(\Delta\)FP indicates additional contaminants.}
\end{table}

\begin{figure}[htbp]
\centering
\includegraphics[width=\textwidth]{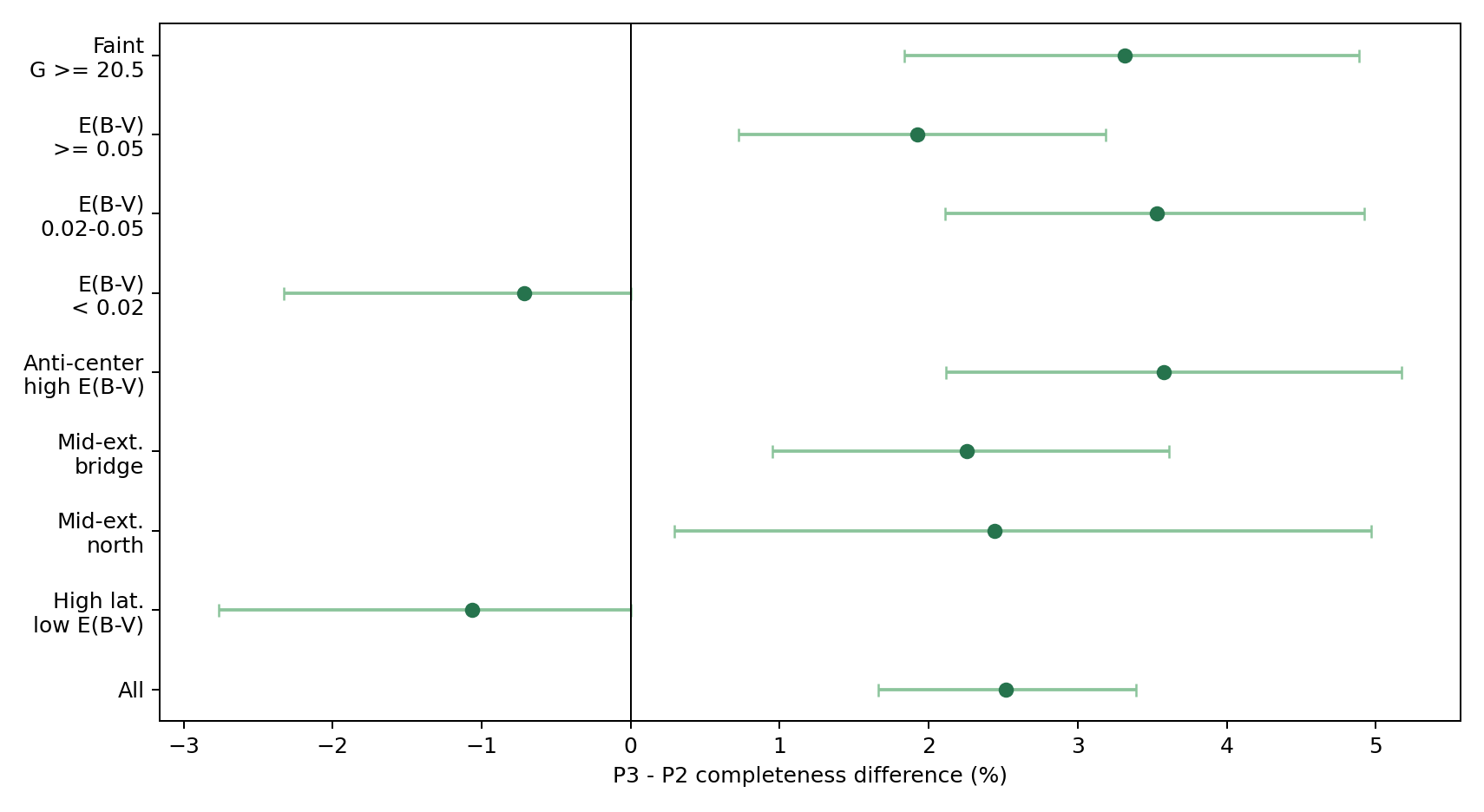}
\caption{Bootstrap uncertainty on the P3--P2 completeness difference at the conservative purity \(\ge 0.98\) operating point. Each point is the bootstrap median P3--P2 completeness difference for one diagnostic slice, and the horizontal bar spans the 2.5--97.5 percentile interval from resampling the frozen test set. Values are shown in percentage points. Intervals entirely to the right of zero support a resolved positive P3 gain in that slice; intervals that cross zero indicate that the available test sample does not resolve a local difference. The most resolved positive differences appear in the anti-center high-extinction field, the high-extinction \(E(B-V)\) bin, the intermediate-extinction bridge, and the faint-source slice. The high-latitude low-extinction slice crosses zero and is therefore treated as a boundary case rather than as evidence for a robust local P3 improvement.}
\label{fig:p3-p2-bootstrap-ci}
\end{figure}

As a complementary uncertainty check, Figure \ref{fig:p3-p2-bootstrap-ci} turns the same P3-P2 comparison into a sample-resampling test. The purpose of this check is not to introduce another model, but to ask whether the measured gain remains positive when the frozen test set is resampled. The vertical line marks zero gain. Points to the right of that line favor P3, and intervals crossing the line indicate slices where the available test sample does not support a resolved positive difference. The figure therefore separates statistically supported P3 gains from slices where the evidence remains suggestive.

The corresponding numerical percentiles are listed in Table \ref{tab:bootstrap-ci}. Together, Table \ref{tab:five-seed-robustness}, Figures \ref{fig:p3-p2-five-seed} and \ref{fig:p3-p2-bootstrap-ci}, and Table \ref{tab:bootstrap-ci} support a bounded claim: P3 improves the conservative high-purity selection function mainly in difficult regimes, rather than globally dominating P2 in every subset.

\begin{table}[htbp]
\centering
\caption{Bootstrap confidence intervals for P3--P2 gains.}
\label{tab:bootstrap-ci}
\papertablesize
\setlength{\tabcolsep}{2.6pt}
\renewcommand{\arraystretch}{1.16}
\begin{tabular*}{0.98\textwidth}{@{\extracolsep{\fill}}lrrcc@{}}
\toprule
Slice & $N$ & QSO & \makecell[c]{$\Delta$ completeness\\2.5/50/97.5\%} & \makecell[c]{$\Delta$ purity\\2.5/50/97.5\%} \\
\midrule
all & 12129 & 2148 & 0.0166 / 0.0252 / 0.0339 & -0.0053 / -0.0021 / 0.0009 \\
\tblleft{1.35in}{anti-center high-extinction} & 5628 & 801 & 0.0212 / 0.0358 / 0.0517 & -0.0036 / 0.0005 / 0.0049 \\
\tblleft{1.35in}{high-latitude low-extinction} & 1354 & 183 & -0.0276 / -0.0106 / 0.0000 & -0.0249 / -0.0056 / 0.0105 \\
\tblleft{1.35in}{mid-extinction bridge} & 3702 & 846 & 0.0095 / 0.0226 / 0.0361 & -0.0102 / -0.0046 / -0.0005 \\
\(E(B-V)\ge0.05\) & 5980 & 990 & 0.0072 / 0.0192 / 0.0319 & -0.0029 / 0.0004 / 0.0039 \\
dark G $\ge$20.5 & 1689 & 784 & 0.0183 / 0.0332 / 0.0489 & -0.0106 / -0.0037 / 0.0020 \\
\bottomrule
\end{tabular*}
\papertablenote{Intervals are the 2.5/50/97.5 percentiles from bootstrap resamples of the frozen test set at target purity \(\ge 0.98\). Gain columns are signed P3--P2 differences.}
\end{table}

\subsection{Comparison with the Gaia QSO Classifier}\label{comparison-with-the-gaia-qso-classifier}

The Gaia comparison anchors the catalog result to a widely used external classifier. Gaia is evaluated with the same threshold-calibration procedure as the proposed student: its score is calibrated on the validation set and evaluated on the same test set. This matched protocol avoids an inflated comparison against an arbitrary default Gaia threshold.

\begin{figure}[htbp]
\centering
\includegraphics[width=\textwidth]{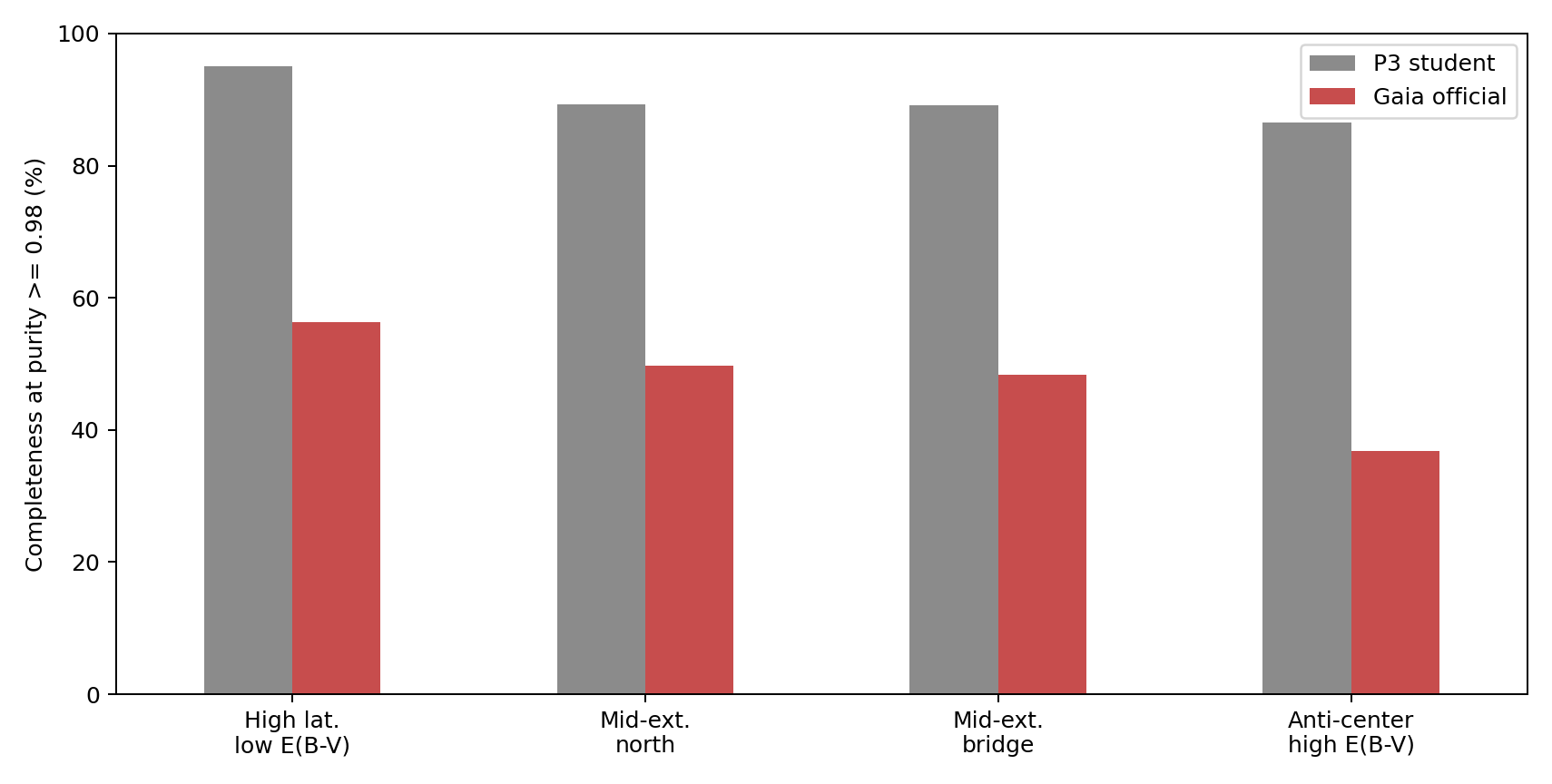}
\caption{Matched-purity regional comparison between the Gaia official QSO classifier and the P3 catalog student. Each bar gives completeness for spectroscopic QSOs in one region after the corresponding score has been calibrated to purity \(\ge 0.98\) under the same validation/test protocol. Gray bars show the P3 catalog student, which uses Gaia measurements, optical/infrared catalog features, and spectrum-teacher supervision during training; red bars show the Gaia official QSO probability used as an external reference classifier. Higher bars therefore mean that more confirmed QSOs are recovered at the same contamination tolerance, not that a different purity threshold has been adopted. The regional pattern supports a Gaia-complement interpretation: Gaia provides a conservative all-sky reference layer, while the selected-field P3 catalog student recovers additional high-purity candidates in the present benchmark fields.}
\label{fig:gaia-vs-p3-regional}
\end{figure}

Figure \ref{fig:gaia-vs-p3-regional} is the main regional comparison against this external reference classifier. The bars compare fixed-purity selections, not fixed score thresholds: both Gaia and P3 are calibrated to the same high-purity operating point, and the ordinate shows how many spectroscopic QSOs are recovered under that constraint. This comparison is central to the catalog interpretation because Gaia provides an all-sky, highly curated classifier, whereas the present work asks whether selected-field, multiwavelength, spectrum-teacher-assisted catalog selection can recover additional QSO candidates for follow-up target selection.

The result is interpreted as a Gaia-complement strategy rather than a replacement claim. In the present fields, Gaia official QSO probability supplies a conservative high-confidence layer, while the P3 catalog student recovers additional candidates at matched high purity by combining Gaia measurements with optical/infrared features and spectrum-teacher supervision. Extensions beyond these fields therefore require empirical calibration against Gaia classification rather than being treated as replacements for Gaia's all-sky product.

\subsection{Selection Function}\label{selection-function}

These diagnostics define the empirical selection function of the catalog. Here the selection function is the probability that a true QSO enters the released candidate sample under a specified threshold policy, as a function of source properties and sky environment. The primary dependencies considered here are apparent magnitude, foreground reddening, and field. These quantities are most likely to affect follow-up target yield for fiber spectroscopic surveys, because they control both the observable contrast between QSOs and contaminants and the density of plausible targets on the sky.

The \(E(B-V)\)-binned threshold policy is intentionally transparent. It is an operating-point correction rather than a physical model of extinction: the model score is produced by the same catalog student, but the score threshold is calibrated separately in broad reddening bins. The observed robustness supports \(E(B-V)\) calibration as the default thresholding rule, while extinction-aware losses or domain adapters remain secondary model variants.

The field-dependent results in Tables \ref{tab:five-seed-robustness} and \ref{tab:bootstrap-ci}, together with Figures \ref{fig:p3-p2-five-seed} and \ref{fig:p3-p2-bootstrap-ci}, show that the P3 gain is part of the selection function rather than a single global number. The largest measured gains occur in the anti-center, mid-extinction bridge, high-extinction, and faint-source regimes, while the high-latitude low-extinction field shows lower within-slice purity and an unresolved local P3-P2 difference. The recommended catalog flags are therefore interpreted together with the accompanying \(E(B-V)\) bin, field label, and threshold policy, especially when the catalog is used to design follow-up observations outside the exact validation sample.

For catalog users, Table \ref{tab:fixed-purity-compact-slices} condenses the same fixed-purity seed-42 result into the slices most relevant for catalog use. The validation-calibrated threshold is fixed first, and the measured completeness and purity are then evaluated inside each slice. The all-sample row gives the main operating point, while the regional, extinction, and faint-source rows show where the catalog gains are concentrated. Because the thresholds are calibrated globally or by broad \(E(B-V)\) bins rather than independently within every diagnostic slice, the measured within-slice purity can deviate from 0.98 in small subsets; this is part of the empirical selection function rather than a separate tuning target.

\begin{table}[htbp]
\centering
\caption{Compact selection-function slice summary.}
\label{tab:fixed-purity-compact-slices}
\papertablesize
\setlength{\tabcolsep}{2.2pt}
\renewcommand{\arraystretch}{1.17}
\begin{tabular*}{0.98\textwidth}{@{\extracolsep{\fill}}lrcccccc@{}}
\toprule
Slice & QSO & \makecell[c]{Gaia\\comp.} & \makecell[c]{P2\\comp.} & \makecell[c]{P3\\comp.} & \makecell[c]{P3\\purity} & \makecell[c]{P3--Gaia\\comp.} & \makecell[c]{P3--P2\\comp.} \\
\midrule
All & 2148 & 44.9\% & 86.2\% & 88.7\% & 98.1\% & +43.8 & +2.5 \\
\tblleft{1.30in}{Anti-center high-extinction} & 801 & 36.8\% & 82.9\% & 86.5\% & 98.6\% & +49.7 & +3.6 \\
\tblleft{1.30in}{Mid-extinction bridge} & 846 & 48.3\% & 86.9\% & 89.1\% & 97.8\% & +40.8 & +2.2 \\
\tblleft{1.30in}{Mid-extinction north} & 318 & 49.7\% & 86.8\% & 89.3\% & 99.3\% & +39.6 & +2.5 \\
\tblleft{1.30in}{High-latitude low-extinction} & 183 & 56.3\% & 96.2\% & 95.1\% & 95.6\% & +38.8 & -1.1 \\
\tblleft{1.30in}{Dark sources, Gaia $G\geq20.5$} & 784 & 10.6\% & 78.2\% & 81.5\% & 97.6\% & +70.9 & +3.3 \\
\(E(B-V)\ge0.05\) & 990 & 36.5\% & 84.0\% & 86.0\% & 98.2\% & +49.5 & +1.9 \\
\bottomrule
\end{tabular*}
\papertablenote{Values are evaluated for the seed-42 run at the recommended target purity \(\ge 0.98\). Completeness is measured against source-grouped spectroscopic QSO labels in each slice; gain columns are absolute completeness differences in percentage points. Five-seed P3--P2 robustness is reported separately in Table \ref{tab:five-seed-robustness}.}
\end{table}

\section{Catalog Description}\label{catalog-description}

Having established performance, the catalog product is organized at the Gaia-source level. Each row represents one candidate source in one of the defined fields and carries the information needed to reproduce the recommended QSO selection: object identifiers, sky position, field name, field layer, available input-feature blocks, calibrated model scores, \(E(B-V)\) bin, threshold policy, recommended conservative selection flag, auxiliary denser selection flag, and provenance/QC indicators. The field-layer column distinguishes core domain-ladder fields, application/stress-test fields, and the COSMOS Extreme Deep layer, so that users do not interpret all rows as carrying the same selection-function evidence. The exact machine-readable column list will be frozen with the public table, but the columns are grouped into identifiers and astrometry, input-coverage flags, scores and thresholds, validation labels where available, and quality-control or provenance flags.

To make the release semantics explicit, Table \ref{tab:release-column-groups} gives the planned release schema at a level directly relevant to catalog use. The table separates columns needed to identify a source, columns needed to determine whether a score is applicable, and columns needed to reproduce or audit a selection flag. The catalog is therefore presented as both a list of selected objects and a documented selection rule. Users can choose the recommended high-purity flag, adopt the denser auxiliary flag, or re-rank candidates using the score and provenance fields under their own observing constraints.

\begin{table}[htbp]
\centering
\caption{Planned machine-readable catalog column groups.}
\label{tab:release-column-groups}
\papertablesize
\setlength{\tabcolsep}{3.0pt}
\renewcommand{\arraystretch}{1.18}
\begin{tabular*}{0.98\textwidth}{@{\extracolsep{\fill}}llll@{}}
\toprule
Group & \tblleft{2.05in}{Representative columns} & \tblleft{2.05in}{Main use} & \tblleft{0.95in}{Release status} \\
\midrule
\tblleft{1.10in}{Identifiers and position} & \tblleft{2.05in}{\texttt{source\_id}, RA, Dec, Galactic $l,b$} & \tblleft{2.05in}{Stable source identity, sky selection, and cross-match reproduction} & Required \\
Field assignment & \tblleft{2.05in}{field name, field layer, footprint flag} & \tblleft{2.05in}{Interpret whether a row belongs to the core ladder, application layer, or COSMOS Extreme Deep layer} & Required \\
Feature coverage & \tblleft{2.05in}{Gaia, optical/IR, variability, and morphology availability flags} & \tblleft{2.05in}{Distinguish non-detections from unavailable feature blocks and audit score applicability} & \tblleft{0.95in}{Required/ recommended} \\
Catalog features & \tblleft{2.05in}{Gaia magnitudes and colors, optical/IR colors, \(E(B-V)\), quality summaries} & \tblleft{2.05in}{Candidate characterization and optional user-side re-ranking} & \tblleft{0.95in}{Required where available} \\
Scores & \tblleft{2.05in}{P3 score, optional P2 score, Gaia reference probability} & \tblleft{2.05in}{Default catalog ranking, teacher-version audit, and external Gaia comparison} & \tblleft{0.95in}{Required/ optional audit} \\
Threshold flags & \tblleft{2.05in}{\(E(B-V)\) bin, purity \(\ge 0.98\) flag, purity \(\ge 0.97\) flag} & \tblleft{2.05in}{Reproduce the recommended conservative selection and the denser auxiliary target list} & Required \\
Validation metadata & \tblleft{2.05in}{spectroscopic label/redshift where available, external support flags} & \tblleft{2.05in}{Benchmark evaluation, COSMOS support checks, and candidate-card construction} & \tblleft{0.95in}{Benchmark or auxiliary product} \\
Provenance and QC & \tblleft{2.05in}{feature version, teacher version, threshold version, QC flag} & \tblleft{2.05in}{Reproducibility, release comparison, and filtering of warning states} & Required \\
\bottomrule
\end{tabular*}
\papertablenote{Representative columns define the intended semantics of the release. The public machine-readable product will freeze exact column names, units, formats, version tags, and null-value conventions.}
\end{table}

For routine scientific use, the default science-facing product is the P3 catalog-student score with \(E(B-V)\)-binned thresholds. The recommended flag is the conservative purity \(\ge 0.98\) selection. A purity \(\ge 0.97\) auxiliary flag is retained as a denser target list for users with larger follow-up capacity, but it is not the operating point used for the main robustness claim. Spectroscopic labels, split assignments, and teacher-cache provenance are included for reproducibility and validation, not as recommended target-selection features. The selection flags and scores define the catalog product, while the label columns provide audit metadata.

In parallel, quality-control checks are part of the reproducibility record. The final QC table verifies that P3 downloads are complete, source grouping and fixed splits are preserved, P2/P3 teacher probabilities are fixed for robustness tests, downstream validation/test Gaia \texttt{source\_id} values are excluded from teacher fitting and checkpoint selection, Gaia official classifier probabilities are excluded from student features, and thresholds are learned only on validation data. FITS complete markers, quarantined-download checks, and auxiliary-catalog row-count summaries are retained so that future releases can be distinguished from the frozen benchmark used in this paper.

The candidate catalog, field-definition table, threshold table, and companion validation/QC products are planned for release through the Chinese National Astronomical Data Center (NADC/China-VO).\footnote{Project data-release landing page: \url{http://101.201.56.194/}.} The release will also include the main intermediate products needed to audit and extend the catalog construction: source-matched feature tables, fixed train/validation/test splits, teacher-logit tables, \(E(B-V)\)-binned threshold and weighting functions, field masks, auxiliary-catalog row-count summaries, and quality-control reports. These products are intended to allow astronomers to assemble target catalogs under their own observing constraints and to support downstream model training, benchmarking, and independent method development on the same source-grouped benchmark.

To preserve rebuildability, the source code used for catalog construction, validation, figure generation, and reproducibility checks will be shared with the data release whenever licensing permits redistribution of the underlying survey products. Large upstream survey files will not be redistributed when the original projects require direct access through their own archives, but the scripts will record the official source, query, or download route needed to rebuild the local feature tables.

\section{Applications}\label{applications}

The applications below distinguish between two outputs of this work: the released candidate catalog and the reusable methodological framework. The catalog product is intended as an input target list for high-purity QSO follow-up in the present Gaia-linked optical/infrared fields. The method, however, is a more general construction framework built around calibrated selection thresholds, provenance-aware source grouping, and cross-survey validation. These methodological elements also motivate extensions to other wavelength regimes, including radio astronomy; we return to this broader use case in the Discussion.

In addition to its direct use as a QSO candidate catalog, the released product may serve as a benchmark resource for multimodal astronomical model development. The fixed source-grouped splits, calibrated scores, teacher and student outputs, field-layer assignments, input-coverage flags, and provenance/QC metadata provide structured training and evaluation material for downstream models that combine astrometry, photometry, catalog context, and spectroscopic supervision.

\subsection{Input Catalogs for LAMOST QSO Follow-up}\label{input-catalogs-for-lamost-qso-follow-up}

The most direct application is target-list construction for LAMOST QSO follow-up. LAMOST is a wide-field, fiber-fed spectroscopic facility in which scientific yield is shaped by the target catalog before any spectrum is taken: finite fibers discretize the focal plane into a limited set of observable positions, and each contaminating target can displace a QSO candidate of potential follow-up value \citep{lamost_2015,ai_2016,dong_2018,yao_2019}. A LAMOST-oriented QSO-selection catalog therefore needs to report not only a classifier score, but also an operating point, target density, field membership, and quality flags that can be used by target preparation or by human follow-up planning.

For this use case, the Gaia-source identity, field layer, \(E(B-V)\)-binned threshold flag, score, and provenance columns define a rankable input list for LAMOST follow-up. The conservative purity \(\ge 0.98\) flag is suitable when fiber resources are scarce or contaminant rejection is the dominant requirement. The purity \(\ge 0.97\) flag provides a denser list for fields where follow-up capacity is larger or where candidate completeness is more valuable. Because the selection rule is catalog-only at inference, it can be applied before spectroscopy and can be re-ranked with local scheduling constraints, fiber-collision constraints, magnitude limits, or field-specific science priorities.

\subsection{Input Catalogs for DESI-like and Future Stage-V Spectroscopic Facilities}\label{input-catalogs-for-desi-like-and-future-stage-v-spectroscopic-facilities}

Beyond LAMOST, DESI provides a depth and target-density reference for interpreting the present catalog, although it is not the primary follow-up application of this release. The DESI instrument uses a wide-field prime-focus corrector feeding 5020 robotic fiber positioners, with ten spectrographs covering approximately 360--980 nm at spectral resolution \(R\sim2000\)--5000 \citep{desi_2016,silber_2023}. Its quasar target selection is a survey-scale example of the same operational problem addressed here: before spectroscopy, candidates are selected from Legacy Surveys \(grz\) photometry and WISE \(W1/W2\) information. The DESI main QSO selection uses a Random Forest classifier over \(16.5<r<23\), with a target density of about 310 deg\(^{-2}\) and more than 200 spectroscopically confirmed QSOs deg\(^{-2}\) expected from the main selection \citep{chaussidon_2023}.

In magnitude space, the selected sources in the present catalog fall within this DESI-like regime. For the conservative purity \(\ge 0.98\) candidate set in the core domain-ladder fields, the Legacy \(r\)-band median is about 20.18 mag, the 95th percentile is about 21.00 mag, and all candidates with valid Legacy \(r\) photometry are brighter than \(r=22\). This comparison has two implications. First, the candidate list is not depth-limited relative to DESI-like QSO spectroscopy. Second, DESI is used in this paper as a spectroscopic teacher, validation resource, and external survey-optimization benchmark rather than as a target to be replaced: DESI's own selection is already a mature survey product, while the present catalog asks how much additional high-purity Gaia-linked recovery can be obtained in selected fields with explicit extinction and field-layer calibration.

The same logic also connects the catalog to the future highly multiplexed facilities discussed in recent MUST papers, where ``Stage-V'' is used as the MUST white-paper terminology for the next step beyond DESI-like surveys \citep{zhao_2026,cai_2026}. MUST is the concrete example: it is a 6.5 m, wide-field spectroscopic telescope under development, with a field of view of order 5 deg\(^2\), more than 20,000 fiber positioners, wavelength coverage of about 370--960 nm, and a first-generation spectrograph concept comparable to DESI but scaled to much higher multiplexing \citep{zhang_2023,zhao_2026,cai_2026}. In the MUST conceptual target-selection plan, the QSO component is described over \(2<z<5\) with a limiting magnitude of \(r<23.5\) and an adopted surface density of 90 deg\(^{-2}\) \citep{zhao_2026}.

Relative to that boundary, the present catalog is a bright, conservative, high-confidence priority layer rather than the full target universe for MUST. The current Gaia-linked candidates are much brighter than the \(r<23.5\) QSO limit considered in the MUST forecast, and the field-layer calibration is tied to Gaia-detected sources and selected spectroscopic labels. A true MUST-depth extension would require a non-Gaia-limited parent sample, deeper imaging features, and a new validation set at the relevant magnitude and redshift range. The contribution of the present work is therefore methodological: it provides a source-grouped, spectrum-teacher, threshold-calibrated selection-function framework that can be reused when deeper Stage-V parent samples and validation spectra become available.

\subsection{Complementing Gaia QSO Classification}\label{complementing-gaia-qso-classification}

The Gaia comparison leads to a second application: using the catalog student as a calibrated augmentation of Gaia-based QSO selection. Gaia official QSO probabilities provide an all-sky, homogeneous, and physically valuable reference. The performance results in Section \ref{performance} show that, after matching the threshold-calibration protocol, Gaia is highly pure but leaves many spectroscopically confirmed QSOs unrecovered in the benchmark fields. The P3 student is therefore interpreted operationally as a high-purity recovery layer for Gaia-linked sources that are not selected by a Gaia-only QSO threshold.

This interpretation is relevant for future all-sky or large-area releases. The method is not presented as a replacement for Gaia classification without all-sky revalidation. Instead, a practical large-area version would publish two linked products: a Gaia-reference selection and a P3-augmented selection with field- or extinction-aware thresholds. The incremental candidates, their score distribution, and their failure modes would then define the scientific return of the larger release. This follows the same logic as Gaia-CatWISE and Quaia, where Gaia provides the astrometric and classification backbone while infrared and external information improve completeness, purity control, or cosmological usability \citep{hughes_2022,storey_fisher_2024}.

\subsection{Precursor Relevance for CSST Engineering and Science}\label{precursor-relevance-for-csst-engineering-and-science}

The same catalog logic is relevant to CSST-era source-classification products, although no CSST observations are used in this work. Published CSST/CSS-OS studies describe a wide-field space survey with a main multiband imaging and slitless-spectroscopy survey camera, MCI, IFS, and other instruments, and they emphasize survey simulation, redshift measurement, photometric-redshift calibration, guide-star and image-stability requirements, sample purity, completeness, and validation of spectroscopic products as central technical requirements \citep{zhan_2011,zhan_2021,gong_2019,yuan_2021,cao_2022,feng_2024,wen_2024,sui_2025,yan_2026,zheng_2026}. The present catalog and method therefore provide two preparatory products for CSST-related work: a high-purity QSO candidate product that can be used as an external bright-source prior, and a source-grouped construction framework that can be reused when CSST imaging, slitless-spectroscopic, MCI, or IFS measurements become available.

For the main survey, the relevant product includes both the candidate list and its provenance record. The catalog records field membership, source provenance, score calibration, \(E(B-V)\)-dependent thresholds, spectroscopic validation status, and quality-control flags. These quantities can support early checks of source classification, slitless-spectrum association, and redshift-product validation by providing bright Gaia-linked QSO candidates with known operating points. They can also define validation subsets in fields where extinction, source density, or compact-galaxy contamination may affect classification performance. In orbit, such sources would not replace the stellar standards used for wavelength or pointing calibration, but they can provide extragalactic point-like and broad-emission-line checks on astrometric consistency, color terms, spectral trace association, and classification failure modes.

The relevance extends beyond the main survey camera. For MCI, whose simultaneous NUV/blue/red imaging is designed for high-precision photometry, weak-signal detection, standard-star-system work, and photometric-redshift calibration, a vetted QSO candidate layer can provide compact extragalactic objects with strong color leverage across the UV--optical bands \citep{cao_2022,zheng_2026}. Such objects provide tests of color-dependent selection, photo-z outliers, AGN-like spectral-energy distributions, and field-dependent photometric systematics. For IFS, published simulations emphasize spatially resolved spectral cubes, realistic instrumental and in-orbit effects, and science near supermassive black holes and AGN-like systems \citep{yan_2026}. A conservative QSO/AGN prior can therefore help select compact validation targets and interpret IFS datacube products, while the same source-provenance framework can track whether a source is validated by external spectra, main-survey slitless spectra, MCI colors, or IFS observations.

For science preparation, the catalog supplies a conservative QSO layer rather than a CSST-depth QSO census. It provides a test of how a multiwavelength, extinction-calibrated QSO prior behaves in selected fields and identifies sources that may support early cross-checks of CSST photometry, slitless spectra, MCI imaging, IFS datacubes, and redshift products. The limitation is explicit: the present denominator is Gaia-linked and calibrated on selected fields, so it cannot be used to infer CSST-depth QSO completeness. A direct CSST extension would require CSST photometry, slitless-spectroscopic measurements, MCI and IFS data products where available, survey-specific masks, and a frozen CSST validation set. Under those conditions, the same source-grouped, spectrum-teacher, and threshold-calibrated framework could be retrained and audited at the CSST survey depth and across multiple CSST instruments.

\subsection{COSMOS as an Extreme Deep Multimodal Testbed}\label{cosmos-as-an-extreme-deep-multimodal-testbed}

COSMOS provides a distinct application test. The field is a classical extragalactic deep field with extensive optical, infrared, X-ray, radio, morphology, photometric-redshift, and spectroscopic resources \citep{scoville_2007,laigle_2016,civano_2016,smolcic_2017,weaver_2022}. It is scientifically attractive for QSO and AGN follow-up, but its depth makes it a boundary case for the present Gaia-linked catalog. The Gaia cone contains 7,968 sources, whereas the COSMOS2020 FARMER table contains 964,506 rows in the downloaded deep-field product, a Gaia-to-COSMOS2020 row ratio of only 0.83\%. The robust subset remains deliberately small: 39 candidates survive the Extreme Deep diagnostic check, 36 of them have X-ray, radio, or spectroscopic support, and 33 have valid redshifts.

\begin{figure}[htbp]
\centering
\includegraphics[width=0.85\textwidth]{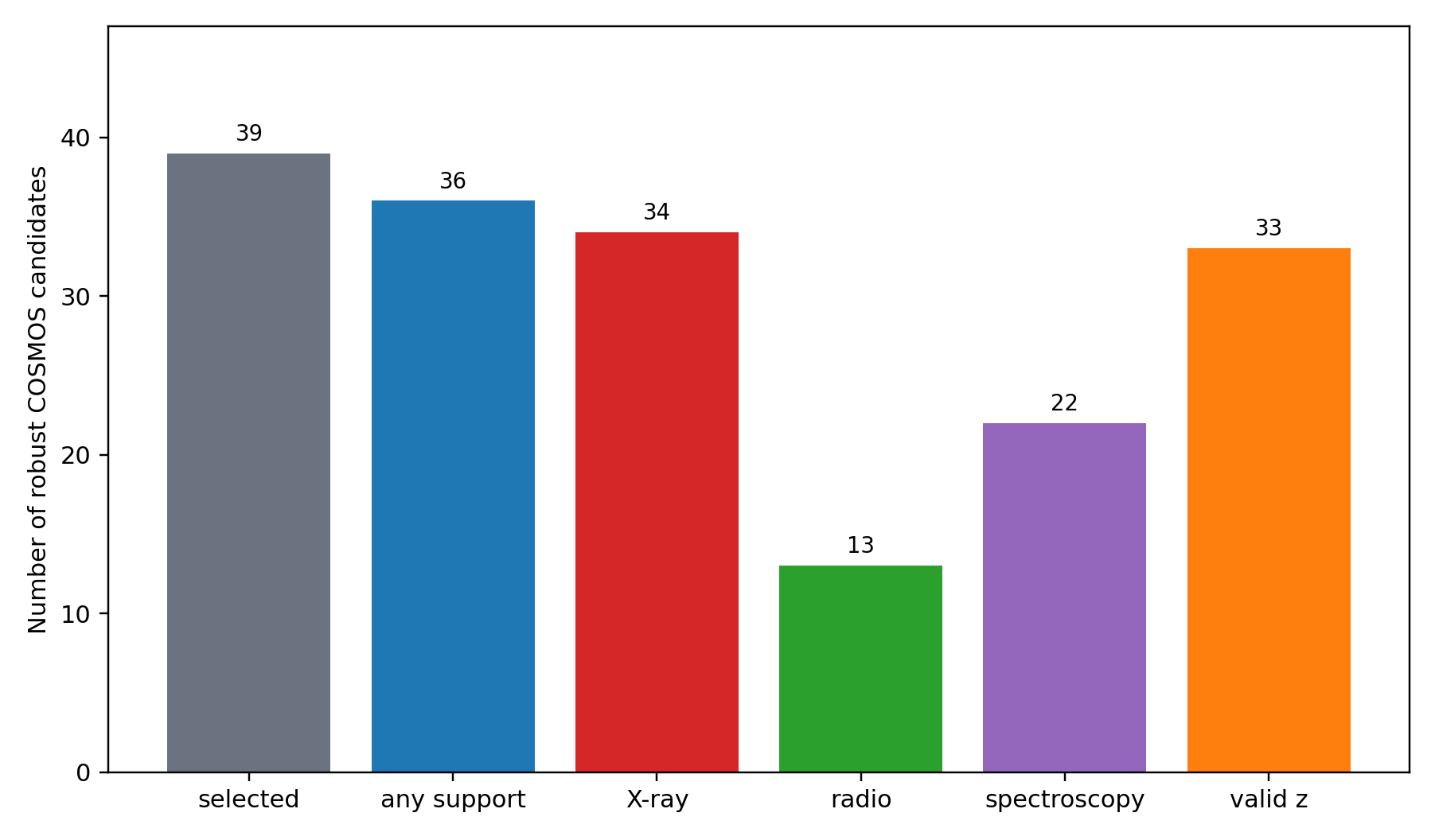}
\caption{External support channels for the 39 robust COSMOS candidates. The first bar gives the retained subset after the Extreme Deep diagnostic check, and the second bar counts candidates with at least one direct support channel among X-ray, radio, and spectroscopy. The remaining bars show the individual support-channel counts and the number of candidates with valid redshift measurements. Because support channels overlap, the individual bars do not sum to the ``any support'' bar. The main result is that most robust COSMOS candidates have independent deep-field evidence, supporting their use as a high-priority follow-up list. The figure does not measure COSMOS completeness, because the denominator is the Gaia-linked robust subset rather than the full COSMOS AGN population.}
\label{fig:cosmos-robust-support}
\end{figure}

The present COSMOS result is therefore interpreted qualitatively and operationally. It identifies a compact high-confidence target list in a field where independent multiwavelength information is unusually rich. Figure \ref{fig:cosmos-robust-support} summarizes the support channels for the robust subset. X-ray support is the largest single channel, as expected for an AGN-rich deep field, while radio and spectroscopic support provide additional independent checks. The ``any support'' bar is not the sum of the individual bars because several candidates have more than one support channel.

\begin{figure}[htbp]
\centering
\includegraphics[width=0.80\textwidth]{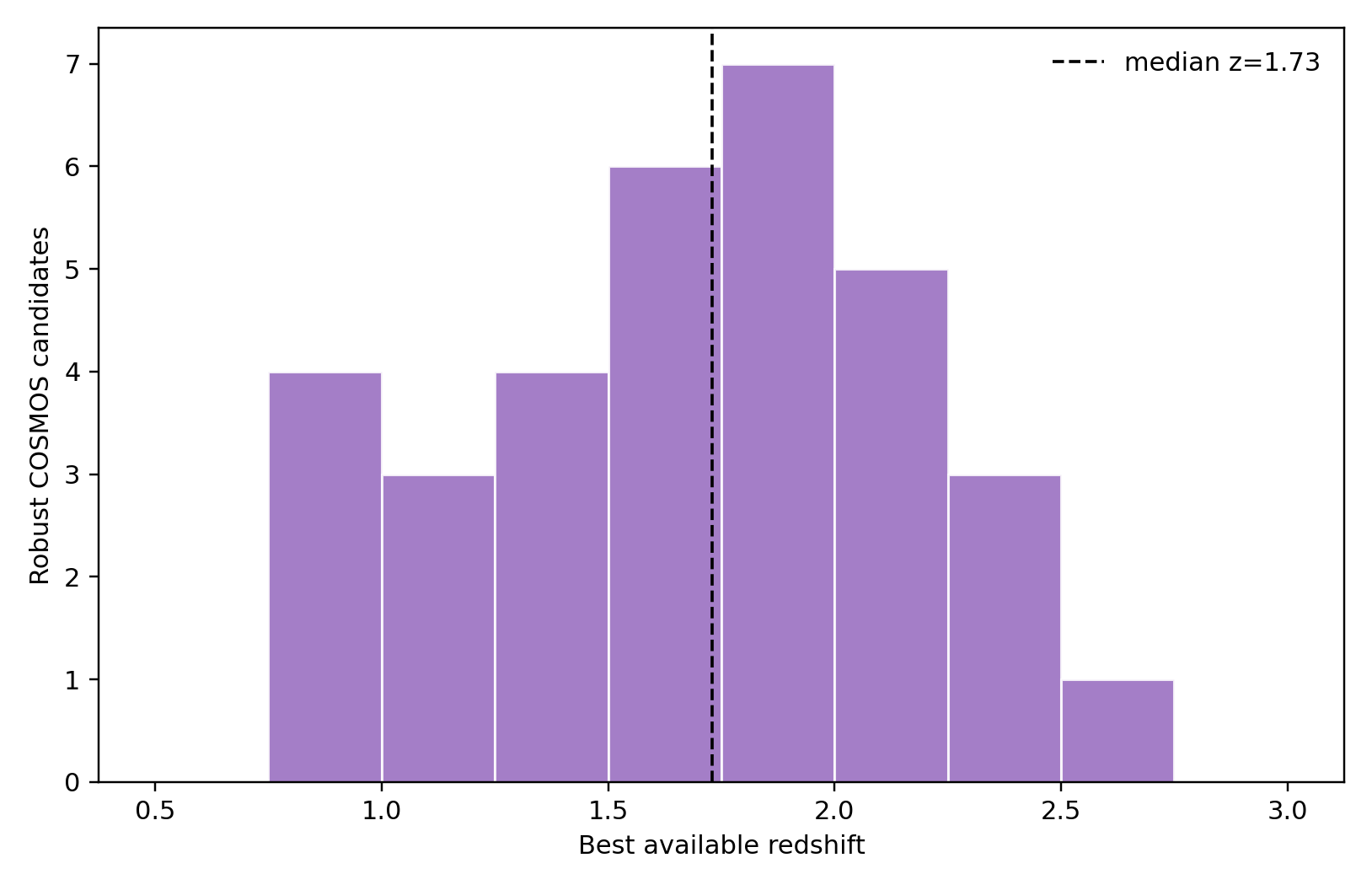}
\caption{Best available redshift distribution for the COSMOS robust candidates with valid redshift information. The histogram includes 33 of the 39 robust candidates, and the dashed vertical line marks the median redshift, \(z=1.729\). The distribution describes the externally supported robust subset and is useful for follow-up planning, especially because it indicates the redshift range over which the priority list already has ancillary support. These redshifts are not used as an input coordinate for selecting previously unobserved candidates, and the histogram should not be interpreted as the redshift distribution of all COSMOS QSOs.}
\label{fig:cosmos-robust-redshift}
\end{figure}

Figure \ref{fig:cosmos-robust-redshift} shows the redshift range for the supported subset with valid redshift information. The median redshift is \(z=1.729\), and the valid range is \(0.851<z<2.735\). This range supports follow-up prioritization because it overlaps the regime where broad-line QSOs and luminous AGN provide strong spectroscopic leverage, but it should not be read as the redshift distribution of COSMOS QSOs. The distribution is conditional on Gaia visibility, the present catalog score, the conservative Extreme Deep diagnostic check, and the availability of external redshift information.

\clearpage

\section{Systematics}\label{systematics}

\subsection{External Reference Samples and Evidence Roles}\label{external-reference-samples-and-evidence-roles}

After the application discussion, the systematics analysis begins with external catalog comparisons. These comparisons are part of the selection-function analysis, rather than only a performance check. Gaia DR3 official non-stellar classification provides the primary baseline because it is all-sky, homogeneous, widely used, and independent of the proposed student model \citep{gaia_2022,delchambre_2022}. Gaia-based QSO catalogs such as Gaia-CatWISE and Quaia further demonstrate that Gaia-linked quasar samples can support large-area and cosmological applications when their angular systematics and selection functions are modeled carefully \citep{hughes_2022,storey_fisher_2024}. For this reason, the central external comparison asks whether the student adds high-purity completeness beyond a calibrated Gaia-only selection, rather than whether it improves on a simple color selection.

To keep these comparisons interpretable, Table \ref{tab:external-comparison-roles} summarizes how the main external resources are used. The purpose is to separate three functions that are often conflated: a calibrated reference classifier, spectroscopic truth labels, and literature selection functions. Only the spectroscopic labels define direct purity and completeness in this paper. The other catalogs help interpret overlap, missed populations, and likely contaminants.

\begin{table}[htbp]
\centering
\caption{External resources used in the systematics analysis.}
\label{tab:external-comparison-roles}
\papertablesize
\setlength{\tabcolsep}{2.8pt}
\renewcommand{\arraystretch}{1.18}
\begin{tabular*}{0.98\textwidth}{@{\extracolsep{\fill}}llll@{}}
\toprule
External resource & \tblleft{1.80in}{Use in this paper} & \tblleft{1.60in}{What it constrains} & \tblleft{1.55in}{Main caution} \\
\midrule
\tblleft{1.45in}{Gaia DR3 QSO classification} & \tblleft{1.80in}{Matched-threshold external reference classifier} & \tblleft{1.60in}{Incremental recovery beyond a homogeneous Gaia-only selection} & \tblleft{1.55in}{Gaia probability is not a truth label and is excluded from student features} \\
\tblleft{1.45in}{SDSS/DESI/LAMOST spectra} & \tblleft{1.80in}{Source-grouped supervised labels and teacher construction} & \tblleft{1.60in}{Direct purity, completeness, false positives, and false negatives} & \tblleft{1.55in}{Spectroscopic target selection is not all-sky random} \\
\tblleft{1.45in}{SDSS/BOSS/eBOSS and DESI target selections} & \tblleft{1.80in}{Survey-optimized comparison context} & \tblleft{1.60in}{Expected target-density and fiber-follow-up trade-offs} & \tblleft{1.55in}{Target definitions reflect survey goals and redshift priorities} \\
\tblleft{1.45in}{WISE, XDQSO/XDQSOz, variability, and literature QSO catalogs} & \tblleft{1.80in}{Overlap and missed-population diagnostics} & \tblleft{1.60in}{Color/infrared/variability channels and likely contaminants} & \tblleft{1.55in}{Agreement is supporting evidence, not independent purity validation} \\
\bottomrule
\end{tabular*}
\papertablenote{The rows define evidence roles in this paper. They are not same-sample performance benchmarks, and only the spectroscopic labels are used for direct purity and completeness measurements.}
\end{table}

Under this structure, the present benchmark answers the Gaia-complement question only for the controlled fields. At matched validation-calibrated operating points, Gaia official QSO probability is highly pure but incomplete relative to the P3 catalog selector: at the validation-calibrated purity \(\ge 0.98\) target, Gaia recovers 0.4493 of spectroscopic QSOs in the frozen test set, while P3 recovers 0.8869 at measured purity 0.9809. We interpret this as a Gaia-complement result. Gaia supplies a conservative all-sky reference selection and a stable source frame; the P3 selector uses Gaia measurements, optical/infrared features, \(E(B-V)\)-binned thresholds, and spectrum-teacher supervision to recover additional candidates in the present fields. The method is therefore used here as an augmentation layer for Gaia-linked candidate selection under this benchmark, not as a claim that Gaia's all-sky classification has been superseded.

Other literature and survey comparisons have different roles. SDSS/BOSS/eBOSS and DESI quasar target selections are survey-optimized input catalogs shaped by fiber allocation, redshift goals, and target-density requirements \citep{richards_2002,ross_2012,myers_2015,desi_2016,chaussidon_2023}. XDQSO/XDQSOz, WISE-based AGN selections, ZTF variability catalogs, and deep-field AGN catalogs define additional selection functions rather than interchangeable truth labels \citep{bovy_2011,bovy_2012,secrest_2015,schmidt_2015,nakoneczny_2025}. Agreement with these catalogs indicates that the candidate population overlaps established QSO-selection channels; disagreement identifies follow-up regions, especially for reddened, host-dominated, variable, or X-ray-selected AGN. Direct purity statements, however, are made only where spectroscopic labels are available under the source-grouped benchmark.

This external-comparison structure increases the portability of the result while keeping the interpretation bounded. A future all-sky or large-area release would report at least two masks: a Gaia-only calibrated selection and a P3-augmented selection. The incremental P3 candidates, their sky dependence, and their overlap with DESI, SDSS, WISE, Quaia, variability, and X-ray AGN samples would then define the large-area selection function and the scientific return of the release.

\subsection{Published Performance Context for High-purity QSO Selection}\label{comparison-context-from-published-qso-selection-studies}

The comparison in this subsection has a different purpose from Table \ref{tab:external-comparison-roles}. Table \ref{tab:external-comparison-roles} defines how external resources enter the evidence chain of this paper. Table \ref{tab:published-qso-selection-context} instead places the purity, completeness, efficiency, target-density, and magnitude-limit language used here in the broader QSO-selection literature, then adds the present operating point as a same-row reference. The purpose is not to rank methods across incompatible survey definitions, but to identify the specific regime in which the present catalog adds value. Published QSO catalogs differ in parent sample, sky area, magnitude limit, redshift range, spectroscopic truth definition, and whether the reported quantity is purity, efficiency, completeness, target density, or catalog reliability. Representative examples include SDSS color and photometric QSO selection, XDQSO/XDQSOz, BOSS/eBOSS targeting, DESI QSO selection, Gaia-CatWISE/Quaia, and recent time-domain QSO selection \citep{richards_2002,richards_2004,bovy_2011,bovy_2012,ross_2012,myers_2015,desi_2016,chaussidon_2023,hughes_2022,storey_fisher_2024,nakoneczny_2025}.

The resulting comparison highlights both the strengths and the boundaries of the present product within its intended domain. First, the baseline is not a simple color cut but Gaia's official QSO probability calibrated under the same validation/test protocol. Second, the main reported gain is a thresholded recovery gain at a high-purity operating point, not only an average ranking metric: P3 recovers 0.8869 of spectroscopic QSOs at measured purity 0.9809, compared with 0.4493 for Gaia under the same protocol. Third, the catalog is packaged as a reproducible selection-function product, with fixed source IDs, field layers, \(E(B-V)\)-binned thresholds, score provenance, and validation metadata. These properties are the practical advantages for selected-field follow-up. By contrast, DESI-, Gaia-/Quaia-, and variability-optimized catalogs provide more appropriate reference products for survey scale, redshift-specific target-density optimization, all-sky uniformity, or time-domain selection functions.

\begin{table*}[htbp]
\centering
\caption{Published and present operating-point context for QSO selection.}
\label{tab:published-qso-selection-context}
\scriptsize
\setlength{\tabcolsep}{1.8pt}
\renewcommand{\arraystretch}{1.08}
\begin{tabular*}{0.98\textwidth}{@{\extracolsep{\fill}}llll@{}}
\toprule
Study or catalog & \tblleft{1.55in}{Parent sample and inputs} & \tblleft{2.25in}{Representative reported operating point} & \tblleft{1.45in}{Interpretation for this work} \\
\midrule
\tblleft{1.15in}{SDSS color target selection} &
\tblleft{1.55in}{SDSS ugriz color space, morphology, and FIRST radio matches} &
\tblleft{2.25in}{Designed for broad spectroscopic targeting: simulated overall completeness \(>90\%\), efficiency \(>65\%\), and about 18 candidates deg\(^{-2}\) to \(i^*=19.1\) for UV-excess QSOs and \(i^*=20.2\) for \(z>3\) QSOs \citep{richards_2002}.} &
\tblleft{1.45in}{Classical baseline showing that efficiency and completeness are always tied to magnitude, color, and redshift domain.} \\
\tblleft{1.15in}{SDSS DR1 photometric QSO catalog} &
\tblleft{1.55in}{Unresolved UV-excess sources in 2099 deg\(^{2}\) of SDSS imaging} &
\tblleft{2.25in}{100,563 candidates to \(g=21\); existing spectra indicated 97.6\% quasars, the catalog efficiency was estimated as 95.0\%, and completeness was 94.7\% for unresolved \(g\lesssim19.5\) UVX quasars in DR1 \citep{richards_2004}.} &
\tblleft{1.45in}{A high-efficiency photometric example, but restricted to UVX unresolved quasars and a specific DR1 reference set.} \\
XDQSO/XDQSOz and BOSS &
\tblleft{1.55in}{Probabilistic SDSS flux-density modeling and BOSS high-redshift QSO targeting} &
\tblleft{2.25in}{BOSS targeted fainter sources for Ly\(\alpha\)-forest science, requiring at least 15 \(z>2.15\) QSOs deg\(^{-2}\); the fixed CORE sample used about 20 targets deg\(^{-2}\), with roughly half confirmed as quasars, and an additional non-uniform BONUS layer used other information \citep{bovy_2011,bovy_2012,ross_2012}.} &
\tblleft{1.45in}{Shows why target density, redshift goals, and homogeneity can matter more than a single global purity number.} \\
\tblleft{1.20in}{BOSS/eBOSS QSO targeting} &
\tblleft{1.55in}{Survey-optimized optical, mid-infrared, and variability-assisted spectroscopy} &
\tblleft{2.25in}{eBOSS CORE was designed to recover at least 58 deg\(^{-2}\) quasars at \(0.9<z<2.2\) from about 90 deg\(^{-2}\) fibers, with SEQUELS-based expectations of about 68.4 deg\(^{-2}\) such quasars and about 95 deg\(^{-2}\) quasars at any redshift \citep{myers_2015}.} &
\tblleft{1.45in}{A fiber-survey example where yield per area and cosmology requirements define the operating point.} \\
\tblleft{1.20in}{DESI QSO target selection} &
\tblleft{1.55in}{Legacy Surveys optical photometry, WISE, and Random Forest targeting} &
\tblleft{2.25in}{The main selection covers \(16.5<r<23\); ultra-deep validation found a target mix of 71\% QSOs, 16\% galaxies, 6\% stars, and 7\% inconclusive spectra, while the DESI quasar catalog built from those spectra reaches \(>99\%\) purity for nominal \(\sim1000\) s exposures. At 310 targets deg\(^{-2}\), the selection yields \(>200\) QSOs deg\(^{-2}\), including about 60 \(z>2.1\) QSOs deg\(^{-2}\) \citep{chaussidon_2023}.} &
\tblleft{1.45in}{Closest large fiber-survey context; DESI optimizes survey yield, while this work optimizes a selected-field Gaia-linked candidate product.} \\
\tblleft{1.20in}{Gaia-CatWISE and Quaia} &
\tblleft{1.55in}{Gaia astrometry/photometry combined with infrared information} &
\tblleft{2.25in}{Gaia-CatWISE work reports quasar purities of 97\% or 96\% when applied to pure Gaia quasar-candidate tables with global or mixed priors \citep{hughes_2022}. Quaia starts from 6.65 million Gaia DR3 candidates, reduces contaminants by about a factor of four with Gaia and unWISE cuts, and releases 1,295,502 \(G<20.5\) quasars plus a cleaner 755,850-object \(G<20.0\) sample with selection-function models \citep{storey_fisher_2024}.} &
\tblleft{1.45in}{Strong all-sky Gaia-linked context; this work is a selected-field complement with explicit source-grouped validation and held-out source controls.} \\
Time-domain QSO catalogs &
\tblleft{1.55in}{Variability information combined with external multi-survey training and validation} &
\tblleft{2.25in}{Recent ZTF-based work illustrates the scale of variability-driven QSO catalog construction, but cadence, light-curve quality, and training-set definitions lead to a different selection function from the static Gaia--optical--infrared catalog used here \citep{nakoneczny_2025}.} &
\tblleft{1.45in}{A complementary channel; agreement supports candidate plausibility but is not a same-denominator performance comparison.} \\
\tblleft{1.20in}{This work} &
\tblleft{1.55in}{Selected Gaia-linked fields with Gaia, optical/infrared catalog features, \(E(B-V)\)-binned thresholds, and spectrum-informed training} &
\tblleft{2.25in}{At the recommended validation-calibrated target purity \(\ge 0.98\), P3 reaches measured test-set purity 0.9809 and spectroscopic-label completeness 0.8869 in the frozen benchmark. Under the same threshold protocol, Gaia official QSO probability gives completeness 0.4493; P3-Gaia completeness gains are +43.8 percentage points overall, +49.7 in the anti-center high-extinction slice, and +70.9 for Gaia \(G\ge20.5\) sources.} &
\tblleft{1.45in}{The main contribution in this comparison is high-purity recovery beyond a calibrated Gaia-only layer in controlled selected fields, with source-level validation and release-level selection-function metadata.} \\
\bottomrule
\end{tabular*}
\papertablenote{The numerical values for published studies are representative operating points reported by the cited works. They are not a direct ranking against the present catalog, because parent samples, magnitude limits, redshift ranges, truth labels, and survey goals differ. The final row gives the present paper's own fixed-benchmark operating point to make the comparison frame explicit.}
\end{table*}

Against this context, the present catalog is a selected-field, Gaia-linked high-purity product rather than a direct replacement for all-sky or survey-optimized QSO catalogs. The relevant comparison is a matched-operating-point Gaia-complement test, not a cross-survey ranking. At a measured test purity just above 0.98, P3 approximately doubles the Gaia-calibrated spectroscopic-label completeness in the frozen benchmark, from 0.4493 to 0.8869, while preserving candidate-realistic inference inputs and excluding Gaia official classifier probabilities from the student feature matrix. This is the central comparison result relative to the published context summarized above: the method provides high-purity recovery beyond a calibrated Gaia-only layer, especially where that layer leaves many follow-up candidates unrecovered, including the anti-center high-extinction and faint-source slices. The axes on which this release is not optimized are equally important: it is not designed to match large surveys in sky coverage, target-density optimization, redshift-specific selection, or all-sky selection-function modeling. The distinctive contribution is therefore a controlled selection-function product for follow-up: the parent sample, field layers, spectroscopic labels, source grouping, \(E(B-V)\)-binned thresholds, and release metadata are all fixed. This makes the method useful as a high-purity Gaia-complement layer in selected fields, while the paper remains explicit that broader superiority claims would require separate large-area validation and selection-function analysis.

\subsection{Contaminants and False Positives}\label{contaminants-and-false-positives}

The next limitation is contamination. The false-positive analysis separates two related but different questions. The first is the astronomical identity of the objects that enter QSO selections when a rule is transferred across sky domains. The second is whether additional morphology or hard-negative filters improve the released high-purity catalog enough to justify their completeness cost. Table \ref{tab:contaminant-diagnostics} summarizes these diagnostics. The sky-transfer tests show that many cross-domain false positives are spectroscopic stars, especially when a high-latitude rule is applied to the anti-center. The morphology diagnostics show a different tail: compact galaxy-like false positives can be reduced, but the largest reductions come with a substantial loss of QSO completeness.

\begin{figure}[htbp]
\centering
\includegraphics[width=0.90\textwidth]{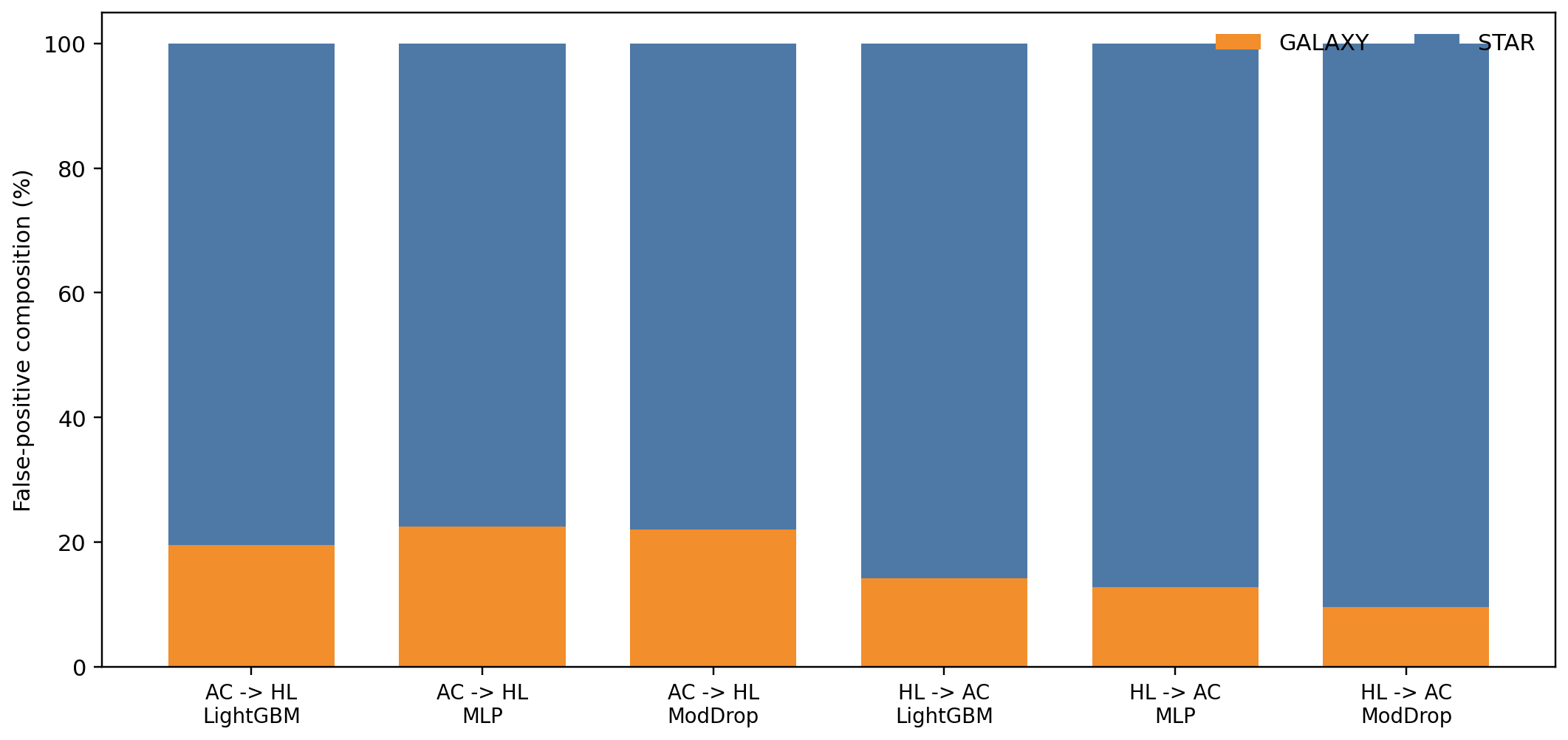}
\caption{False-positive composition in sky-transfer diagnostics. Bars show the spectroscopic labels of false positives when models trained in one sky regime are evaluated in another; AC denotes the anti-center field and HL denotes the high-latitude field. The diagnostic isolates domain-transfer failures and should not be read as the final catalog contaminant mixture. Stellar contaminants dominate the transfer failures, especially when a high-latitude-trained rule is applied to the anti-center field, while galaxy-labeled contaminants remain smaller but persistent. This separation motivates treating stellar foreground contamination and compact-galaxy contamination as distinct systematics in the release products.}
\label{fig:false-positive-taxonomy}
\end{figure}

The false positives and false negatives also reflect overlap in the non-spectroscopic catalog feature space. To quantify this effect without using test labels to define the neighborhoods, we compute a held-out nearest-neighbor diagnostic. Training-set sources in the four-field frozen benchmark define a standardized catalog feature space using only inference-time quantities: Gaia photometry and astrometry, optical/infrared fluxes and derived colors, and foreground reddening. Sky coordinates, spectroscopic quantities, labels, and Gaia official classifier probabilities are excluded. Each test-set source is then assigned the local QSO-label fraction among its 50 nearest training-set neighbors. A local fraction \(\geq0.80\) is called QSO-like, a fraction \(\leq0.20\) is called non-QSO-like, and intermediate values define a mixed catalog-space region. This is not an intrinsic physical separability limit; it is a conditional diagnostic for the present feature set, labels, fields, and measurement errors.

\begin{table}[htbp]
\centering
\caption{Catalog-space overlap diagnostic in the frozen test set.}
\label{tab:catalog-space-overlap}
\papertablesize
\setlength{\tabcolsep}{3.0pt}
\renewcommand{\arraystretch}{1.16}
\begin{tabular*}{0.98\textwidth}{@{\extracolsep{\fill}}llrrrr@{}}
\toprule
Spectroscopic label & \tblleft{1.40in}{Local catalog neighborhood} & \(N\) & \makecell[c]{Within label\\(\%)} & \makecell[c]{P3 selected\\\(N\) (\%)} & \makecell[c]{P3 error\\\(N\) (\%)} \\
\midrule
QSO & QSO-like & 2002 & 93.2 & 1864 (93.1) & FN 138 (6.9) \\
QSO & Mixed & 118 & 5.5 & 35 (29.7) & FN 83 (70.3) \\
QSO & Non-QSO-like & 28 & 1.3 & 6 (21.4) & FN 22 (78.6) \\
Non-QSO & QSO-like & 98 & 1.0 & 33 (33.7) & FP 33 (33.7) \\
Non-QSO & Mixed & 172 & 1.7 & 3 (1.7) & FP 3 (1.7) \\
Non-QSO & Non-QSO-like & 9711 & 97.3 & 1 (0.0) & FP 1 (0.0) \\
\bottomrule
\end{tabular*}
\papertablenote{Neighborhoods are defined by the local QSO-label fraction among the 50 nearest training-set sources in standardized inference-time catalog features. Test labels are used only after the neighborhoods are assigned. The P3 errors are false negatives for spectroscopic QSOs and false positives for spectroscopic non-QSOs. Percentages in the final two columns are within each row.}
\end{table}

\begin{figure}[htbp]
\centering
\includegraphics[width=0.90\textwidth]{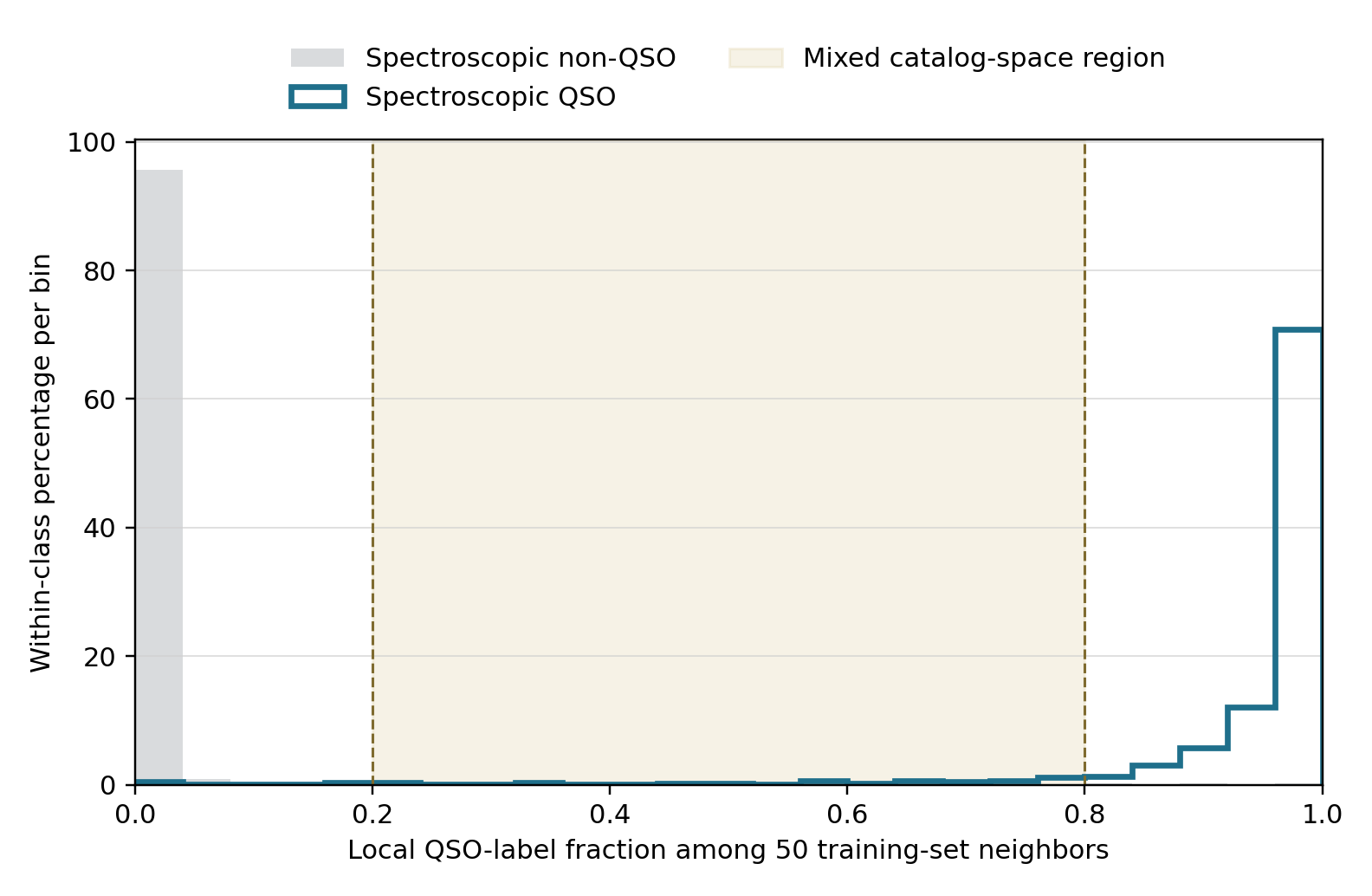}
\caption{Held-out catalog-space overlap diagnostic for the frozen test set. The horizontal axis gives the local QSO-label fraction among the 50 nearest training-set neighbors in standardized inference-time catalog features. The shaded interval marks the mixed region between 0.20 and 0.80; values above 0.80 are QSO-like and values below 0.20 are non-QSO-like. Most spectroscopic QSOs and non-QSOs occupy opposite ends of this catalog space, but the small overlapping tail accounts for a large fraction of the threshold errors: mixed or non-QSO-like spectroscopic QSOs account for 105 of 243 P3 false negatives, while QSO-like or mixed non-QSOs account for 36 of 37 P3 false positives.}
\label{fig:catalog-space-overlap}
\end{figure}

This diagnostic helps explain why the high-purity operating point has a measurable completeness cost even with a calibrated score. Most non-QSOs lie in non-QSO-like neighborhoods and almost never enter the P3 selection. Conversely, the rare non-QSOs that lie in QSO-like or mixed neighborhoods account for most of the remaining false positives. On the QSO side, 146 of 2148 spectroscopic QSOs, or 6.8\%, lie in mixed or non-QSO-like neighborhoods under the present catalog features, and these sources contribute 105 of the 243 false negatives. These objects are not necessarily physically unclassifiable; rather, they are difficult for the current Gaia--optical--infrared catalog projection. Deeper photometry, variability, morphology, X-ray, radio, or spectra could move some of them into a more separable representation.

\begin{table}[htbp]
\centering
\caption{Contaminant and auxiliary-filter diagnostics.}
\label{tab:contaminant-diagnostics}
\papertablesize
\setlength{\tabcolsep}{3.0pt}
\renewcommand{\arraystretch}{1.18}
\begin{tabular*}{0.98\textwidth}{@{\extracolsep{\fill}}lll@{}}
\toprule
Diagnostic & \tblleft{2.15in}{Main numerical result} & \tblleft{2.25in}{Interpretation for the release} \\
\midrule
\tblleft{1.65in}{Anti-center to high-latitude transfer} & \tblleft{2.15in}{STAR false positives 77.6--80.4\%; GALAXY false positives 19.6--22.4\% across LightGBM, MLP, and modality-dropout variants} & \tblleft{2.25in}{Stellar contaminants dominate this transfer failure mode; the result diagnoses foreground mismatch rather than compact-galaxy contamination alone.} \\
\tblleft{1.65in}{High-latitude to anti-center transfer} & \tblleft{2.15in}{STAR false positives 85.9--90.5\%; GALAXY false positives 9.5--14.1\%} & \tblleft{2.25in}{Moving a high-latitude rule into the anti-center greatly increases stellar contamination, consistent with the field-design motivation.} \\
\tblleft{1.65in}{Morphology penalty diagnostics} & \tblleft{2.15in}{Available auxiliary tests give only small operating-point changes after accounting for the default selection rule} & \tblleft{2.25in}{The effect supports a diagnostic flag but is not large enough to redefine the default rule without a dedicated morphology rerun under the same frozen split protocol.} \\
\tblleft{1.65in}{Morphology reranker diagnostics} & \tblleft{2.15in}{Aggressive morphology filtering can remove galaxy-like contaminants, but it also removes many true QSOs in the available tests} & \tblleft{2.25in}{The completeness cost is too large for the default catalog; morphology remains an optional inspection and future-filter axis.} \\
\tblleft{1.65in}{Hard-negative variant diagnostics} & \tblleft{2.15in}{The tested hard-negative variants diagnose difficult contaminants but do not provide a stable improvement over the default rule} & \tblleft{2.25in}{Hard-negative products are retained as auxiliary information rather than promoted into the released selection flag.} \\
\bottomrule
\end{tabular*}
\papertablenote{The diagnostics support keeping morphology and hard-negative products as optional flags or auxiliary information rather than as the default catalog selection rule.}
\end{table}

The resulting release strategy is to expose these diagnostics separately from the main rule. For applications requiring the most conservative high-purity subset, Gaia IPD/PSF-LSF quantities, Legacy Tractor morphology, and hard-negative auxiliary scores can be used as optional filters or as candidate-inspection features. The default release, however, keeps the P3 \(E(B-V)\)-calibrated selection unchanged because the present ablations either provide only modest gains or reduce completeness too strongly.

\FloatBarrier

\subsection{Extinction and Latitude Dependence}\label{extinction-and-latitude-dependence}

Foreground structure is handled more conservatively. Extinction enters the present selection function through threshold calibration rather than through an explicit physical correction to every feature. The three-bin \(E(B-V)\) policy is straightforward to audit and is supported by the selection-function checks in Section \ref{selection-function}. It is nevertheless an operating-point correction tied to the present fields and labels. Low-latitude and structured-foreground fields can change the balance of stellar contaminants, compact galaxies, infrared blending, and photometric uncertainties, so candidate densities in Taurus/Perseus and Orion/Monoceros are not used here as all-sky QSO surface-density measurements.

\subsection{Selection Function and Training-set Biases}\label{selection-function-and-training-set-biases}

These foreground and label issues set the scope of the selection function. The benchmark is not an all-sky unbiased QSO density measurement. Spectroscopic labels inherit SDSS/DESI/LAMOST selection functions, and the domain ladder was chosen for controlled comparison rather than completeness over the whole sky. The SDSS-DESI merge reduces random label sparsity and improves teacher stability, but it does not remove spectroscopic selection bias. DESI contributes most of the final P3 teacher cache, SDSS contributes a smaller but historically important reference component, and both surveys preferentially observe sources selected by their own target definitions. The resulting labels are therefore suitable for a controlled source-grouped benchmark and for high-purity candidate selection, but not for estimating an absolute QSO surface density without additional selection-function modeling. The P3 gain is also not uniform: it is modest in the five-seed mean, most apparent in several high-purity, higher-extinction, anti-center, and faint-end diagnostics, and unresolved or purity-limited in some smaller slices.

The released selection function is therefore empirical and conditional: it is defined by the input catalogs, field boundaries, source-grouped labels, score model, \(E(B-V)\)-binned thresholds, and recommended operating point. This definition is sufficient for follow-up candidate selection and for controlled comparison across the current fields, but absolute QSO number-density inference would require additional selection-function modeling.

\subsection{The COSMOS Deep-field Boundary Layer}\label{capability-boundaries-and-the-cosmos-deep-field-layer}

\begin{figure}[htbp]
\centering
\includegraphics[width=\textwidth]{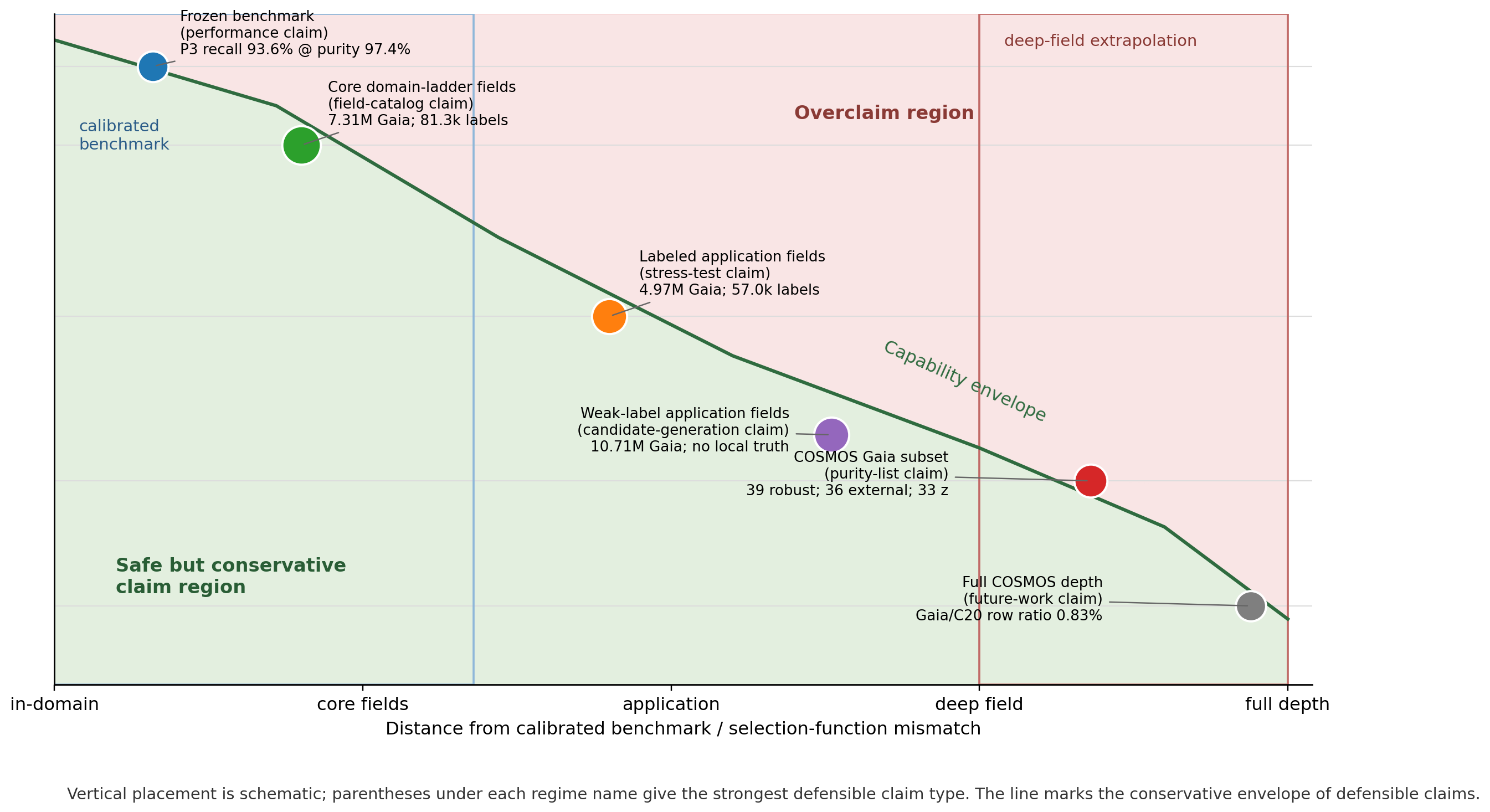}
\caption{Interpretation-domain envelope for the present catalog. The horizontal axis orders regimes by increasing distance from the frozen benchmark and by increasing selection-function mismatch. The vertical placement is schematic, not a measured performance axis; the parenthetical text under each regime names the most direct inference supported by the available evidence. Points near the upper left correspond to calibrated benchmark performance, where purity, completeness, and P3--P2 gain are directly measured. Application fields move rightward and downward because they support stress testing and candidate ranking but have weaker local truth coverage. The COSMOS Gaia subset lies farther to the right and supports a high-purity follow-up-priority list with independent deep-field evidence, not a local completeness measurement. The full COSMOS-depth point marks the COSMOS2020-scale population, which is largely outside the Gaia-linked source universe used by the present selector. The green region denotes conservative or directly supported interpretations at the indicated domain shift, whereas the pink region marks unsupported extrapolations, such as transferring benchmark completeness to an uncalibrated deep-field population.}
\label{fig:capability-envelope}
\end{figure}

The COSMOS analysis makes the deep-field selection boundary explicit. The same score can be applied to Gaia-linked sources, but the evidentiary meaning changes because COSMOS contains many AGN/QSO-like objects below the Gaia detection limit and has a different external-catalog mix. Figure \ref{fig:capability-envelope} summarizes this distinction as an interpretation-domain diagram rather than as another performance curve, and Table \ref{tab:cosmos-extrapolation-summary} gives the corresponding numerical boundary for this COSMOS run.

In this diagram, the frozen benchmark and the core domain ladder lie in the performance-calibration regime: spectroscopic labels, validation-derived thresholds, and fixed splits support direct purity/completeness statements. Labeled application fields still permit stress tests, but their role is more limited because they are not the frozen domain ladder. Weak-label application fields support candidate generation and external QC rather than formal performance statistics.

COSMOS lies beyond these regimes. The robust COSMOS subset has independent deep-field support: 36 of the 39 robust candidates have X-ray, radio, or spectroscopic evidence, and 33 have valid redshifts. This supports follow-up prioritization, but it also limits the interpretation. The Gaia cone samples only a small, bright subset of the full COSMOS source population, and the main-field optical feature coverage is shifted. The available evidence therefore supports a conservative Gaia-linked priority list in COSMOS, but not a measurement of COSMOS QSO completeness or total AGN surface density.

\begin{table}[htbp]
\centering
\caption{COSMOS Extreme Deep boundary summary.}
\label{tab:cosmos-extrapolation-summary}
\papertablesize
\setlength{\tabcolsep}{3.0pt}
\renewcommand{\arraystretch}{1.18}
\begin{tabular*}{0.98\textwidth}{@{\extracolsep{\fill}}lll@{}}
\toprule
Quantity & Value & \tblleft{3.15in}{Interpretation} \\
\midrule
\tblleft{2.15in}{Gaia DR3 sources in the COSMOS cone} & 7,968 & \tblleft{3.15in}{Candidate universe for the Gaia-linked catalog product.} \\
COSMOS2020 FARMER rows & 964,506 & \tblleft{3.15in}{Deep-field photometric source population; much deeper than Gaia.} \\
Gaia/COSMOS2020 row ratio & 0.83\% & \tblleft{3.15in}{Quantifies why COSMOS cannot be interpreted as a complete Gaia-based AGN census.} \\
Legacy matched Gaia sources & 45.2\% & \tblleft{3.15in}{Indicates partial availability of main-field optical survey features.} \\
CatWISE matched Gaia sources & 91.6\% & \tblleft{3.15in}{Provides broad mid-infrared support for Gaia-linked candidates.} \\
\tblleft{2.15in}{First global purity $\geq$ 0.97 candidates} & 59 & \tblleft{3.15in}{Initial high-purity COSMOS priority list from the main catalog score.} \\
\tblleft{2.15in}{Robust after Extreme Deep diagnostic check} & 39 & \tblleft{3.15in}{Conservative subset retained after COSMOS2020 HSC g/r/z diagnostic projection.} \\
\tblleft{2.15in}{External-supported robust candidates} & 36/39 & \tblleft{3.15in}{Candidates with X-ray, radio, or spectroscopic support.} \\
\tblleft{2.15in}{Robust candidates with valid redshift} & 33/39 & \tblleft{3.15in}{Redshift-supported subset, with median $z=1.729$ and range $0.851<z<2.735$.} \\
\bottomrule
\end{tabular*}
\papertablenote{These quantities define COSMOS as a high-purity follow-up-priority case. They do not measure completeness for the full COSMOS AGN/QSO population.}
\end{table}

The first three rows of the COSMOS summary table quantify the denominator mismatch: the method operates on Gaia-linked sources, whereas the deep COSMOS catalogs probe a far larger and fainter population. The middle rows describe the available feature domain, and the final rows summarize the validated subset. The robust 39-object subset is suitable for fiber follow-up, detailed candidate inspection, and a pilot application of the multimodal framework; it is not a completeness measurement for the COSMOS QSO population.

This distinction also guides future releases. A deeper COSMOS-specific extension could define a new denominator, for example Gaia-detected and multiwavelength-quality-selected sources, or a separate non-Gaia deep-field AGN candidate universe. Until such a denominator and local validation set are frozen, COSMOS is treated here as a scientifically informative boundary case and as a source of high-value follow-up candidates.

\subsection{Limitations and Future Improvements}\label{limitations-and-future-improvements}

Taken together, this release has four main limitations. First, the parent sample is Gaia-linked and restricted to selected fields. The catalog is therefore suitable for high-purity candidate selection and controlled field-to-field comparison, but it is not an all-sky QSO census and it does not define the full AGN population in deep fields such as COSMOS. Second, the validation labels are spectroscopic but not survey-random. They inherit the target-selection functions, depth limits, and class-assignment conventions of SDSS, DESI, and LAMOST; the resulting benchmark is well defined for the present source-grouped comparison, but not for absolute QSO surface-density inference. Third, the input feature coverage is heterogeneous. Legacy/WISE availability, foreground extinction, crowding, and compact-galaxy contamination all vary by field, so the recommended threshold flags must be interpreted together with the field label, \(E(B-V)\) bin, and input-coverage flags. Fourth, the released classifier is deliberately binary, QSO versus non-QSO. It does not attempt a final STAR/GALAXY/QSO public taxonomy.

Future larger-area and deep-field extensions require new sample definitions rather than only additional sky coverage. A larger-area Gaia-linked release would need a frozen parent sample, sky mask, feature-availability mask, and validation protocol before reporting selection-function quantities. It would also need to publish at least two linked products: a calibrated Gaia-reference selection and a P3-augmented selection with field- or extinction-aware thresholds. A deep-field release would require a different denominator, such as a COSMOS2020-anchored or multiwavelength-quality-selected parent sample, with local validation labels and depth-dependent thresholds. In this form, the present framework can be extended without mixing the calibrated Gaia-linked catalog product with a different deep-field selection problem.

COSMOS provides the natural test case for such a deep-field extension. A COSMOS-depth implementation would no longer be anchored to Gaia as the parent sample; it would instead begin from COSMOS2020 or an equivalent deep-field detection table. Gaia, Legacy, CatWISE, X-ray, radio, morphology, variability, photometric-redshift, and spectroscopic information would then enter as separate modality blocks. The target product would also be broader than a binary QSO flag, including QSO/AGN priority, evidence flags, redshift-quality information, compactness diagnostics, X-ray/radio support, contaminant flags, and selection-function metadata. Spectroscopic classifications, X-ray and radio AGN identifications, high-quality literature AGN catalogs, and deep-field photometric-redshift information would need to be treated as heterogeneous supervision sources rather than collapsed into a single unqualified truth label. Thresholds would then be recalibrated across magnitude, morphology, photometric-redshift range, depth, masks, and wavelength coverage.

\section{Discussion}\label{discussion}

The central result is that spectrum-informed training can improve a catalog-only QSO selector without requiring spectra for future candidates. This separation is important for a survey catalog: the spectroscopic information is used to shape the score during training, while the released selection rule remains applicable to sources with Gaia and multiwavelength catalog measurements alone. In the fixed-split evaluation, the gains are modest but useful at the stricter high-purity operating point, especially in several higher-extinction or faint-end diagnostics. These gains should be read as evidence that expanded spectrum-informed supervision can improve the downstream catalog score under a fixed protocol, not as a claim that model architecture is the primary explanation for the P3-P2 difference.

Operationally, the result is most relevant to fiber-spectroscopic follow-up. A QSO candidate catalog is an input to a finite observing system in which target priorities, fiber assignment, field geometry, magnitude limits, and contaminant rates all determine scientific yield. The catalog-student score and threshold flags are therefore intended to support target-list construction as well as retrospective accuracy reporting. In that context, the Gaia comparison provides a reference level: a corresponding catalog comparison needs to quantify how many additional candidates are supplied beyond a calibrated Gaia-only selection and at what purity cost.

Scientifically, the field design is also part of the result. The core ladder supports the main validation claim, and the application fields test the method where foregrounds and label coverage are less uniform. COSMOS adds a third role: it is an Extreme Deep layer that tests how the Gaia-linked catalog behaves when the external validation resources are excellent but the full astrophysical source population is much deeper than Gaia. This layered structure is more informative than a single pooled metric because it separates the fields that define the validated selection function from those used to test candidate generation and from the field used to map the boundary of the present catalog.

Methodologically, the present evidence does not justify adding all diagnostic products to the default catalog. Gaia IPD, Legacy morphology, and hard-negative reranking are valuable for understanding the false-positive population, but the current ablations show only modest default gains or substantial completeness losses at stricter cuts. They are therefore retained as optional flags and future high-purity filters rather than incorporated into the main selection rule.

A further prospective use of this work is in the training and evaluation of astronomical machine-learning models. Because the framework connects heterogeneous survey measurements with spectrum-informed supervision while preserving source-grouped splits, calibration metadata, and selection-function information, it provides a controlled setting for studying multimodal representation learning, teacher-student distillation, domain adaptation, missing-modality robustness, and calibrated candidate ranking. However, the catalog should not be treated as an unbiased ground-truth training set: its use for model development must account for the field definition, spectroscopic incompleteness, extinction-dependent thresholds, input-coverage patterns, and the distinction between high-purity candidate selection and population-level completeness.

Beyond the optical/infrared Gaia-linked setting explored in this work, the method has a well-defined extension path to radio astronomy, but that extension should be viewed as a new validated application rather than a direct reuse of the present trained selector. Modern radio surveys and facilities, including the VLA/VLASS, LOFAR/LoTSS, ASKAP/EMU, MeerKAT/MIGHTEE, and future SKA programs, operate with observables that are partly analogous to optical/infrared catalog quantities and partly distinct from them \citep{lacy_2020_vlass,shimwell_2022_lotss,norris_2011_emu,jarvis_2016_mightee,braun_2015_ska}. Table \ref{tab:radio-domain-extension} summarizes the most relevant quantities. The practical opportunity is not simply to match radio and optical sources by position. Rather, once counterpart association is handled with explicit astrometric and morphological uncertainty, radio measurements could enter as an additional modality block alongside optical, infrared, X-ray, spectroscopic, and astrometric information \citep{budavari_2008}. Such a framework could support radio-source classification, QSO-counterpart prioritization, anomaly searches, and selection of radio-detected QSOs useful for reference-frame or follow-up programs. The main limitations are equally important: radio angular resolution and morphology differ strongly between surveys, observations are often non-simultaneous, spectral indices depend on frequency coverage, and extended or multi-component radio sources can have ambiguous optical counterparts. Dedicated radio-domain validation is therefore required before reporting radio-selection purity, completeness, or surface-density quantities.

\begin{table*}[htbp]
\centering
\caption{Radio-domain quantities relevant to a possible framework extension.}
\label{tab:radio-domain-extension}
\papertablesize
\setlength{\tabcolsep}{1.0pt}
\renewcommand{\arraystretch}{1.18}
\begin{tabular*}{\textwidth}{@{\extracolsep{\fill}}llll@{}}
\toprule
\tblleft{0.90in}{Quantity} & \tblleft{1.35in}{Native radio measurement} & \tblleft{2.05in}{Connection to optical/infrared QSO work} & \tblleft{1.95in}{Required caution} \\
\midrule
Flux density & \tblleft{1.35in}{Jy, mJy, or $\mu$Jy at a stated frequency} & \tblleft{2.05in}{Optical/infrared magnitudes can be converted to flux density for SED or radio-loudness comparisons, while retaining the original magnitude columns.} & \tblleft{1.95in}{Bandpass, variability, and K-correction differences prevent a one-to-one comparison with optical brightness.} \\
Spectral index & \tblleft{1.35in}{$S_\nu \propto \nu^\alpha$ from multi-frequency radio data} & \tblleft{2.05in}{Provides a radio-continuum analogue of color information and helps separate flat-spectrum compact cores from steep-spectrum lobes.} & \tblleft{1.95in}{Requires matched frequencies and compatible angular resolution; non-simultaneous data can bias $\alpha$.} \\
Morphology & \tblleft{1.35in}{Component size, axis ratio, compactness, and multi-component structure} & \tblleft{2.05in}{Can identify compact radio-QSO cores or extended radio galaxies that are not evident from optical colors alone.} & \tblleft{1.95in}{Extended lobes, blended components, and survey-dependent deblending make source-level association nontrivial.} \\
\makecell[l]{Polarization/\\compactness} & \tblleft{1.35in}{Fractional polarization, Stokes parameters, or brightness-temperature proxies} & \tblleft{2.05in}{Can add evidence for jet-dominated AGN and compact reference-frame sources.} & \tblleft{1.95in}{Availability is survey-specific and often shallower than total-intensity catalogs.} \\
Variability & \tblleft{1.35in}{Epoch-to-epoch radio flux changes} & \tblleft{2.05in}{Complements optical variability for AGN activity and follow-up prioritization.} & \tblleft{1.95in}{Cadence, calibration systematics, and non-simultaneity with optical/infrared data must be modeled.} \\
Counterpart association & \tblleft{1.35in}{Radio component-to-source grouping and optical/IR/X-ray match probability} & \tblleft{2.05in}{Defines the source identity before any joint representation or teacher-student training is meaningful.} & \tblleft{1.95in}{This is itself a statistical inference problem, especially for extended or multi-component radio sources.} \\
\bottomrule
\end{tabular*}
\papertablenote{The table lists native radio units explicitly because radio flux density and optical/infrared magnitudes are measured on different systems. Magnitude-to-flux conversion is useful for cross-domain comparison, but the original survey units, bandpasses, and epochs should be preserved in any released catalog.}
\end{table*}

\section{Summary}\label{summary}

We have presented an extinction-calibrated QSO candidate catalog, selection function, and reproducible catalog-construction pipeline for selected fields. The framework uses spectra for source-grouped supervision and teacher construction, while the released candidate score is computed from Gaia and multiwavelength catalog measurements alone.

\begin{enumerate}
\def\labelenumi{\arabic{enumi}.}
\tightlist
\item
  The input data model separates standard survey measurements, spectroscopic truth labels, and value-added literature catalogs. Gaia DR3 defines the source frame, SDSS/DESI/LAMOST provide the supervised spectroscopic backbone, and Gaia official QSO probability is retained as an external reference rather than as a student feature.
\item
  The field design is layered. The core domain ladder contains four fields, 7.31 million Gaia sources, 81,347 spectroscopic labels, and 14,677 QSO labels, and defines the main calibrated performance baseline. Four application/stress-test fields add 15.68 million Gaia sources and 57,015 labels for portability tests. COSMOS is treated separately as an Extreme Deep layer, because its deep multiwavelength source population is not defined by the Gaia-linked parent sample.
\item
  The catalog construction is organized at the Gaia-source level. Cross-matching, train/validation/test splitting, repeated-spectrum handling, union-best teacher construction, threshold calibration, and QC/provenance checks are all tied to the same source identity, making the released selection rule reproducible and auditable.
\item
  At the recommended conservative operating point, calibrated to a validation-set purity target of 0.98, the P3 catalog selector reaches spectroscopic-label completeness 0.8869 at measured test-set purity 0.9809, compared with Gaia official spectroscopic-label completeness 0.4493 under the same calibration protocol. The corresponding compact slice summary gives P3-Gaia spectroscopic-label completeness gains of +43.8 percentage points for the full benchmark, +49.7 percentage points in the anti-center high-extinction slice, and +70.9 percentage points for Gaia G \(\ge 20.5\) sources.
\item
  Fixed-seed and bootstrap checks show that the P3-P2 gain is modest at the stricter operating point and should be interpreted with slice-level uncertainty. The five-seed all-sample P3-P2 completeness gain is \(0.0090 \pm 0.0097\), with the largest measured improvements in the anti-center, higher-extinction, mid-extinction bridge, and faint-source diagnostics and a small purity/false-positive cost.
\item
  The released catalog is intended as a catalog and selection-function product. It provides identifiers and positions, field-layer assignments, input-feature coverage, catalog features, calibrated scores, \(E(B-V)\)-binned threshold flags, validation metadata, and provenance/QC information. The associated intermediate products and source code are intended to support independent catalog assembly and method development.
\item
  Relative to Gaia official QSO classification, the method is interpreted here as a calibrated recovery layer. Gaia supplies the all-sky reference classifier, while the P3 selector identifies additional high-purity candidates in selected fields where optical/infrared information, source-grouped spectrum supervision, and local threshold calibration add value.
\item
  In COSMOS, the present catalog is interpreted as a purity-oriented priority list for Gaia-linked sources. The Gaia-linked parent sample contains 7,968 sources, whereas COSMOS2020 FARMER contains 964,506 rows in the same field; this scale difference defines the present completeness boundary. The robust subset contains 39 candidates, of which 36 have X-ray, radio, or spectroscopic support and 33 have valid redshifts, supporting follow-up prioritization rather than a complete COSMOS QSO/AGN census.
\item
  A catalog-space overlap diagnostic explains part of the remaining error floor. About 6.8\% of spectroscopic QSOs in the frozen test split occupy locally mixed or non-QSO-like neighborhoods in the deployable catalog feature space; this subset contains 105 of 243 false negatives and 36 of 37 false positives at the conservative P3 operating point. This indicates that some errors reflect limited separability in the available pre-spectroscopic measurements, not only model optimization.
\item
  The remaining false-positive population has both stellar-foreground and compact-galaxy components. Sky-transfer diagnostics are star-dominated, while morphology/IPD and hard-negative diagnostics identify a compact-galaxy-like high-purity tail. These diagnostics can support optional filtering and future releases, but they are not promoted into the default selection rule because the available auxiliary tests do not yet provide a stable improvement under the same frozen split protocol.
\end{enumerate}

\begin{acknowledgments}
We thank China Manned Space Engineering and the Technology and Engineering Center for Space Utilization, Chinese Academy of Sciences, for providing data used in this study. This work was supported by the National Key Research and Development Program of China (grant Nos. 2025YFF0511000 and 2025YFF0510602), the National Natural Science Foundation of China (grant Nos. 12273075 and 12473070), the China Manned Space Project (grant Nos. CMS-CSST-2025-A19, CMS-CSST-2021-A12, and CMS-CSST-2021-B10), the International Partnership Program of the Chinese Academy of Sciences (grant No.~018GJHZ2023110GC), and the Talent Plan of the Shanghai Branch, Chinese Academy of Sciences (No.~CASSHB-QNPD-2023-016). We also acknowledge Zhejiang Lab, Zhejiang Province, China, for support of this work.

During manuscript preparation, the authors used ChatGPT \citep{openai_chatgpt_2026} for language editing and readability improvement only.
\end{acknowledgments}

\appendix

\section{Auxiliary Galaxy Classification Test}\label{appendix-auxiliary-galaxy-classification-test}

This appendix reports a framework-extension test for a GALAXY-versus-non-GALAXY task. It uses the same source-grouped, catalog-only student protocol as the main QSO selector and the same P3 spectrum-teacher cache, with downstream validation/test sources excluded from teacher fitting and checkpoint selection. The teacher galaxy probability is used only for downstream training rows. The purpose is to test whether the controlled training and threshold-calibration protocol can support an auxiliary galaxy-candidate layer in the same selected fields. It is not intended as a definitive public STAR/GALAXY/QSO taxonomy, nor is it presented as a replacement for purpose-built galaxy catalogs. COSMOS is intentionally excluded from this auxiliary test and from the auxiliary STAR/GALAXY catalog products.

The common held-out-source check for the auxiliary GALAXY and STAR tasks is summarized in Table \ref{tab:aux-star-galaxy-heldout-source}. The check uses the frozen Gaia-linked label table with 138,362 rows and the P3 spectrum-teacher cache with 100,143 source-grouped teacher rows. In both auxiliary tasks, the overlap between the teacher cache and the downstream benchmark is 43,606 sources, but all downstream validation and test \texttt{source\_id} values are excluded from teacher fitting and checkpoint selection. Teacher probabilities for downstream validation and test rows are retained only for diagnostics, not as student-training targets. The student feature matrix also excludes Gaia official classifier probabilities; \texttt{classprob\_dsc\_combmod\_galaxy} and \texttt{classprob\_dsc\_combmod\_star} are used only as Gaia DR3 reference baselines.

\begin{table}[htbp]
\centering
\caption{Auxiliary STAR/GALAXY held-out-source check.}
\label{tab:aux-star-galaxy-heldout-source}
\papertablesize
\setlength{\tabcolsep}{3.0pt}
\renewcommand{\arraystretch}{1.15}
\begin{tabular*}{0.98\textwidth}{@{\extracolsep{\fill}}lrrrrr@{}}
\toprule
Target & \makecell[c]{COSMOS\\rows} & \makecell[c]{Downstream\\rows} & \makecell[c]{Teacher\\overlap} & \makecell[c]{Val/test in\\teacher fit} & \makecell[c]{Soft-target\\train rows} \\
\midrule
GALAXY & 0 & 138362 & 43606 & 0 & 30454 \\
STAR & 0 & 138362 & 43606 & 0 & 30454 \\
\bottomrule
\end{tabular*}
\papertablenote{The check confirms that no downstream validation/test Gaia \texttt{source\_id} rows appear in teacher train/validation splits. Soft-target train rows are the downstream training rows for which the P3 teacher probability is available and used by the student.}
\end{table}

Table \ref{tab:aux-galaxy-gaia-baseline} compares the auxiliary GALAXY student with the Gaia DR3 official galaxy probability under the same validation-calibrated \(E(B-V)\)-binned threshold protocol. The threshold-policy label \texttt{purity\_ge\_098} denotes the validation-set target, not a guaranteed test-set purity. On the frozen test split, the P3 auxiliary galaxy student reaches test purity 0.9756 and spectroscopic-label completeness 0.8531 in the pooled benchmark, compared with Gaia DR3 completeness 0.3419 at purity 0.9688. The primary signal is therefore increased recovery at comparable purity, not a demonstrated 0.98-purity galaxy catalog. Across five downstream seeds, the P3 auxiliary galaxy student gives mean test purity 0.9776, mean spectroscopic-label completeness 0.8467, and mean F1 0.9075 in the pooled benchmark.

\begin{table*}[htbp]
\centering
\caption{Auxiliary GALAXY performance compared with the Gaia DR3 galaxy baseline.}
\label{tab:aux-galaxy-gaia-baseline}
\papertablesize
\setlength{\tabcolsep}{4.0pt}
\renewcommand{\arraystretch}{1.15}
\begin{tabular*}{0.98\textwidth}{@{\extracolsep{\fill}}llrrrrrrrr@{}}
\toprule
Model & Slice & $N$ & Galaxy & Purity & Comp. & F1 & Selected & FP & FN \\
\midrule
Gaia DR3 & all & 20578 & 2907 & 0.9688 & 0.3419 & 0.5055 & 1026 & 32 & 1913 \\
Hard-label student & all & 20578 & 2907 & 0.9750 & 0.8442 & 0.9049 & 2517 & 63 & 453 \\
P3 student & all & 20578 & 2907 & 0.9756 & 0.8531 & 0.9103 & 2542 & 62 & 427 \\
Gaia DR3 & core ladder & 12129 & 1354 & 0.9685 & 0.3183 & 0.4792 & 445 & 14 & 923 \\
Hard-label student & core ladder & 12129 & 1354 & 0.9764 & 0.8264 & 0.8952 & 1146 & 27 & 235 \\
P3 student & core ladder & 12129 & 1354 & 0.9741 & 0.8323 & 0.8977 & 1157 & 30 & 227 \\
Gaia DR3 & application & 8449 & 1553 & 0.9690 & 0.3625 & 0.5276 & 581 & 18 & 990 \\
Hard-label student & application & 8449 & 1553 & 0.9737 & 0.8596 & 0.9131 & 1371 & 36 & 218 \\
P3 student & application & 8449 & 1553 & 0.9769 & 0.8712 & 0.9210 & 1385 & 32 & 200 \\
\bottomrule
\end{tabular*}
\papertablenote{Metrics use seed 42, \(E(B-V)\)-binned validation thresholds, and the \texttt{purity\_ge\_098} validation target. Comp. denotes spectroscopic-label completeness within the frozen Gaia-linked benchmark. The application slice contains the non-COSMOS application/stress fields present in the frozen label table.}
\end{table*}

The corresponding auxiliary galaxy product contains 17,752 selected benchmark rows, stored in both Parquet and compressed CSV formats. Each row retains the Gaia \texttt{source\_id}, sky position, field label, frozen split, spectroscopic provenance where available, P3 galaxy score, applied \(E(B-V)\)-bin threshold, and the auxiliary high-purity flag. Because the product is selected from the frozen benchmark table, the train and validation rows support reproducibility and calibration checks, whereas the test rows provide the independent benchmark evidence.

\section{Auxiliary Star Classification Test}\label{appendix-auxiliary-star-classification-test}

The same protocol was repeated for a STAR-versus-non-STAR task. This auxiliary test has a different role from the GALAXY test because stars dominate the labeled sample and are the main foreground contaminant in several QSO transfer diagnostics. The auxiliary STAR score is therefore useful both as a possible stellar-candidate layer and as a foreground-screening diagnostic for future releases. COSMOS remains excluded here, as in Appendix \ref{appendix-auxiliary-galaxy-classification-test}.

Table \ref{tab:aux-star-gaia-baseline} compares the auxiliary STAR student with the Gaia DR3 official star probability under the matched validation-threshold protocol. In the pooled frozen test split, Gaia DR3 is already highly pure, with purity 0.9825, but recovers only 0.5567 of the spectroscopic STAR labels at the validation-calibrated high-purity threshold. The P3 auxiliary STAR student reaches purity 0.9830 and spectroscopic-label completeness 0.9890 in the same pooled benchmark. The five-seed summary is similarly stable: the P3 auxiliary STAR student has mean test purity 0.9827, mean spectroscopic-label completeness 0.9895, and mean F1 0.9861 in the pooled benchmark. The STAR task is therefore the more stable of the two auxiliary tests; the method recovers a substantially larger fraction of spectroscopic stars than Gaia at comparable purity.

\begin{table*}[htbp]
\centering
\caption{Auxiliary STAR performance compared with the Gaia DR3 star baseline.}
\label{tab:aux-star-gaia-baseline}
\papertablesize
\setlength{\tabcolsep}{4.0pt}
\renewcommand{\arraystretch}{1.15}
\begin{tabular*}{0.98\textwidth}{@{\extracolsep{\fill}}llrrrrrrrr@{}}
\toprule
Model & Slice & $N$ & Star & Purity & Comp. & F1 & Selected & FP & FN \\
\midrule
Gaia DR3 & all & 20578 & 14325 & 0.9825 & 0.5567 & 0.7107 & 8117 & 142 & 6350 \\
Hard-label student & all & 20578 & 14325 & 0.9820 & 0.9897 & 0.9858 & 14437 & 260 & 148 \\
P3 student & all & 20578 & 14325 & 0.9830 & 0.9890 & 0.9860 & 14413 & 245 & 157 \\
Gaia DR3 & core ladder & 12129 & 8627 & 0.9877 & 0.4852 & 0.6508 & 4238 & 52 & 4441 \\
Hard-label student & core ladder & 12129 & 8627 & 0.9852 & 0.9882 & 0.9867 & 8653 & 128 & 102 \\
P3 student & core ladder & 12129 & 8627 & 0.9865 & 0.9879 & 0.9872 & 8640 & 117 & 104 \\
Gaia DR3 & application & 8449 & 5698 & 0.9768 & 0.6650 & 0.7913 & 3879 & 90 & 1909 \\
Hard-label student & application & 8449 & 5698 & 0.9772 & 0.9919 & 0.9845 & 5784 & 132 & 46 \\
P3 student & application & 8449 & 5698 & 0.9778 & 0.9907 & 0.9842 & 5773 & 128 & 53 \\
\bottomrule
\end{tabular*}
\papertablenote{Metrics use seed 42, \(E(B-V)\)-binned validation thresholds, and the \texttt{purity\_ge\_098} validation target. Comp. denotes spectroscopic-label completeness within the frozen Gaia-linked benchmark.}
\end{table*}

\begin{figure*}[htbp]
\centering
\includegraphics[width=\textwidth]{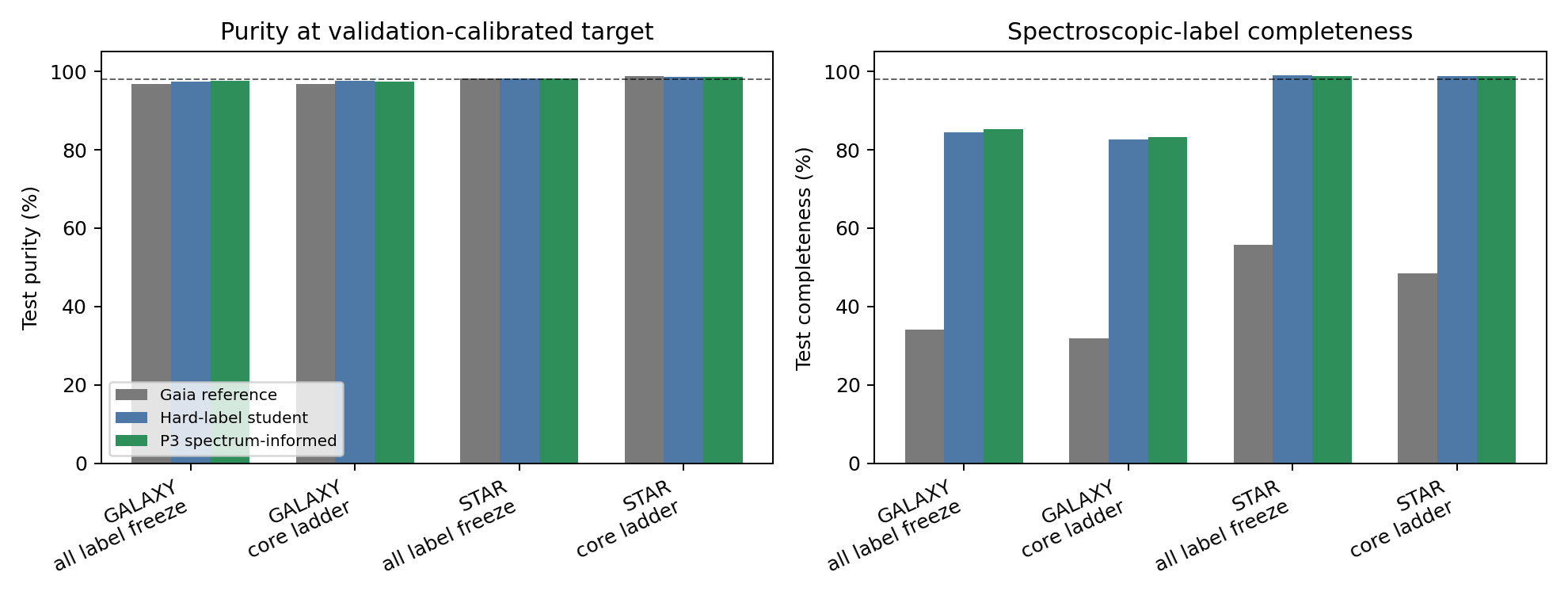}
\caption{Auxiliary GALAXY and STAR performance at the validation-calibrated \texttt{purity\_ge\_098} operating point. The left panel gives measured test purity in percent, and the right panel gives spectroscopic-label completeness in percent. The dashed horizontal line marks 98\% for visual reference. Gaia DR3 denotes the official Gaia class probability for the corresponding class, used only as an external reference baseline; the hard-label and P3 students use catalog-level features and exclude Gaia official classifier probabilities from the feature matrix. The principal gain is the increased recovery of spectroscopic labels at comparable purity, especially for STAR; the GALAXY result should be described conservatively because its measured test purity remains below 98\% in the pooled and core-ladder slices.}
\label{fig:aux-star-galaxy-performance}
\end{figure*}

Figure \ref{fig:aux-star-galaxy-performance} summarizes the matched-threshold comparison for the auxiliary GALAXY and STAR tasks. Purity and completeness are plotted in percent. The comparison indicates that the principal gain over Gaia DR3 is recovery at comparable purity, rather than a large purity increase. This distinction is important for interpretation: the auxiliary products are candidate-priority and benchmark products, not population-complete catalogs.

\begin{figure*}[htbp]
\centering
\includegraphics[width=\textwidth]{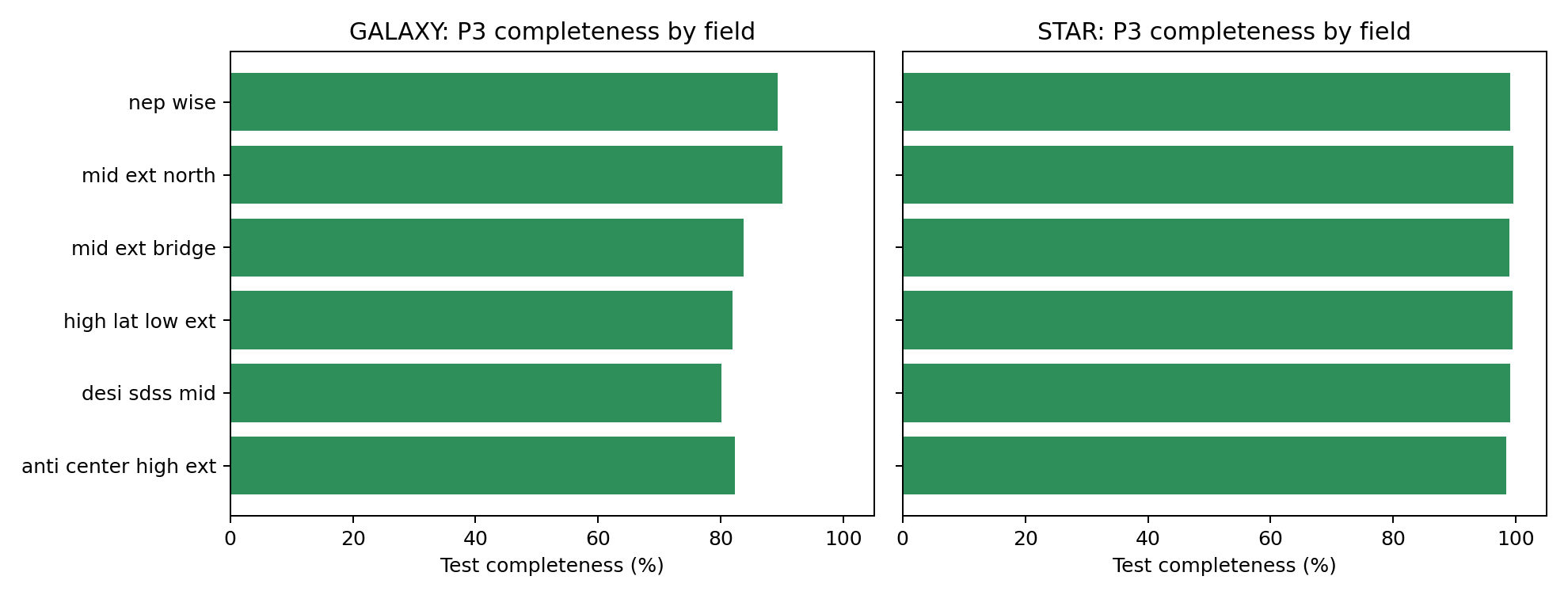}
\caption{Field-level spectroscopic-label completeness for the P3 auxiliary GALAXY and STAR students. Values are plotted in percent at the same \(E(B-V)\)-binned validation-calibrated operating point as Figure \ref{fig:aux-star-galaxy-performance}. The GALAXY task shows useful but field-dependent recovery, while the STAR task reaches high completeness across the non-COSMOS fields in the frozen benchmark.}
\label{fig:aux-star-galaxy-regional-completeness}
\end{figure*}

The auxiliary star product contains 96,667 selected benchmark rows in Parquet and compressed CSV formats, with the same schema style as the auxiliary galaxy product. Both auxiliary products retain source identifiers, field-layer tags, split labels, score and threshold metadata, and spectroscopic provenance fields where available. They can support future release checks and contaminant diagnostics, but they do not change the main paper's default scientific product, which remains the high-purity QSO candidate catalog and its empirical selection function.

\clearpage

\bibliographystyle{aasjournalv7}
\bibliography{references}

\end{document}